\documentclass[twocolumn]{aastex631} %for arxiv
\usepackage{amsmath}
\usepackage{amssymb}
\usepackage{hyperref}
\usepackage{soul}

\begin{document}

%\title{Revisiting Early TeV gamma-ray Afterglow of GRB221009A: External Inverse Compton Scattering off Prompt Emission}
%\title{Effect of External Inverse Compton Scattering off the Prompt Emission on the early TeV  Afterglow of GRB221009A}
\title{On the External Inverse Compton Scattering off the Prompt Emission in GRB 221009A}

\author[0000-0002-0170-0741]{Cui-Yuan Dai}
\affiliation{School of Astronomy and Space Science, Nanjing University, Nanjing 210023, China; ryliu@nju.edu.cn; xywang@nju.edu.cn}
\affiliation{Key Laboratory of Modern Astronomy and Astrophysics (Nanjing University), Ministry of Education, Nanjing
210023, China}

\author[0000-0001-5751-633X]{Jian-He Zheng}
\affiliation{School of Astronomy and Space Science, Nanjing University, Nanjing 210023, China; ryliu@nju.edu.cn; xywang@nju.edu.cn}
\affiliation{Key Laboratory of Modern Astronomy and Astrophysics (Nanjing University), Ministry of Education, Nanjing
210023, China}

\author[0000-0003-3659-4800]{Xiao-Hong Zhao}
\affiliation{Yunnan Observatories, Chinese Academy of Sciences Kunming, 650011, People’s Republic of China}
\affiliation{Key Laboratory for the Structure and Evolution of Celestial Objects, Chinese Academy of Science Kunming, 650011, People’s Republic of China }
\affiliation{Center for Astronomical Mega-Science, Chinese Academy of Science 20A Datun Road, Chaoyang District 
Beijing 100012, People’s Republic of China}

\author[0000-0003-1576-0961]{Ruo-Yu Liu}
\affiliation{School of Astronomy and Space Science, Nanjing University, Nanjing 210023, China; ryliu@nju.edu.cn; xywang@nju.edu.cn}
\affiliation{Key Laboratory of Modern Astronomy and Astrophysics (Nanjing University), Ministry of Education, Nanjing
210023, China}

\author[0000-0002-5881-335X]{Xiang-Yu Wang}
\affiliation{School of Astronomy and Space Science, Nanjing University, Nanjing 210023, China; ryliu@nju.edu.cn; xywang@nju.edu.cn}
\affiliation{Key Laboratory of Modern Astronomy and Astrophysics (Nanjing University), Ministry of Education, Nanjing
210023, China}

\begin{abstract}

The light curve of the TeV emission in GRB 221009A displays a smooth transition from an initial rapid rise to a slower rise and eventually a decay phase. The smooth temporal profile of the TeV emission suggests that it mainly results from an external shock. The temporal overlap between the prompt KeV-MeV emission and the early TeV afterglow indicates that external inverse Compton scattering (EIC) between the prompt KeV-MeV photons and the afterglow electrons is inevitable. Since the energy density of the prompt emission is much higher than that of the afterglow during the early phase, the EIC process dominates the cooling of afterglow electrons. The EIC scattering rate is influenced by the anisotropy of the seed photon field, which depends on the radii of the internal dissipation ($R_{\rm dis}$), where the prompt emission is produced, and that of the external shock ($R_{\rm ext}$), where the afterglow emission is produced. We investigate the EIC process for different values of $R_{\rm dis}/R_{\rm ext}$. We find that, for varying \( R_{\rm dis}/R_{\rm ext} \), the EIC scattering rate can differ by a factor of $\sim 2$. For GRB 221009A, the EIC emission is dominated during the early rising phase of the TeV afterglow. It then transitions to a phase dominated by the synchrotron self-Compton (SSC) emission  as the intensity of the prompt emission decreases. Additionally, we investigate the effect of $\gamma \gamma$ absorption in the TeV afterglow caused by prompt MeV photons and find that it is insufficient to explain the early rapid rise in the TeV afterglow, even in the case of $R_{\rm dis}/R_{\rm ext} \sim 1$.

% {\bf The approximate transition time from EIC to SSC dominance in the TeV light curve of GRB 221009A can be inferred from MeV observations.}

\end{abstract}

\section{Introduction} \label{sec:intro}

The brightest-of-all-time (BOAT) GRB 221009A, triggered on 2022 October 9 at 13:16:59.99 UTC ($T_0$), exhibits bright MeV emission from approximately $T_0 + 175\, \rm s$ to $T_0 + 330 \, \rm s$ \citep{An2023,Lesage2023,Frederiks2023,LHAASO2023}. The TeV afterglow, characterized by a smooth light curve, begins at $T^* = T_0 + 226 \, \rm s$ and initially displays a rapid rise, transitioning to a slower rise around $T^*+5 \, \rm s$ until peaking at $T^*+18 \, \rm s$. Following the peak, the afterglow enters a gradual decay phase \citep{LHAASO2023}. The time overlap between the early TeV afterglow and  the prompt MeV emission suggests that the external shock is immersed by the prompt keV/MeV photons during this period. Hence, keV/MeV photons are inevitably scattered by the relativistic electrons accelerated by the external shock, a process known as EIC \citep{wang2006,fan2006,Fan2008,Panaitescu2008,Murase2010,Murase2011}. This additional photon field dominates the electron cooling early on because its energy density is higher than that of the magnetic field and the radiation field of the synchrotron afterglow. As a result, the energy of relativistic electrons is mainly transferred to photons through the EIC process instead of the synchrotron and SSC radiation.

In contrast to the SSC process, where the incident photons are always isotropically distributed (in the jet's comoving frame), the anisotropy of the prompt emission is influenced by the ratio between the internal dissipation radius $R_{\rm dis}$ and the external shock radius $R_{\rm ext}$, which in turn affects the EIC scattering rate. However, $R_{\rm dis}$ is highly uncertain. In internal shock models (e.g., \citep{Meszaros1994, Kobayashi1997, Daigne1998}), the typical dissipation radius is $\sim 10^{14} \, {\rm cm} \left( \frac{\Gamma_{\rm dis}}{500} \right)^2 \left( \frac{\Delta t_{\rm v}}{0.01 \, \rm s} \right)$, where $\Gamma_{\rm dis}$ is the Lorentz factor of the faster shell involved in the collision and $\Delta t_{\rm v}$ is the variability time of the light curve. On the other hand, in magnetic field dissipation scenarios (such as the Internal-Collision-induced Magnetic Reconnection and Turbulence (ICMART) model, \cite{zhang2011}) or magnetic turbulence scenarios \citep{Lyutikov2003, Lyutikov2006, Kumar2009}, the prompt emission is generated at a much larger radius $R_{\rm dis} \gtrsim 10^{15-16} \, \rm cm$ \citep{zhang2011}, or possibly even at the deceleration radius of the external shock $R_{\rm dec}$ \citep{Lyutikov2003, Lyutikov2006, Kumar2009}. In this paper, we compare the EIC scattering rate across different radii, ranging from $R_{\rm dis} \ll R_{\rm ext}$ to $R_{\rm dis} \sim R_{\rm ext}$. We find that the scattering rate along the LOS differs by only a factor of \(\sim 2\) for different values of \(R_{\rm dis}/R_{\rm ext}\). Furthermore, our analysis shows that the incident photons can be treated as quasi-anisotropic if $R_{\rm dis} < 0.3 R_{\rm ext}$ and isotropic if $R_{\rm dis} > 0.85 R_{\rm ext}$ in the comoving frame of the external shock, with both approximations resulting in less than 10\% error in the scattering rate.

In addition, in the case of large prompt emission radii, the irradiation of the prompt emission can lead to $\gamma \gamma$ absorption on the TeV photons at early times. We will study the absorption effect and how it affects the TeV light curve. 

%We will show that this effect could naturally account for the ''rapid rising'' observed in the early TeV afterglow light curve between $T^* - T^*+5 \, \rm s$, as reported by LHAASO \citep{LHAASO2023}.

Even though many recent works have studied the TeV and lower-energy afterglows of GRB 221009A \citep{Gill2023, Sato2023, O'Connor2023, ren2023, Ren2024, Zheng2024}, a comprehensive study on the EIC and $\gamma\gamma$ absorption processes from the prompt photons is still lacking. The main motivation of this work is to perform a comprehensive calculation of the EIC process and to understand how the prompt keV/MeV photons affect the early afterglow of GRB 221009A.

Our paper is organized as follows: in Sect.\ref{sec_EIC_in_jet}, we describe the model used for EIC calculations, including different angular distributions of photon number density, and discuss the resulting EIC properties. Sect.\ref{Sec_case_study} presents a case study of EIC in GRB 221009A. In Sect.\ref{sec_gg_abs}, we examine the effects of absorption caused by prompt MeV emission under both internal shock and magnetic field dissipation/turbulence scenarios. A discussion is provided in Sect.\ref{sec_discussion}, with our conclusions summarized in Sect.\ref{sec_conclusion}.

\section{Radiative properties of EIC in relativistic jets}

\label{sec_EIC_in_jet}

Before discussing the EIC process in GRB 221009A as a case study, we explore the EIC process in a relativistic jet as a general study, which helps understand the underlying physics.

\subsection{model description}
\label{sec_model_descripition}

\begin{figure} [h]
    \centering
    \includegraphics[width = 1\linewidth]{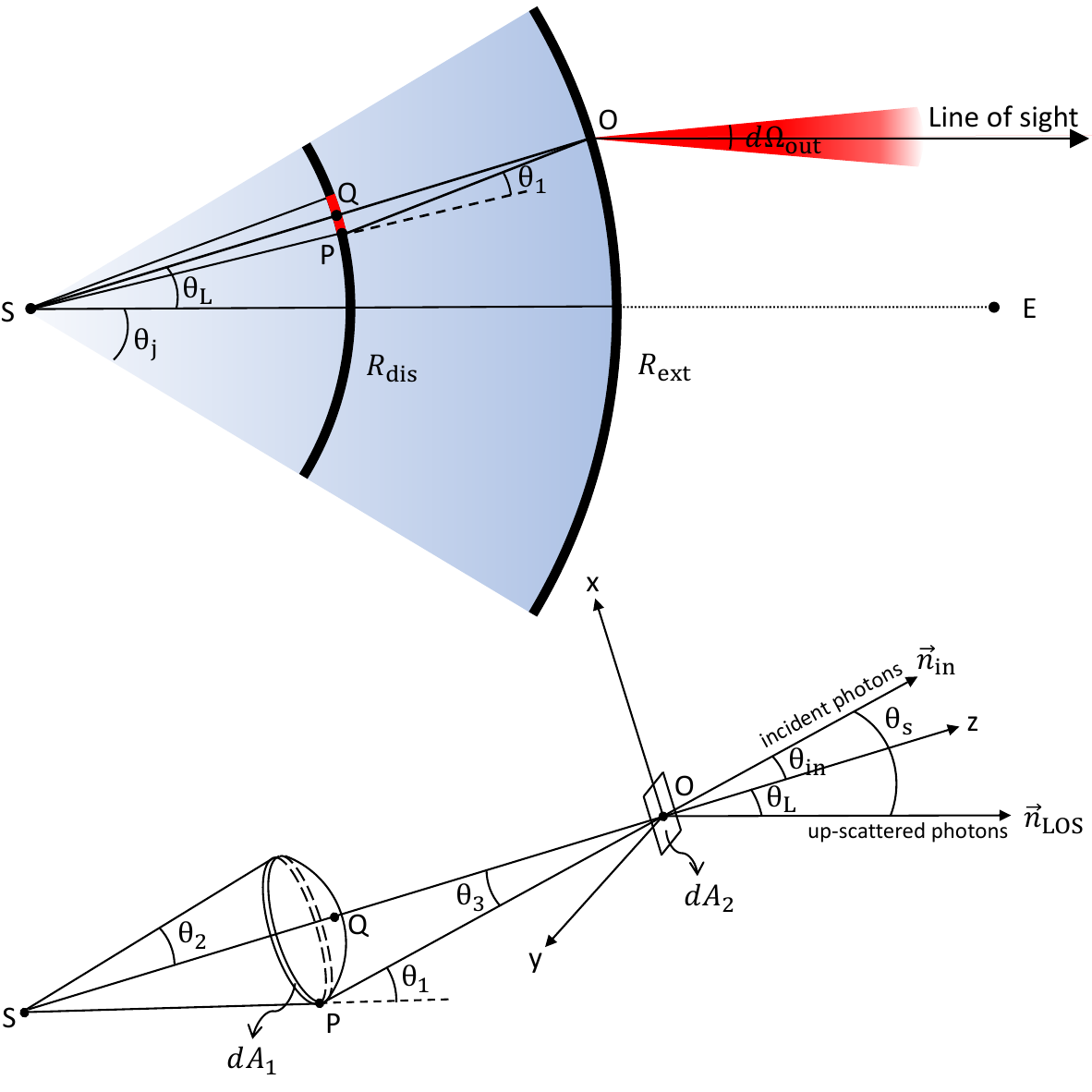}
    \caption{The upper panel illustrates the EIC process in the two-zone scenario. The incident photons are emitted as a "flash" at $R_{\rm dis}$ and scattered by electrons accelerated at $R_{\rm ext}$ in the direction of the LOS. Due to relativistic beaming, the scattering at point $O$ is primarily contributed by incident photons within $\theta_1 \leqslant 1/\Gamma_{\rm dis}$. The lower panel provides a zoomed-in view of this region, focusing on the radiation cone with $\theta_1 \leqslant 1/\Gamma_{\rm dis}$. In this diagram, $\vec{n}_{\rm in}$ and $\vec{n}_{\rm LOS}$ represent the unit vectors along the directions of the incident photons and the LOS, respectively. The vector $\vec{n}_{\rm LOS}$ lies in the XOZ plane.}
    \label{fig_jet}
\end{figure}

In the standard afterglow model, relativistic electrons are accelerated by the external shock \citep{Meszaros1997, Meszaros2006, Piran2004, Kumar2015}. We assume a uniform distribution of accelerated electrons in an outward-expanding thin shell. We focus on the EIC process from this thin shell at a distance $R_{\rm ext}$ from the GRB explosion center (see Fig.\ref{fig_jet}).

Let us designate the explosion center by $S$ and the observer by $E$, and consider an annulus region on the external shock of symmetry for $\overrightarrow{SE}$ (see Fig.\ref{fig_jet}). The volume of the region is denoted by $dV$. The electron spectrum within this volume is described as $dN'_{e, \rm j}(\gamma_e') = N'_{e, \rm j}(\gamma_e') \times \frac{d\Omega_{\rm L}}{\Omega_{\rm j}}$, where $N'_{e, \rm j}(\gamma_e')$ (i.e., $\equiv dN_{e, \rm j}'/d\gamma_e'$) describes the distribution of relativistic electrons in the thin shell located at $R_{\rm ext}$, (we note that quantities denoted by a prime ($'$) are measured in the jet comoving frame, while those without a prime are in the source frame. Additionally, for a physical quantity \(y\), which is a function of \(x_1, x_2, \dots, x_n\), we define \(y(x_1, x_2, \dots, x_{n-1}) = \int y(x_1, x_2, \dots, x_n) \, dx_n\) unless stated otherwise). In this case, $d\Omega_{\rm L} = 2\pi \sin\theta_{\rm L} d\theta_{\rm L}$ represents the solid angle of the annular region, while $\Omega_{\rm j} = 2\pi(1-\cos\theta_{\rm j})$ corresponds to the solid angle of the entire thin shell at $R_{\rm ext}$, both measured from the center of the shock. Here, $\theta_{\rm j}$ denotes the jet's half-opening angle, and $\theta_{\rm L}$ refers to the latitude of the afterglow emission region. 

The scattering rate $\rm (photon  \, Hz^{-1} \, s^{-1} \, sr^{-1})$ resulting from the interaction between isotropic electrons and incident photons moving in a fixed direction, with a scattering angle $\theta'_{\rm s}$, can be written as \citep{Aharonian1981}
\begin{align}
\label{Eq_sc_basic}
    &\frac{dN_{\gamma}'}{d\nu'_{\rm out}d\Omega'_{\rm out}dt'} = \int \, d\gamma'_e  \int  \, d\nu'_{\rm in}  \\ \notag
    &\times \frac{3\sigma_{\rm T} c }{16 \pi {\gamma'_e}^{2}\nu'_{\rm in}}  f_{\rm ani}(\nu'_{\rm in}, \nu'_{\rm out}, \gamma_e', \theta'_{\rm s}) n'_{\rm ph}(\nu'_{\rm in}) N'_e(\gamma'_e), 
\end{align}
where $\nu'_{\rm in}$ denotes the frequency of incident photons and $\nu'_{\rm out}$ that of the up-scattered photons. The spectra of electrons and incident photons are represented by $N'_e(\gamma'_e)$ and $n'_{\rm ph}(\nu'_{\rm in})$ (i.e., $\equiv dn'_{\rm ph}/d\nu'_{\rm in} \, (\rm photon \, Hz^{-1} \, cm^{-3})$), respectively. The function $f_{\rm ani}$ reads
\begin{equation}
\label{Eq_f_ani_spec}
    f_{\rm ani} = 1 + \frac{{\xi'}^2}{2(1-\xi')} - \frac{2\xi'}{b'_\theta(1-\xi')} + \frac{2{\xi'}^2}{{b'_\theta}^2(1-\xi')^2}
\end{equation}
for $h\nu'_{\rm in} \ll h\nu'_{\rm out} \leqslant \gamma'_e m_e c^2 b'_{\theta}/(1+b'_{\theta})$, where $\xi' \equiv h\nu'_{\rm out}/(\gamma'_e m_e c^2)$ and $b'_{\theta} \equiv 2(1-\cos\theta'_{\rm s})\gamma'_e h\nu'_{\rm in}/(m_e c^2)$.

By replacing $N'_{e}(\gamma_e')$ with $N'_{e, \rm j}(\gamma_e') \times \frac{d\Omega_{\rm L}}{\Omega_{\rm j}}$ in Eq.(\ref{Eq_sc_basic}), and considering that the incident photons are distributed as $n'_{\rm ph}(\nu'_{\rm in}, \Omega'_{\rm in})$ $(\rm photon \, Hz^{-1} \, s^{-1} \, sr^{-1})$ rather than propagating in a fixed direction, we can derive the scattering rate $\mathcal{S}(\nu_{\rm out}, \Omega_{\rm L}) d\Omega_{\rm L} \, (\rm photon  \, Hz^{-1} \, s^{-1} \, sr^{-1})$ along LOS in the observer frame (for simplicity, the cosmological effect is not included in our analysis), contributed by electrons within a solid angle element $d\Omega_{\rm L}$ (or volume element $dV$):
\begin{align}
\label{Eq_eic_jet_total}
    &\mathcal{S}(\nu_{\rm out}, \Omega_{\rm L}) d\Omega_{\rm L} = D_{\rm ext, out}^2 \frac{d\Omega_{\rm L}}
    {\Omega_{\rm j}} \int \, d\gamma'_e \int \, d\nu'_{\rm in} \int \, d\Omega'_{\rm in} \\ \notag
    &\times \frac{3\sigma_{\rm T} c}{16\pi{\gamma'_e}^{2}\nu'_{\rm in}}f_{\rm ani}(\nu'_{\rm in}, \nu'_{\rm out}, \gamma'_e, \theta'_{\rm s}) n'_{\rm ph}(\nu'_{\rm in}, \Omega'_{\rm in}) N'_{e, \rm j}(\gamma_e'),
\end{align}
where $D_{\rm ext, out} = \frac{1}{\Gamma_{\rm ext}(1-\beta_{\rm ext}\cos \theta_{\rm L})}$, with $\Gamma_{\rm ext}$ being the Lorentz factor of the external shock and $\beta_{\rm ext} = \sqrt{1 - \Gamma_{\rm ext}^{-2}}$. Note that all scattering rates in the following discussion are calculated along the LOS.

In the numerical code, continuous emission is treated as a series of "flashes", each emitted at a corresponding time grid. Thus, we begin by discussing the simplest case, the two-zone model: as shown in Fig.\ref{fig_jet}, the internal dissipation is emitted as a "flash" at $R_{\rm dis}$ and scattered by electrons accelerated by the external shock at $R_{\rm ext}$. The angular distribution of the prompt emission photon number density \(n_{\rm ph}(\nu_{\rm in}, \Omega_{\rm in})\) at \(R_{\rm ext}\) is dependent on both \(R_{\rm dis}\) and \(R_{\rm ext}\). Once these two radii are known, the angular distribution of the incident photon number density, $n_{\rm ph}(\nu_{\rm in}, \Omega_{\rm in})$, can be derived, allowing us to calculate the EIC scattering rate. Due to relativistic beaming, the incident photons are constrained within a cone. For instance, in the case of EIC occurring at point $O$, the prompt emission photons are incident within a half-apex angle $\theta_3$ (see Fig.\ref{fig_jet}), where $\theta_1 \leqslant 1/\Gamma_{\rm dis}$. Thus, \(n_{\rm ph}(\nu_{\rm in}, \Omega_{\rm in})\) incident to \(R_{\rm ext}\) at different latitudes \(\theta_{\rm L}\) can be considered the same, as long as \(\theta_{\rm j} \gg 1/\Gamma_{\rm dis}\). However, the scattering rate varies with \(\theta_{\rm L}\) as it influences the scattering angle \(\theta_{\rm s}\). A complete calculation of the EIC scattering rate is presented in Sect.\ref{sec_exa_EIC}.

Considering the uncertainties in the dissipation radius and the need to account for the full evolution of the external shock (i.e., calculating the EIC at \(R_{\rm ext}(t)\) for each time step using the two-zone model described in Sect.\ref{sec_exa_EIC}) would be time-consuming. Therefore, we start by analyzing two limiting cases: (1) full anisotropic scattering as an approximation for the $R_{\rm dis} \ll R_{\rm ext}$ scenario, and (2) full isotropic scattering as an approximation for the $R_{\rm dis} \sim R_{\rm ext}$ scenario. In the anisotropic EIC limit, the incident photons propagate strictly radially, and their number density is $n'_{\rm ph}(\nu'_{\rm in}, \Omega'_{\rm in}) = - n'_{\rm ph0} \mathcal{B}' \left( \frac{\nu'_{\rm in}}{\nu'_{\rm p}} \right)\delta (\cos\theta'_{\rm in} - 1)/(2\pi \nu'_{\rm p})$, while in the isotropic limit, the incident photon number density is given by $n'_{\rm ph}(\nu'_{\rm in}, \Omega'_{\rm in}) = n'_{\rm ph0} \mathcal{B}'\left( \frac{\nu'_{\rm in}}{\nu'_{\rm p}} \right)/(4\pi \nu'_{\rm p})$, where $\theta_{\rm in}$ is the angle between the radial direction (or the motion direction of the emission site) and the incident photon propagation direction, $n'_{\rm ph0}$ is the normalization factor, $\mathcal{B}(x)$ follows the form of the Band function \citep{band1993}, $\nu'_{\rm p}$ is the peak frequency, and $\int \mathcal{B}(x) \, dx = 1$. The accuracy of these two limits depends on $\theta_{\rm in} = \theta_3 \simeq \theta_1 \frac{R_{\rm dis}}{R_{\rm ext}}$ as illustrated in Fig.\ref{fig_jet}. If $R_{\rm dis} \ll R_{\rm ext}$, $\theta_{\rm in}$ approaches 0, leading to an almost mono-directional angular distribution of incident photons in the comoving frame of the external shock. If $R_{\rm dis} \sim R_{\rm ext}$, the angle $\theta_{\rm in} \sim \theta_1 \leqslant 1/\Gamma_{\rm dis}$. Hence, the scattering angle \(\theta_{\rm s}\) can increase to \(\sim \theta_{\rm L} + 1/\Gamma_{\rm dis}\), larger than in the anisotropic case, where \(\theta_{\rm s} = \theta_{\rm L}\). According to the Lorentz transformation, $\theta_{\rm in} \sim 1/\Gamma_{\rm ext}$ corresponds to $\theta'_{\rm in} \sim \pi/2$ if $\Gamma_{\rm dis} = \Gamma_{\rm ext}$. Thus, for $R_{\rm dis} \sim R_{\rm ext}$ and $\Gamma_{\rm dis} \sim \Gamma_{\rm ext}$, the scattering angle in the comoving frame of the external shock becomes much larger compared to the anisotropic scenario, particularly at lower latitudes. For instance, if \(R_{\rm dis} \sim R_{\rm ext}\) and \(\Gamma_{\rm dis} = \Gamma_{\rm ext}\), the scattering angle \(\theta'_{\rm s}\) can approach \(\sim \pi/2\) at \(\theta_{\rm L} = 0\) and reach \(\sim \pi\) at \(\theta_{\rm L} = 1/\Gamma_{\rm ext}\). Thus, the isotropic limit assumption is reasonable if $R_{\rm dis} \sim R_{\rm ext}$ and $\Gamma_{\rm dis} \sim \Gamma_{\rm ext}$.

\subsection{The properties of EIC emission in anisotropic and isotropic limits}
\label{sec_EIC_properties}

% Before comparing the scattering rates of the anisotropic and isotropic scenarios, we first discuss the typical properties of the {\bf EIC ??}  emissions in these two limiting cases. 

In the anisotropic EIC limit, as we demonstrate in Sect.\ref{sec_th_anisotropic}, the EIC scattering rate $\mathcal{S}(\nu_{\rm out}, \Omega_{\rm L})$ is mainly influenced by the minimum incident photon frequency $\nu'_{\rm min}(\theta'_{\rm s})$ if $\theta_{\rm L}\leqslant 1/\Gamma_{\rm ext}$. The minimum incident frequency is given by \(\nu'_{\rm in, min}\left( \theta'_{\rm s} \right) = \frac{m_e c^2}{2h\gamma'_{e,0}(1-\cos \theta'_{\rm s})} \frac{\xi'}{1-\xi'}\), as determined by the conservation of energy and momentum during the scattering process (i.e., \(h\nu'_{\rm out} \leqslant \frac{\gamma'_e m_e c^2 b'_{\theta}}{1 + b'_{\theta}}\)). A larger $\theta_{\rm L}$ leads to a smaller $\nu'_{\rm min}$, which allows the participation of more incident photons in the scattering and hence increases the scattering rate. On the other hand, for $\theta_{\rm L} > 1/\Gamma_{\rm ext}$, $\nu'_{\rm in, min}$ becomes less sensitive to $\theta_{\rm L}$ and $\mathcal{S}_{\rm ani}(\nu_{\rm out}, \Omega_{\rm L})$ will mainly change through Doppler factor $D_{\rm ext, out}$. Analytic derivation (see Eq.(\ref{Eq_eic_ani_ana})) shows that $\mathcal{S}_{\rm ani}(\nu_{\rm out}, \Omega_{\rm L}) \propto (1-\cos\theta'_{\rm L})^{\alpha} D_{\rm ext, out}^{3+p-\alpha}$, indicating that the scattering rate decreases rapidly when $\theta_{\rm L} > 1/\Gamma_{\rm ext}$ (i.e., $\theta'_{\rm L} > \pi/2$), where $\alpha$ and $p$ are the spectral indices of the incident photons and electrons, respectively. 

Thus, the observable EIC emission is mainly concentrated within $\theta_{\rm L} \sim 1/\Gamma_{\rm ext}$, and its scattering rate $\mathcal{S}_{\rm ani}(\nu_{\rm out}, \Omega_{\rm L})$ reaches the maximum value at $\theta_{\rm L}\sim 1/\Gamma_{\rm ext}$. The delay time of the emission from $\theta_{\rm L}=1/\Gamma_{\rm ext}$ with respect to the emission from $\theta_{\rm L}=0$ is approximately $R_{\rm ext}/(c\Gamma_{\rm ext}^2)$, which is close to the dynamical time of the external shock. We therefore expect that this delayed emission of EIC will smooth the light curve in the anisotropic EIC limit. 
  
In the isotropic EIC limit, as discussed in Sect.\ref{sec_th_isotropic}, $\mathcal{S}_{\rm iso}(\nu_{\rm out}, \Omega_{\rm L})$ depends on $\theta_{\rm L}$ only through the Doppler factor $D_{\rm ext, out}$, which decreases slowly with $\theta_{\rm L}$ for $\theta_{\rm L} \leqslant 1/\Gamma_{\rm ext}$ and decreases rapidly for $\theta_{\rm L} > 1/\Gamma_{\rm ext}$. Therefore, in the isotropic EIC limit, the EIC will not exhibit a significant delayed emission, in contrast to the anisotropic EIC limit. Moreover, the scattering rate in the isotropic EIC limit is primarily contributed by electrons within $\theta_{\rm L} \leqslant 1/\Gamma_{\rm ext}$, similar to that in the anisotropic EIC limit.

\subsection{The ratio of EIC scattering in the anisotropic EIC limit to that in the isotropic limit}

\label{sec_ratio_ani_iso_EIC}

%(3) Anisotropic scattering suppresses the observed flux by $\sim 0.3$ compared to isotropic scattering:

To discuss the suppression of anisotropic scattering in a relativistic jet, we ignore the minor contribution from electrons accelerated at locations with latitude \(\theta_{\rm L} > 1/\Gamma_{\rm ext}\) for simplicity. In the simplest case, both the incident photons ($h\nu'_{\rm in0}$) and electrons ($\gamma'_{e0} m_e c^2$) are mono-energetic, and the ratio of the scattering rates, integrated over the latitude, between the anisotropic and isotropic scattering cases (i.e., the suppression factor) is (see Sect.\ref{sec_compare_ani_iso}):
\begin{align}
    \chi & \equiv \frac{\int \mathcal{S}_{\rm ani} (\nu_{\rm out}, \Omega_{\rm L}) \, d\Omega_{\rm L}}{\int \mathcal{S}_{\rm iso}(\nu_{\rm out}, \Omega_{\rm L}) \, d\Omega_{\rm L}} \\ \notag
    & \simeq \frac{\int_{\cos\theta_{\rm s, min}}^{\cos\frac{1}{\Gamma_{\rm ext}}} f_{\rm ani}(\nu'_{\rm in0}, \nu'_{\rm out}, \gamma'_{e0}, \theta'_{\rm L}) \, d\cos\theta_{\rm L}}{\int_1^{\cos\frac{1}{\Gamma_{\rm ext}}} f_{\rm iso}(\nu'_{\rm in0}, \nu'_{\rm out}, \gamma'_{e0}) \, d\cos\theta_{\rm L}}, \, \nu'_{\rm in0} \geqslant \nu'_{\rm in, min}\left(\frac{\pi}{2} \right),
\end{align}
where the minimum scattering angle is given by $\cos\theta'_{\rm s, min} = 1 - \frac{m_e c^2}{2\gamma'_{e0} h\nu'_{\rm in0}} \frac{\xi'}{1-\xi'}$, which is derived from the  condition $h\nu'_{\rm out} \leqslant \frac{\gamma'_e m_e c^2 b'_{\theta}}{1 + b'_{\theta}}$.

Since $f_{\rm iso}(\nu'_{\rm in}, \nu'_{\rm out}, \gamma'_e) = \int^{\cos\theta'_{\rm s, min}}_{-1} f_{\rm ani}(\nu'_{\rm in}, \nu'_{\rm out}, \gamma'_e, \theta'_{\rm s}) \, d\cos\theta'_{\rm s}/2$, and $f_{\rm ani}$ almost does not change with $\theta'_{\rm s}$ (see Sect.\ref{sec_fani_approximation}), we have
\begin{align}
    \chi&\sim \frac{f_{\rm ani}}{f_{\rm iso}}\frac{\cos\frac{1}{\Gamma_{\rm ext}} - \cos\theta_{\rm s, min}}{\cos\frac{1}{\Gamma_{\rm ext}} - 1} \\ \notag
    &\sim \frac{2}{\cos\theta'_{\rm s, min} + 1}\frac{\cos\frac{1}{\Gamma_{\rm ext}} - \cos\theta_{\rm s, min}}{\cos\frac{1}{\Gamma_{\rm ext}} - 1}, \, \nu'_{\rm in0} \geqslant \nu'_{\rm in, min}\left(\frac{\pi}{2} \right).
\end{align}
At this point, we can clearly see that the difference between the anisotropic and isotropic scattering rates consists of two terms if $\nu'_{\rm in0} \geqslant \nu'_{\rm in, min}\left(\frac{\pi}{2} \right)$. The first term comes from the difference between $f_{\rm ani}$ and $f_{\rm iso}$. For $\theta_{\rm s, min} \leqslant \theta_{\rm L} \leqslant 1/\Gamma_{\rm ext}$, this term $\in (1, 2]$. The second term arises from the lower limit of the integration over \(\theta_{\rm L}\) in the anisotropic EIC limit, where \(\theta_{\rm s} = \theta_{\rm L}\), and \(\theta_{\rm s}\) is constrained by the conservation condition \(h\nu'_{\rm out} \leqslant \frac{\gamma'_e m_e c^2 b'_{\theta}}{1 + b'_{\theta}}\). Hence, in the anisotropic EIC limit, electrons with a given Lorentz factor \(\gamma'_{e0}\) located at latitude \(\theta_{\rm L} \leqslant \theta_{\rm s, min}\) cannot scatter an incident photon of energy \(h\nu'_{\rm in0}\) to \(h\nu'_{\rm out}\). This implies that fewer electrons (only those accelerated at $\theta_{\rm L} \geqslant \theta_{\rm s, min}$) can participate in the scattering process compared to the isotropic EIC limit. As $\nu'_{\rm in}$ increases, $\theta'_{\rm s, min}$ rapidly approaches 0, causing the difference in the scattering rate from both the first term and the second term to disappear, leading to $\chi \sim 1$. Thus, the main difference in the scattering rate between the anisotropic and isotropic EIC limits arises from the minimum energy of the incident photons. In the anisotropic EIC limit with $\theta_{\rm L} \leqslant 1/\Gamma_{\rm ext}$, we have $\theta'_{\rm s} = \theta'_{\rm L} \leqslant \pi/2$, and the minimum incident photon energy is $\nu'_{\rm in, min}\left( \frac{\pi}{2} \right)$. In the isotropic EIC limit, the minimum incident photon energy is irrelevant with $\theta_{\rm L}$ and it becomes $\nu'_{\rm in, min}(\pi) = 0.5\nu'_{\rm in, min}\left( \frac{\pi}{2} \right)$. This is illustrated in the left panel of Fig.\ref{fig_12}, which shows $S_{\rm ani}(\nu_{\rm out})$ ($=\int S_{\rm ani}(\nu_{\rm out}, \Omega_{\rm L}) \, d\Omega_{\rm L}$) and $S_{\rm iso}(\nu_{\rm out})$ ($=\int S_{\rm iso}(\nu_{\rm out}, \Omega_{\rm L}) \, d\Omega_{\rm L}$) for different values of $\nu'_{\rm in0}$. The primary difference between the two is the value of $\nu'_{\rm in, min}$, once $\nu'_{\rm in0} > \nu'_{\rm in, min}$, $\chi \simeq 1$, as discussed above.

For a more realistic estimate, we consider that the incident photons have a spectrum being described by the Band function, though still assuming mono-energetic electrons. Consequently, the scattering rates are directly proportional to Eq.(\ref{eq_eic_me_mph_ani}) and Eq.(\ref{eq_eic_me_mph_iso}), multiplied by the Band function (see the right panel in Fig.\ref{fig_12}), and then integrated over $\nu'_{\rm in}$. By performing the integration over $\nu'_{\rm in}$, we find that $\chi \in (0.3, 0.6)$ for $\alpha_{\rm l} \in (1, 2)$. This result also applies to the scenario where the electron spectrum follows a power-law form. We note that the value of $\chi$ slightly depends on the spectral indices of the Band function and the spectrum of electrons accelerated by the external shock. $\chi$ is hardly affected whether the scattering proceeds in the Thomson regime or the Klein-Nishina (KN) regime, as this only results in a minor change when approximating $f_{\rm ani}$ according to Sect.\ref{sec_fani_approximation}. However, if the synchrotron self-absorption frequency is higher than $\nu'_{\rm in, min}\left( \frac{\pi}{2} \right)$, the minimum energy of the incident photons becomes the same for both the isotropic and anisotropic EIC limits.This results in $\chi \sim 1$ after integrating over $\nu'_{\rm in}$ in Eqs. (\ref{eq_eic_me_mph_ani}) and (\ref{eq_eic_me_mph_iso}), multiplied by the Band function.

\subsection{The accuracy of the anisotropic EIC limit for $R_{\rm dis} \ll R_{\rm ext}$ and the isotropic EIC limit for $R_{\rm dis} \sim R_{\rm ext}$}

\label{sec_EIC_condition}
In this section, we quantitatively compare the EIC scattering rates derived from the precise calculations in Sect.\ref{sec_exa_EIC} (\(\mathcal{S}_0\)) with those obtained in the anisotropic (\(\mathcal{S}_{\rm ani}\)) and isotropic (\(\mathcal{S}_{\rm iso}\)) EIC limits. For comparison among the three scenarios, we normalize the incident photon number density $n'_{\rm ph}(\nu'_{\rm in})(= \int n'_{\rm ph}(\nu'_{\rm in}, \Omega'_{\rm in}) \, d\Omega'_{\rm in}$) to the same value. Since $n'_{\rm ph}(\nu'_{\rm in})$ varies with the arrival time $t_{\rm s}$ of incident photons in precise calculations (see Sect.\ref{sec_exa_EIC}), we consider the time-averaged photon number density $\overline{n'_{\rm ph}(\nu'_{\rm in})} = \frac{\int n'_{\rm ph}(\nu'_{\rm in}, t'_{\rm s})\, dt'_{\rm s}}{\int \, dt'_{\rm s}}$ over the interval $\Delta t_{\rm s}'$. $\Delta t_{\rm s}'$ refers to the time delay between the first and last photons arriving at point $O$ from the external shock, both of which are emitted simultaneously from points $Q$ and $P$ in the internal dissipation region (see Fig.\ref{fig_jet}).

\begin{figure} [h]
    \centering
    \includegraphics[width = 1\linewidth]{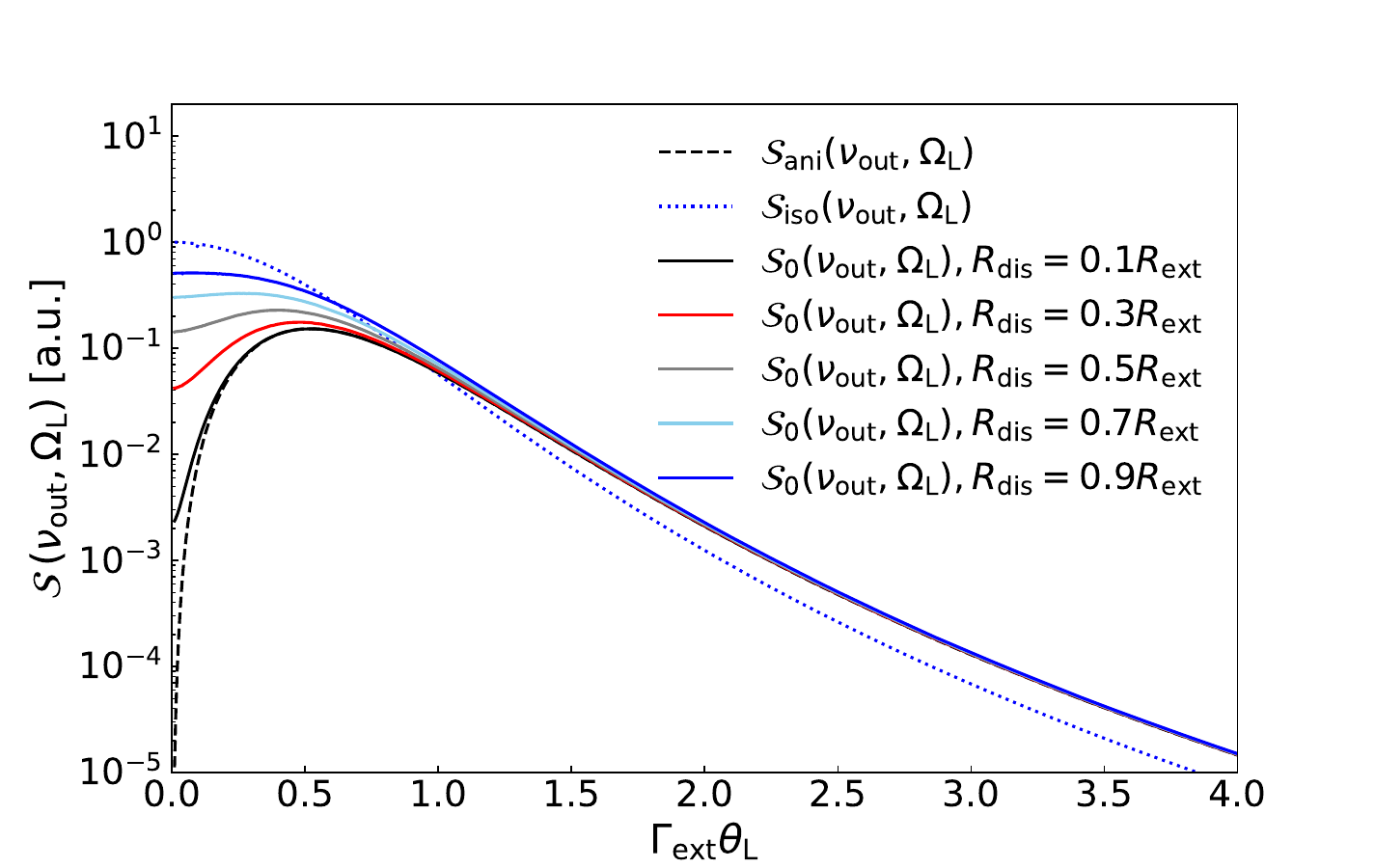}
    \includegraphics[width = 1\linewidth]{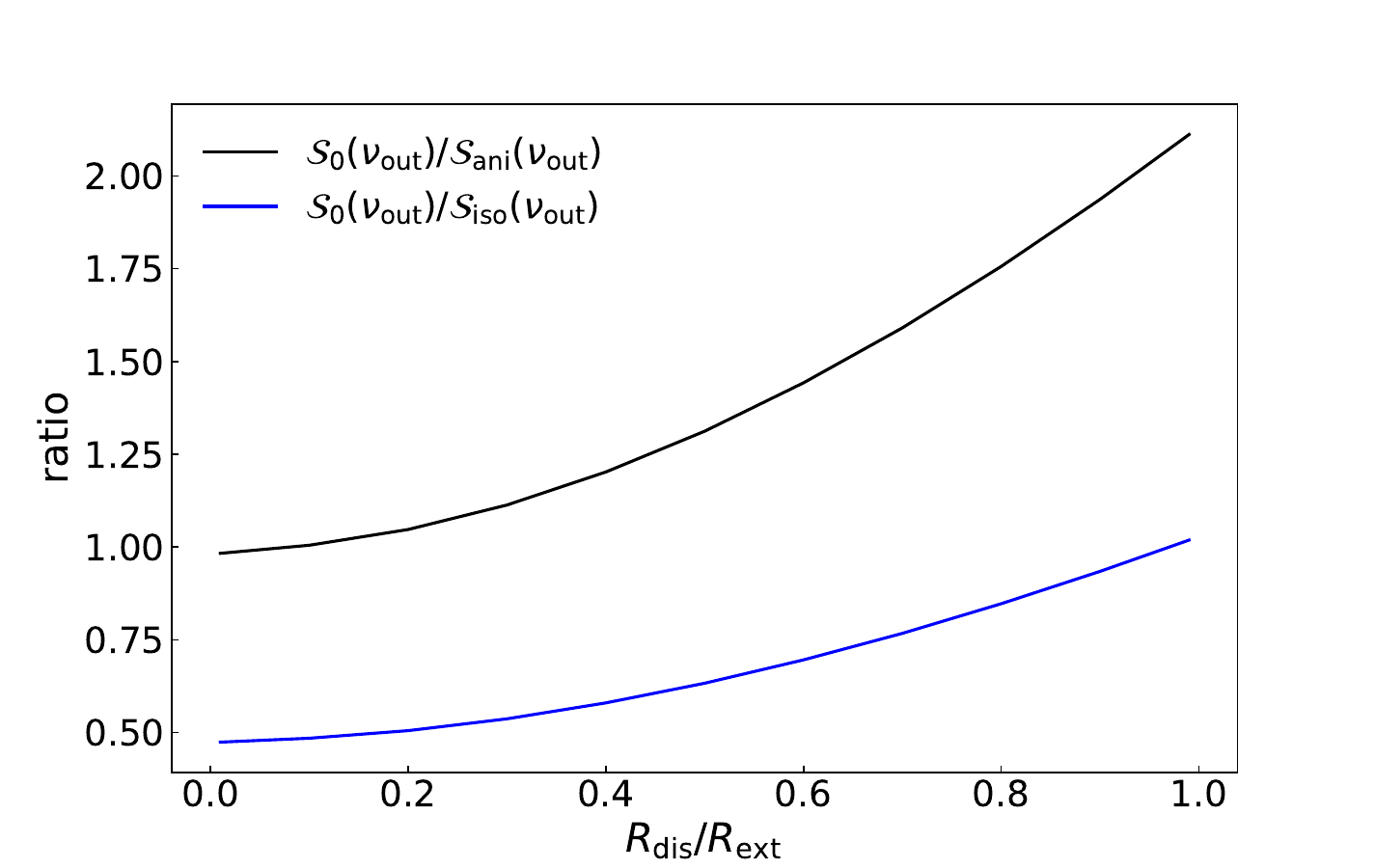}
    \caption{The upper panel displays the scattering rate per solid angle $\mathcal{S}(\nu_{\rm out}, \Omega_{\rm L})$ at $h\nu_{\rm out} = 1 \, \rm TeV$ for the precise calculation ($\mathcal{S}_0$), as well as for the anisotropic ($\mathcal{S}_{\rm ani}$) and isotropic ($\mathcal{S}_{\rm iso}$) EIC limit. The lower panel shows the ratio of the total scattering rate $\mathcal{S}(\nu_{\rm out}) = \int \mathcal{S}(\nu_{\rm out}, \Omega_{\rm L}) \, d\Omega_{\rm L}$ at $h\nu_{\rm out} = 1 \, \rm TeV$ obtained from the precise calculation compared to those obtained in the anisotropic and isotropic EIC limits. In the precise calculation, $\mathcal{S}_0$ is computed according to the method described in Sect.\ref{sec_exa_EIC}, with the incident photons following a Band function characterized by a low-energy index $\alpha_{\rm l} = 1$, a high-energy index $\alpha_{\rm h} = 2.2$, and a peak energy $h\nu_{\rm p} = 1 \, \rm MeV$, injected at $R_{\rm dis}$ and scattered by power-law electrons with index $p = 2.2$ accelerated at $R_{\rm ext} = 10^{17} \, \rm cm$, assuming $\Gamma_{\rm dis} = \Gamma_{\rm ext} = 500$. For the anisotropic and isotropic EIC limits, the photon number density $n'_{\rm ph}(\nu'_{\rm in})$ and spectral indices of the incident photons and electrons are the same as those used in the precise calculation, but the angular distributions of the incident photons differ (see text for details).}
    \label{fig_scattering_rate}
\end{figure}

The results are presented in Fig.\ref{fig_scattering_rate}. The upper panel of the figure shows that the anisotropic EIC limit closely matches the precise calculation when \(R_{\rm dis} \ll R_{\rm ext}\). This consistency is expected, as the incident photons are almost aligned with the radial direction (i.e., \(\theta'_{\rm in} = \theta'_3\) approaches 0 (see Eq.\ref{A_Eq_costhe_3})) for $R_{\rm dis} \ll R_{\rm ext}$.

For $R_{\rm dis} \sim R_{\rm ext}$, the scattering rate under the isotropic EIC limit, $\mathcal{S}_{\rm iso}(\nu_{\rm out}, \Omega_{\rm out})$, does not exactly match $\mathcal{S}_0(\nu_{\rm out}, \Omega_{\rm out})$. For instance, if we assume $\Gamma_{\rm dis} = \Gamma_{\rm ext}$, at $\theta_{\rm L} = 0$, the maximum scattering angle is $\theta'_{\rm s} = \theta'_{\rm 3} \leqslant \pi/2$ in the precise calculation (i.e., without head-on collision between photons and electrons), resulting in $\mathcal{S}_0(\nu_{\rm out}, \Omega_{\rm out}) < \mathcal{S}_{\rm iso}(\nu_{\rm out}, \Omega_{\rm out})$. On the other hand, for $\theta_{\rm L} = 1/\Gamma_{\rm ext}$, the scattering angle $\theta'_{\rm s}$ can vary from 0 to $\pi$ in the precise calculation over the time interval $t_{\rm s} \in [0, \Delta t_{\rm s}]$. However, even if the angular distribution of photon density is relatively uniform in the source frame, it becomes non-uniform in the comoving frame of external shock, as $d\Omega_{\rm in}/d\Omega'_{\rm in} = D_{\rm ext, in}^{-2} \propto (1 - \beta_{\rm ext} \cos \theta_{\rm in})^2$, causing it to deviate from the isotropic assumption. When $\theta_{\rm L} \gtrsim 1/\Gamma_{\rm ext}$, $\mathcal{S}_{\rm iso}(\nu_{\rm out}, \Omega_{\rm out})$ may exceed $\mathcal{S}_0(\nu_{\rm out}, \Omega_{\rm out})$ because, in the precise calculation, the average scattering angle of the incident photons is larger than that in the isotropic case. This behavior can be seen in the upper panel of Fig.\ref{fig_scattering_rate}. Nevertheless, the total scattering rates, $\mathcal{S}_0(\nu_{\rm out})$ and $\mathcal{S}_{\rm ani}(\nu_{\rm out})$, integrated over $\Omega_{\rm L}$, only differ by less than 10\% as long as $R_{\rm dis} > 0.85 R_{\rm ext}$ (see the lower panel of Fig.\ref{fig_scattering_rate}).

It is worth noting that if \(\Gamma_{\rm dis} \gg \Gamma_{\rm ext}\), the Doppler boosting of \(\theta_{\rm s}\) is reduced in the comoving frame of the external shock compared to the case where \(\Gamma_{\rm dis} \sim \Gamma_{\rm ext}\). Nevertheless, even in this case, the discrepancy between the isotropic EIC limit and the precise calculation remains within a factor of 2.

In conclusion, the obtained EIC fluxes in both the anisotropic and isotropic EIC limits deviate from that of the precise calculation by less than a factor of 2, with the degree of accuracy depending on the ratio $R_{\rm dis}/R_{\rm ext}$. Notably, for \(R_{\rm dis}/R_{\rm ext} < 0.3 \) and \(R_{\rm dis}/R_{\rm ext} > 0.85 \), the scattering rate $S(\nu_{\rm out})$ errors in both limits remain within 10\% of the precise calculations, given \(\Gamma_{\rm dis} \sim \Gamma_{\rm ext}\) (see the lower panel of Fig.\ref{fig_scattering_rate}).

\section{A case study of EIC in GRB 221009A}

Following the analysis of EIC properties, this section examines the EIC process in GRB 221009A as a case study. We begin by evaluating the cooling effects of EIC and how they modify the electron and radiation spectra in the early afterglow, followed by modeling the light curve through numerical calculations.

\label{Sec_case_study}

\subsection{the cooling of electrons caused by EIC}
\label{sec_ana_shock_EIC_cooling}

To calculate the influence of EIC cooling on the electron spectrum, we need to incorporate the dynamical evolution of the external shock and relevant parameters into the model. We first employ an analytical description of the shock's hydrodynamics \citep{Blandford1976,Sari98}. For simplicity, the cosmology effect is ignored since GRB 221009A has a low redshift of $z=0.151$  \citep{Postigo2022, Castro-Tirado2022}.

%\subsubsection{hydrodynamic and }

%\label{sec_ana_shock_EIC_cooling}

Before the external shock decelerates, its Lorentz factor remains roughly constant. After deceleration, the Lorentz factor of the external shock follows the self-similar solution \citep{Blandford1976,Sari98}. Therefore, the Lorentz factor of the shock can be expressed as

\begin{equation}
\begin{array}{ll}
\label{Eq_Gamma}
\Gamma_{\rm ext} \simeq \left \{
    \begin{array}{ll}
       &\Gamma_{\rm ext, 0} \,\,\,\,\,\,\,\,\,\,\,\,\,\,\,\,\,\,\,\,\,\, , t \leqslant t_{\rm d} \\
       &\Gamma_{\rm ext, 0}(\frac{t}{t_{\rm d}})^{-3/8}, t > t_{\rm d}
   \end{array}
   \right.
\end{array}
\end{equation}
where $\Gamma_{\rm ext, 0}$ is the initial Lorentz factor of the external shock and $t$ is the time measured in the observer frame. The deceleration time of the external shock is $t_{\rm d} = \left( \frac{3E_{\rm k0, iso}}{32\pi n m_p c^5 \Gamma_{\rm ext, 0}^8} \right)^{1/3}$, where $E_{\rm k0, iso}$ and $n$ are the initial isotropic kinetic energy of the external shock and the circumburst density, respectively. In addition, The radius of the external shock can be expressed as
\begin{equation}
\label{Eq_R_ana}
\begin{array}{ll}
R_{\rm ext} \simeq \left \{
    \begin{array}{ll}
       &2\Gamma_{\rm ext, 0}^2 c t \,\,\,\,\,\,\,\,\,\,\,\,\,\,\,\,\,\,\,\,\,\,\,\,\,\, , t \leqslant t_{\rm d} \\
       &2\Gamma_{\rm ext, 0}^2 c t_{\rm d} \left( \frac{t}{t_{\rm d}} \right)^{1/4}. t > t_{\rm d}
   \end{array}
   \right.
\end{array}
\end{equation}

The energy density of the magnetic field in the shock comoving frame is

\begin{equation}
    U'_B = 4 m_p \epsilon_B n(\Gamma_{\rm ext} c)^2,
\end{equation}
where $\epsilon_B$ is the fraction of the energy from the shock that is converted into magnetic field energy. And the energy density of the prompt radiation can be estimated as \footnote{In the Lorentz transform of photon energy density, we have $\nu = 2 \Gamma_{\rm ext} \nu'$ for each photon's
energy and $n_{\rm ph} = 2\Gamma_{\rm ext} n'_{\rm ph}$ for the photon number density ($\rm photons \, cm^{-3}$). Therefore, we derive $U_{\gamma} = 4\Gamma_{\rm ext}^2 U'_{\gamma}$. In some papers, the approximation $U_{\gamma} = \Gamma^2_{\rm ext} U'_{\gamma}$ is employed.}

\begin{align}
    U'_{\gamma} &= \frac{L_{\gamma}}{16\pi \Gamma_{\rm ext}^2 R_{\rm ext}^2 c} \\ \notag
    &\sim \frac{E_{\gamma, \rm iso}}{ 16\pi \Gamma_{\rm ext}^2 R_{\rm ext}^2c\Delta t_{\gamma}} \\ \notag
    &= \frac{\eta_{\gamma}E_{\rm k0, iso}}{16\pi (1-\eta_{\gamma}) \Gamma_{\rm ext}^2 R_{\rm ext}^2c\Delta t_{\gamma}},
\end{align}
where $L_{\gamma}$, $E_{\gamma, \rm iso}$, and $\Delta t_{\gamma}$ are the luminosity, isotropic-equivalent energy, and duration of the prompt emission, respectively. For simplicity, we assume $L_{\gamma}$ is constant during the prompt emission period for magnitude estimation. The radiation efficiency is defined as $\eta_{\gamma} \equiv \frac{E_{\gamma, \rm iso}}{E_{\gamma, \rm iso} + E_{\rm k0, iso}}$. Therefore, the ratio of $U'_{\gamma}$ to $U'_B$ can be written as

\begin{equation}
\begin{array}{ll}
&\frac{U'_{\gamma}}{U'_B} = 44 \left(\frac{\eta_{\gamma}}{1-\eta_{\gamma}}\right) \left( \frac{E_{\rm k0,iso}}{5\times 10^{54} \, \rm erg} \right) \left( \frac{n}{0.1 \, \rm cm^{-3}} \right)^{-1} \left( \frac{\epsilon_B}{10^{-3}} \right)^{-1} \\ \notag
&\times \left( \frac{\Gamma_{\rm ext,0}}{500} \right)^{-8} \left( \frac{\Delta t_{\gamma}}{20 \, \rm s} \right)^{-1} \left( \frac{t}{t_{\rm d}} \right)^{-2} \\ \notag
\end{array}
\end{equation}
for $t \leqslant t_{\rm d}$, and
\begin{equation}
\begin{array}{ll}
\frac{U'_{\gamma}}{U'_B} = 44 \left(\frac{\eta_{\gamma}}{1-\eta_{\gamma}}\right) \left( \frac{\epsilon_B}{10^{-3}} \right)^{-1} \left( \frac{\Delta t_{\gamma}}{20 \, \rm s} \right)^{-1} \left( \frac{t}{t_{\rm d}} \right)
\end{array}
\end{equation}
for $t > t_{\rm d}$.

% If $t < t_{\rm d}$, $U'_B$ remains constant while $U'_{\gamma}$ decreases with time due to the expanding radius. After $t > t_{\rm d}$, $U'_B \propto \Gamma_{\rm ext}$ decreases with time, and $U'_{\gamma}$ increases with time because $\Gamma_{\rm ext}$ decreases in $U'_{\gamma} = U_{\gamma}/(4\Gamma_{\rm ext}^2)$.

Thus, the presence of the prompt emission can lead to additional cooling of electrons if $U'_{\gamma}$ is comparable to or greater than $U'_B$. The cooling of the electrons is independent of the direction of the incident photons as long as the electrons themselves are isotropic. Furthermore, according to the discussion in \cite{Khangulyan2014}, in our case, EIC cooling would not cause an anisotropic angular distribution of electrons, despite the anisotropy of the incident photons. This is because the electron cooling time is much longer than the electron cyclotron timescale, and hence electrons are isotropized by the magnetic field.

\begin{figure} [h]
    \centering
    \includegraphics[width = 1\linewidth]{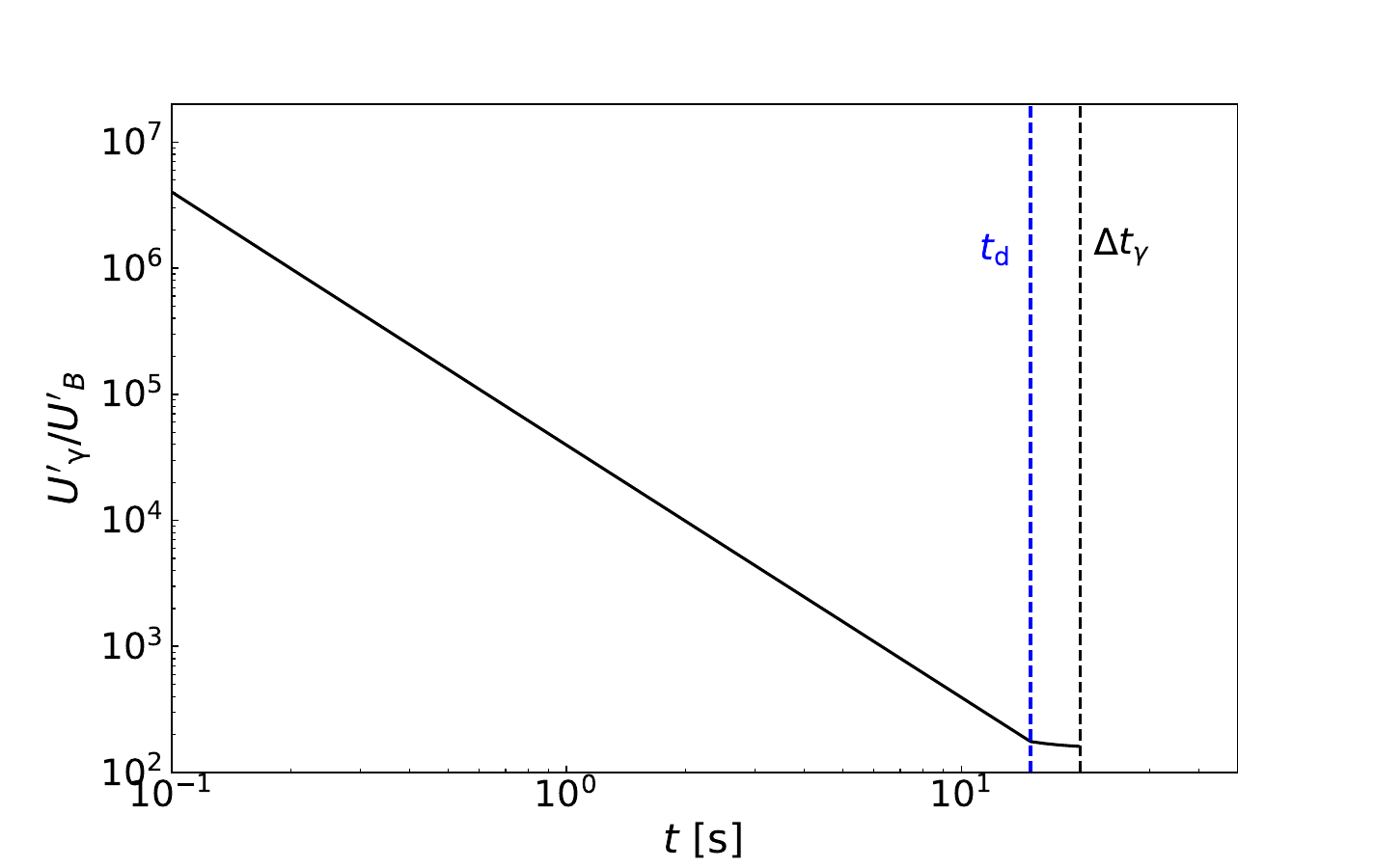}
    \caption{The ratio of the prompt radiation density $U'_{\gamma}$ to the magnetic field density $U'_{B}$. The blue and black vertical dashed lines represent $t = t_{\rm d}$ and $t = \Delta t_{\gamma}$, respectively. The following parameters are adopted: $E_{k0, \rm iso} = 5 \times 10^{54} \, \rm erg$, $\Gamma_{\rm ext, 0} = 500$, $n = 0.1 \, \rm cm^{-3}$, $\epsilon_B = 1 \times 10^{-3}$, $\eta_{\gamma} = 0.8$, and $\Delta t_{\gamma} = 20 \, \rm s$. }
    \label{fig_ratio_Uph_UB}
\end{figure}

Thus, the cooling rate for both EIC and SSC, \(\dot{\gamma}'_{e, \rm IC}\), including the KN effect, can be expressed as \citep{Jones1968, Blumenthal1970}:
\begin{equation}
    {\dot{\gamma}}'_{e, \rm IC} = -\int \, d\nu'_{\rm out} \int \, d\nu'_{\rm in} \frac{3\sigma_{\rm T} h\nu'_{\rm out}}{4m_e c {\gamma'_e}^2 \nu'_{\rm in}} f_{\rm iso}(\nu'_{\rm in}, \nu'_{\rm out}, \gamma'_e) n'_{\rm ph}(\nu'_{\rm in}),
\end{equation}

where
\begin{equation}
    f_{\rm iso} = 2q\ln{q}+(1+2q)(1-q)+\frac{1}{2}\frac{(4gq)^2}{1+4gq}(1-q)
\end{equation}
with $h\nu'_{\rm in}/(\gamma'_e m_e c^2) \leqslant f \leqslant 4g/(1 + 4g)$, $g \equiv \gamma'_e h\nu'_{\rm in}/(m_e c^2)$, $f \equiv h\nu'_{\rm out}/(\gamma'_e m_e c^2)$, and $q \equiv f/[4g(1 - f)]$. The Thomson limit applies when $g \ll 1$. In this limit, ${\dot{\gamma}}'_{e, \rm IC}$ simplifies to $-\frac{4\sigma_{\rm T} }{3m_e c}{\gamma'_e}^{2} U'_{\rm ph}$, where $U'_{\rm ph}$ is the energy density of incident photons. In addition, the synchrotron cooling rate can be written as ${\dot{\gamma}}'_{e, \rm SYN} = -\frac{4\sigma_{\rm T} }{3m_e c}{\gamma'_e}^{2} U'_B$ \citep{Rybicki1979}. And the total cooling rate of the electrons ${\dot{\gamma}}'_e = {\dot{\gamma}}'_{e, \rm SYN} + {\dot{\gamma}}'_{e, \rm SSC} + {\dot{\gamma}}'_{e, \rm EIC}$. The electron spectrum accelerated by the external shock is given by \citep{Sari98, Nakar2009}: in the fast-cooling regime, the electron spectrum $N'_e({\gamma'_e})$, scales as $\frac{{\gamma'_e}^{-2}}{1 + Y'(\gamma'_e)} \quad (\gamma'_{\rm c} < \gamma'_e < \gamma'_{\rm m})$, and $\frac{{\gamma'_e}^{-p-1}}{1 + Y'(\gamma'_e)} \quad (\gamma'_{\rm m} < \gamma'_e)$, where \(Y' \equiv \frac{L'_{\rm SSC} + L'_{\rm EIC}}{L'_{\rm SYN}}\), and \(L'_{\rm SSC}\), \(L'_{\rm EIC}\), and \(L'_{\rm SYN}\) represent the luminosities of SSC, EIC, and synchrotron radiation, respectively. In the slow-cooling regime, it scales as ${\gamma'_e}^{-p} \quad (\gamma'_{\rm m} < \gamma'_e < \gamma'_{\rm c})$, and $\frac{{\gamma'_e}^{-p-1}}{1 + Y'(\gamma'_e)} \quad (\gamma'_{\rm c} < \gamma'_e)$. The minimum Lorentz factor of electrons is given by: $\gamma'_{\rm m} = \epsilon_e \left( \frac{p-2}{p-1} \right) \frac{m_p}{m_e} \Gamma_{\rm ext}, $ and $\gamma'_{\rm c}$ is derived from the condition $t' = t'_{\rm c}$, where $\epsilon_e$ is the fraction of the shock energy converted into electron energy, $p$ is the power-law index of the electron spectrum, and $t'_{\rm c} = \gamma'_e/\dot{\gamma}'_e$ is the electron cooling time. Using the method described above, we derive the following results:

(1) The influence of EIC on the electron and emission spectra: Fig.\ref{fig_ratio_Uph_UB} shows how $U'_B$ and $U'_{\gamma}$ change over time. Initially, the ratio $U'_{\gamma}/U'_B$ decreases with $t^{-2}$ as the expanding radius reduces $U'_{\gamma}$. The ratio reaches its minimum value at approximately $t \sim t_{\rm d}$, where $U'_{\gamma} \sim 100 U'_B$. After this point, the ratio increases linearly with time until the prompt emission ends (We use $R_{\rm ext} = c\int 2\Gamma^2_{\rm ext}(t) \, dt$ instead of Eq.(\ref{Eq_R_ana}), which makes $U'_{\gamma}/U'_B$ transition smoothly after $t_{\rm d}$ and does not increase with $t$ immediately, as shown in Fig.\ref{fig_ratio_Uph_UB}). This increase is due to the decreasing $\Gamma_{\rm ext}$, which enhances the energy density of the prompt emission in the comoving frame of the external shock. The result in Fig.\ref{fig_ratio_Uph_UB} shows that the energy density of prompt emission is much larger than that of the magnetic field at early times, making it essential to consider the cooling from EIC, particularly in the Thomson regime, where $\dot{\gamma}_{e, \rm SYN}/\dot{\gamma}_{e, \rm EIC} \propto U'_B/U'_{\gamma}$.

\begin{figure} 
    \centering
    \includegraphics[width = 1\linewidth]{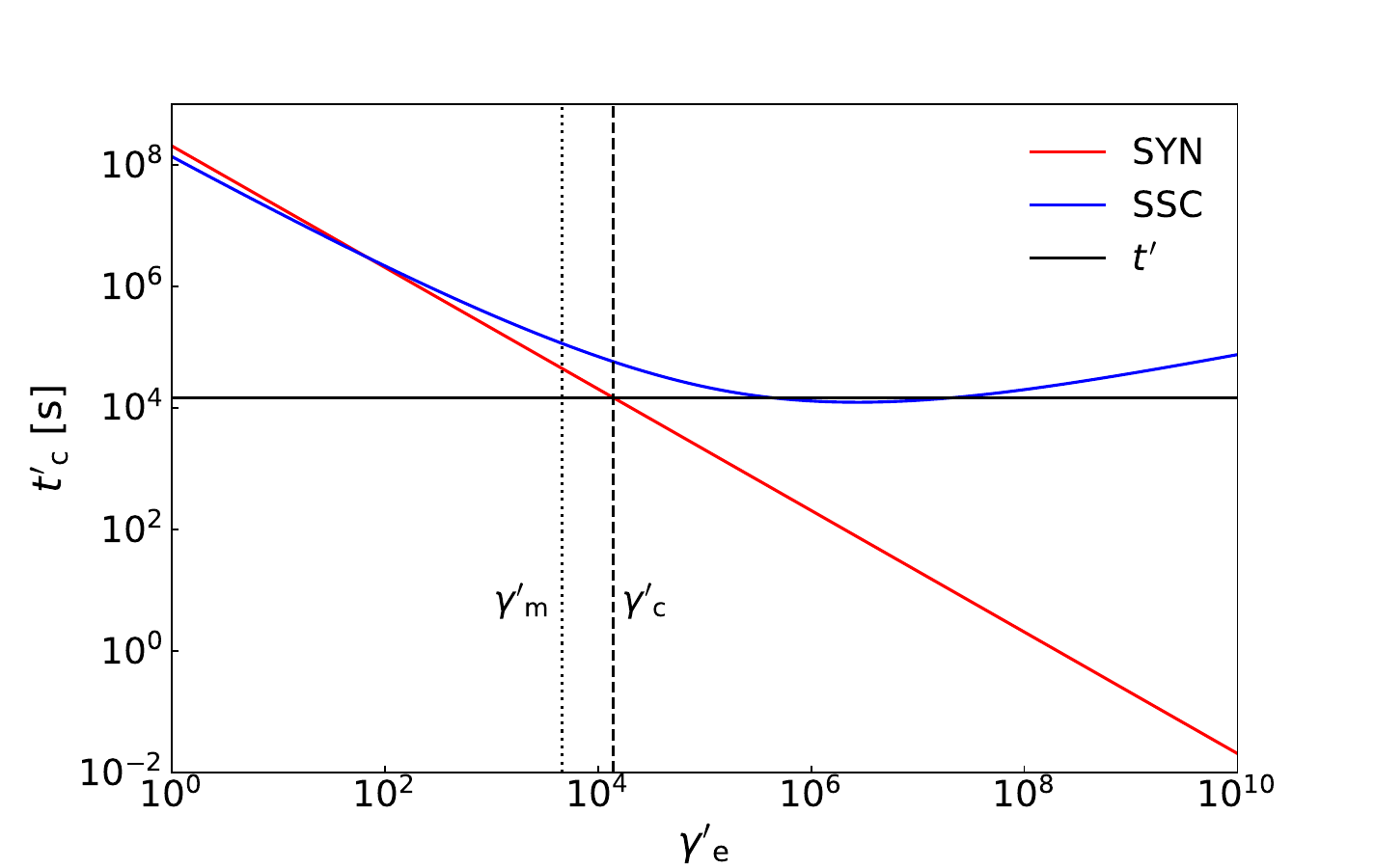}
    \includegraphics[width = 1\linewidth]{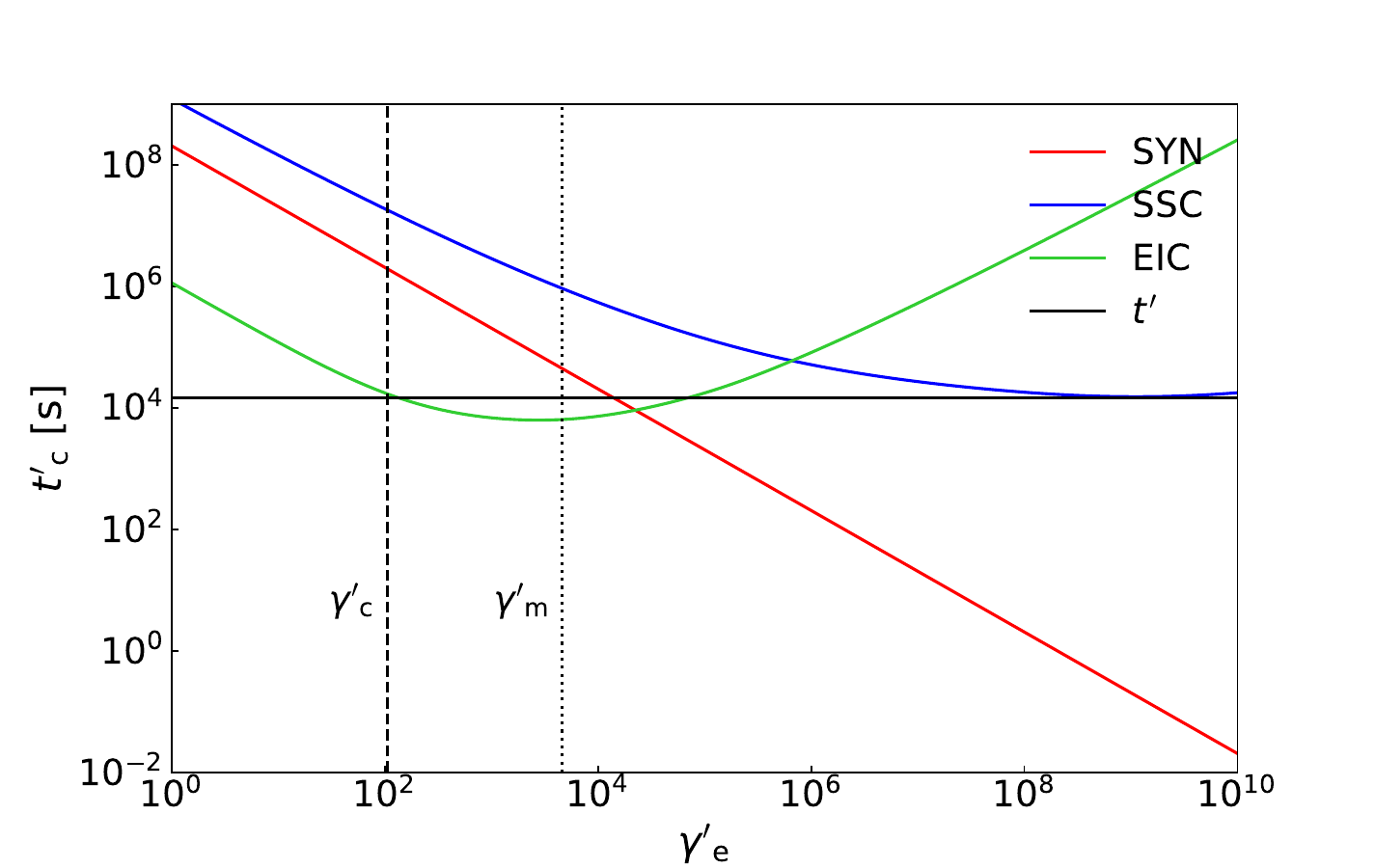}
    \caption{The electron cooling time is evaluated at \(t = t_{\rm d}\). The black vertical dashed and dotted lines represent $\gamma'_e = \gamma'_{\rm c}$ and $\gamma'_e = \gamma'_{\rm m}$, respectively. The upper panel excludes EIC cooling, while the lower panel includes it. We assume $\epsilon_e = 3 \times 10^{-2}$, $p = 2.2$, $\alpha_{\rm l} = 1$, $\alpha_{\rm h} = 2.2$, and $h\nu_{\rm p} = 1 \, \rm MeV$. All other parameters are the same as in Fig.\ref{fig_ratio_Uph_UB}.}
    \label{fig_tc}
\end{figure}

\begin{figure} [h]
    \centering
    \includegraphics[width = 1\linewidth]{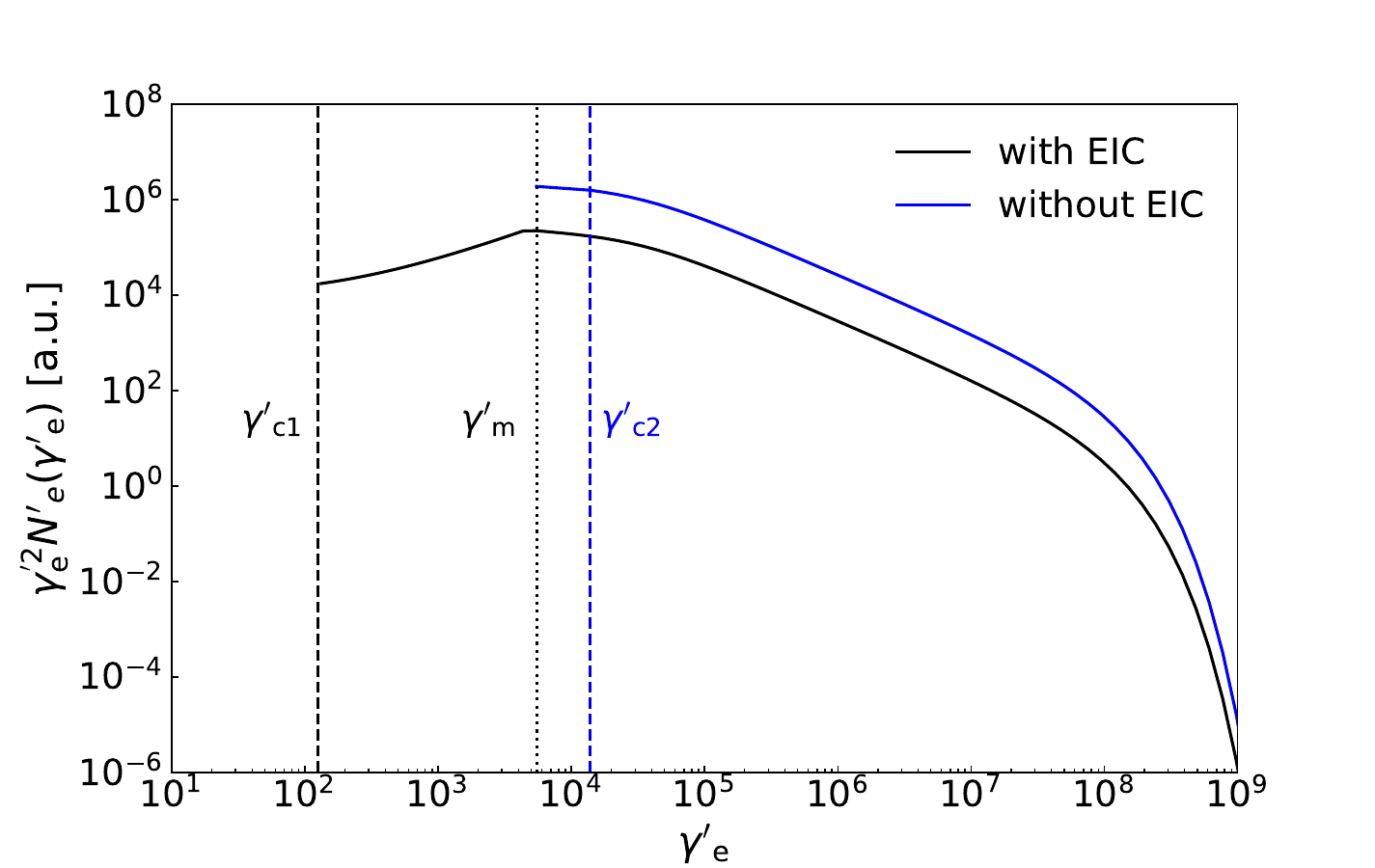}
    \includegraphics[width = 1\linewidth]{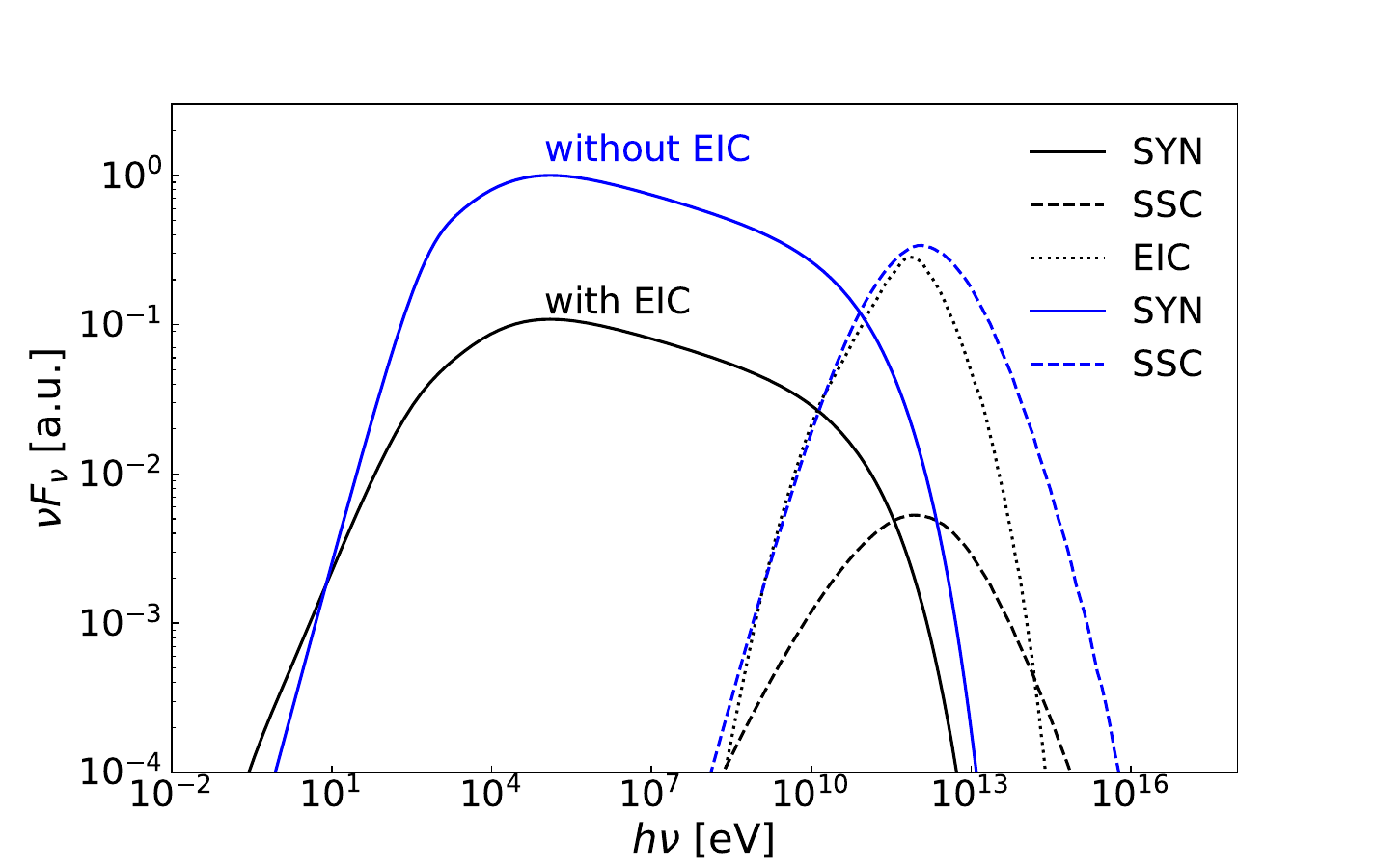}
    \caption{The upper panel shows the electron spectrum at deceleration time $t_{\rm d}$. The black vertical dashed line, blue vertical dashed line, and black vertical dotted line represent $\gamma'_{\rm c1}$, $\gamma'_{\rm c2}$, and $\gamma'_{\rm m}$, respectively. $\gamma'_{\rm c1}$ and $\gamma'_{\rm c2}$ denote the critical Lorentz factors with and without EIC cooling, while $\gamma'_{\rm m}$ is the minimum Lorentz factor, unaffected by EIC. The lower panel displays the radiation spectrum emitted by the electrons shown in the upper panel. The blue line represents the spectrum without EIC cooling, while the black line includes EIC cooling. For simplicity, isotropic scattering is assumed for the EIC process; anisotropic scattering would introduce a suppression factor of approximately 2. For a more detailed and realistic analysis, refer to the numerical results in Sect.\ref{sec_modeling}. The parameters used are the same as in Fig.\ref{fig_tc}.}
    \label{fig_emission}
\end{figure}

Utilizing the same parameters as in Fig.\ref{fig_ratio_Uph_UB}, we compute the electron cooling times for synchrotron, EIC, and SSC processes at the deceleration time \footnote{Before $t_{\rm d}$, the ratio $U'_{\gamma}/U'_B$ is higher, resulting in more significant EIC cooling and emission. After $t_{\rm d}$, $U_{\gamma}$ decreases rapidly with time, as observed in the MeV light curve \citep{An2023}, and the EIC will be surpassed by SSC. For further details, refer to the discussion in Sect.\ref{sec_modeling_result}.}. As shown in Fig.\ref{fig_tc}, EIC cooling significantly reduces the critical Lorentz factor, $\gamma'_{\rm c}$, and causes the electron cooling to transition from the slow cooling regime to the fast cooling regime. From Fig.\ref{fig_emission}, we see that the number of high-energy electrons decreases significantly, shifting towards the lower energy band due to EIC cooling. This shift leads to a reduction in the total power of synchrotron and SSC. However, EIC is enhanced because of the high energy density of prompt MeV photons. As a result, the Compton parameter $Y'$, which measures the ratio of power radiated via inverse Compton (IC) to that radiated via synchrotron emission, will increase significantly.

(2) The scattering regime of EIC: to discuss the EIC regime, where the incident photon spectrum follows the Band function, it is useful to define two Lorentz factors: $\gamma'_{\rm KN1} h \nu'_{\rm p} = \gamma'_{\rm KN2} h \nu'_{\rm min} = m_e c^2$, where $h\nu'_{\rm min}$ represents the minimum cut-off energy of the prompt emission. These Lorentz factors, $\gamma'_{\rm KN1}$ and $\gamma'_{\rm KN2}$, determine the strength of the KN effect. Since the energy density of the prompt emission $U'_{\gamma}$ is primarily contributed by photons with energy $\sim h\nu'_{\rm p}$, electrons with Lorentz factors $\gamma'_e < \gamma'_{\rm KN1}$ remain within the Thomson regime. For $\gamma'_e \gtrsim \gamma'_{\rm KN1}$, while $h\nu'_{\rm p}$ enters the KN regime, the majority of the photon energy density $U'_{\gamma}$, which is dominated by photons with $h\nu'_{\rm in} \lesssim h\nu'_{\rm p}$, still lies within the Thomson regime. This leads to a relatively mild KN effect on electron cooling. As $\gamma'_e$ continues to increase, the scattering gradually shifts deeper into the KN regime. Finally, for $\gamma'_e \geqslant \gamma'_{\rm KN2}$, the minimum cut-off energy of the prompt emission spectrum, $h\nu'_{\rm min}$, enters the KN regime, leading to scattering in the deep KN regime.

For the case of GRB 221009A, $h\nu'_{\rm p} \sim 1 \, \rm MeV / (2\Gamma_{\rm ext})$. Before deceleration, $\Gamma_{\rm ext} \sim 500$, and $h\nu'_{\rm p} \sim 1 \, \rm keV$, implying that electrons with Lorentz factors $\gamma'_e < \gamma'_{\rm KN1} \sim 500$ remain within the Thomson regime. For $\gamma'_e > \gamma'_{\rm KN2} \sim  10^4$, the minimum cut-off energy of the prompt emission spectrum, $h\nu'_{\rm min} = 20 \, \rm keV / (2\Gamma_{\rm ext})$, enters the KN regime \footnote{The spectrum is assumed to follow a single power-law from $\sim 20 \, \rm keV$ to $1 \, \rm MeV$, and no observations are available below $20 \, \rm keV$ \cite{An2023}. Therefore, we adopt $ h\nu_{\rm min} = 20 \, \rm keV $ as the minimum energy of the prompt emission. If the actual value is lower, $\gamma'_{\rm KN2}$ will increase and surpass $10^4$.}. At larger $\gamma'_e$, the scattering process shifts into the deep KN regime. As shown in Fig.\ref{fig_tc}, the EIC cooling time is proportional to ${\gamma'_e}^{-2}$ up to $\gamma'_e > \gamma'_{\rm KN1} \sim 500$, consistent with the prediction from IC cooling in the Thomson regime. Beyond this point, EIC cooling gradually transitions into the KN regime, and for $\gamma'_e > \gamma'_{\rm KN2} \sim 10^4$, EIC occurs entirely in the deep KN regime. The observed TeV photons are mainly contributed by electrons with $\gamma'_e \gtrsim 10^3 \sim \gamma'_{\rm KN1}$, indicating that the EIC TeV light curve remains only mildly affected by the KN regime before shock deceleration. After deceleration, both \(h\nu'_{\rm out} = \frac{1 \, \rm TeV}{2\Gamma_{\rm ext}}\) and \(\nu'_{\rm p} = \frac{\nu_{\rm p}}{2\Gamma_{\rm ext}}\) increase over time, as \(\Gamma_{\rm ext} \propto t^{-3/8}\) under the assumption of adiabatic expansion. This results in a more significant KN effect. In this phase, $\gamma'_{\rm KN1} = \frac{2m_e c^2}{h\nu_{\rm p}} \Gamma_{\rm ext} \propto t^{-3/8}$ and $\gamma'_{\rm KN2} = \frac{2m_e c^2}{h\nu_{\rm min}} \Gamma_{\rm ext} \propto t^{-3/8}$.

%(4) the peak of EIC?

%(5) the spectra of EIC?

\subsection{multi-band afterglow modeling}

\label{sec_modeling}

In this section, we use numerical approaches to model the light curve and spectrum of the early afterglow assuming a top-hat jet, which is valid at an early time when the influence of the wide wing of the structured jet can be ignored \citep{Zheng2024, ZhangBing2024}. The circumburst density is assumed to be roughly uniform \citep{Zheng2024, Ren2024}. %After that we examine how EIC affects the early afterglow light curve.

\subsubsection{numerical code for modeling}

\label{sec_modeling_method}

The numerical code employed is based on \cite{Liu2013}, with the dynamics of the external shock described by \cite{Huang1999}. The calculation of electron spectra follows the method outlined in Sect.\ref{sec_ana_shock_EIC_cooling}, incorporating cooling effects from synchrotron, SSC, and EIC processes. Emission from the external shock is modeled as a series of "flashes", each corresponding to a specific time grid. Additionally, the equal arrival time surface (EATS) has been integrated into the code (see also \cite{Granot1999, Fan2008, Murase2011}): for a single "flash," the relationship between the emitted photon flux \( \mathcal{S}(\nu_{\rm out}, \Omega_{\rm L}) d\Omega_{\rm L} \Delta t_{\rm g} \, (\rm photon \, Hz^{-1} \, sr^{-1}) \) at $R_{\rm ext}$ and the observed flux \( F_{\rm obs}(\nu_{\rm out}) \, (\rm photon \, Hz^{-1} \, s^{-1} \, cm^{-2}) \) at point $E$ (see Fig.\ref{fig_jet}) is governed by the conservation of photon numbers:
\begin{equation}
\label{eq_eats}
    \mathcal{S}(\nu_{\rm out}, \Omega_{\rm L}) d\nu_{\rm out}  d\Omega_{\rm L}  d\Omega_{\rm out}  \Delta t_{\rm g} = F_{\rm obs}(\nu_{\rm out}) d\nu_{\rm out}  dt dA_{\rm obs},
\end{equation}
where $\Delta t_{\rm g}$ is the width of the time grid, $dA_{\rm obs}$ is the area of the receiver at point $E$, $d\Omega_{\rm out} = \frac{dA_{\rm obs}}{D_{\rm L}^2}$ is the solid angle of $dA_{\rm obs}$ as seen from the source and $D_{\rm L}$ is the luminosity distance. The time delay for photons arriving from $\theta_{\rm L}$ to $\theta_{\rm L} + d\theta_{\rm L}$ is given by $dt = -\frac{R_{\rm ext} d\cos\theta_{\rm L}}{c}$.

The jet composition of GRB 221009A has been suggested to be dominated by Poynting flux \citep{Dai2023}, which favors a relatively large internal dissipation radius $R_{\rm dis}$ (e.g. in the ICMART model), potentially reaching $\sim 10^{16} \, \rm cm$ and even comparable to the deceleration radius $R_{\rm dec}$ of the external shock during the early coasting phase. In this scenario, we may calculate the EIC flux with the approximation in the isotropic limit, as discussed in Sect.\ref{sec_EIC_condition}. On the other hand, if the internal shock scenario works for the prompt emission of this GRB, we would expect $R_{\rm dis} \ll R_{\rm ext}$. For comparison, we also calculate the EIC flux in the anisotropic limit, which can give a reasonable approximation in this scenario as discussed in Sect.\ref{sec_EIC_condition}. As will be shown below, the model parameters obtained in these two scenarios do not show significant differences.

\subsubsection{the results of the modeling}

\label{sec_modeling_result}

\begin{figure} [h]
    %\centering
    \includegraphics[width = 1\linewidth]{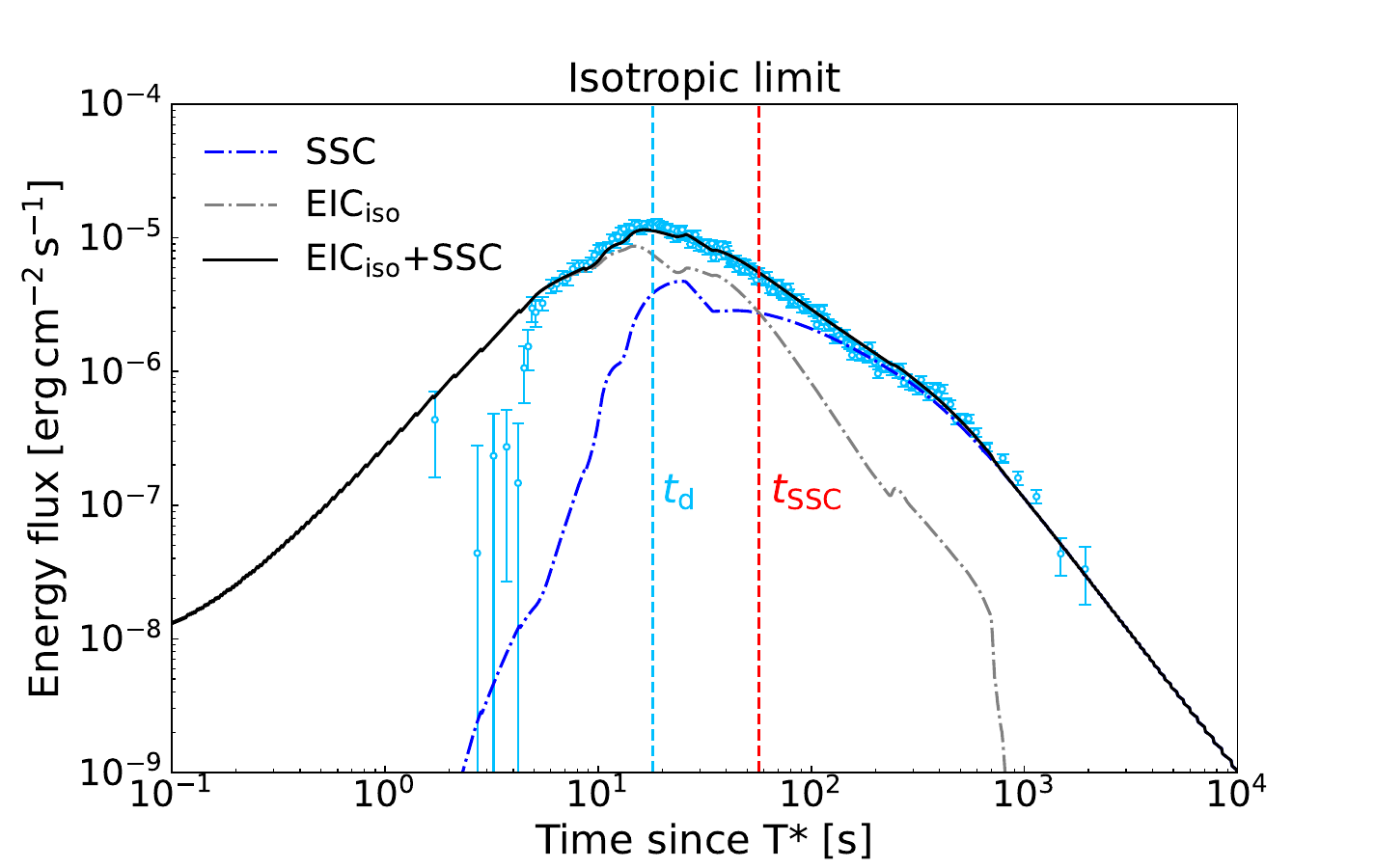} \\
    \includegraphics[width = 1\linewidth]{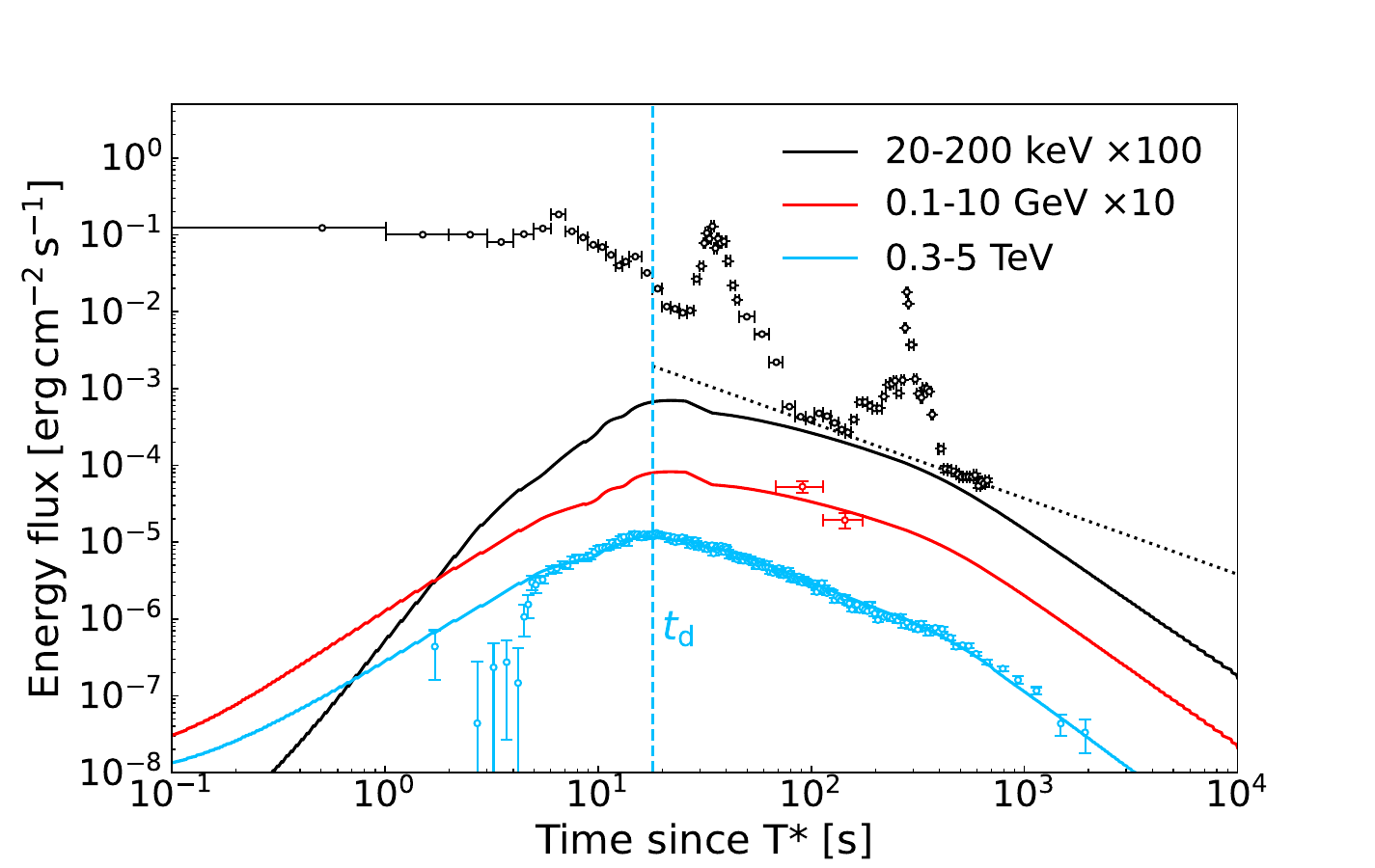} \\
    \includegraphics[width = 1\linewidth]{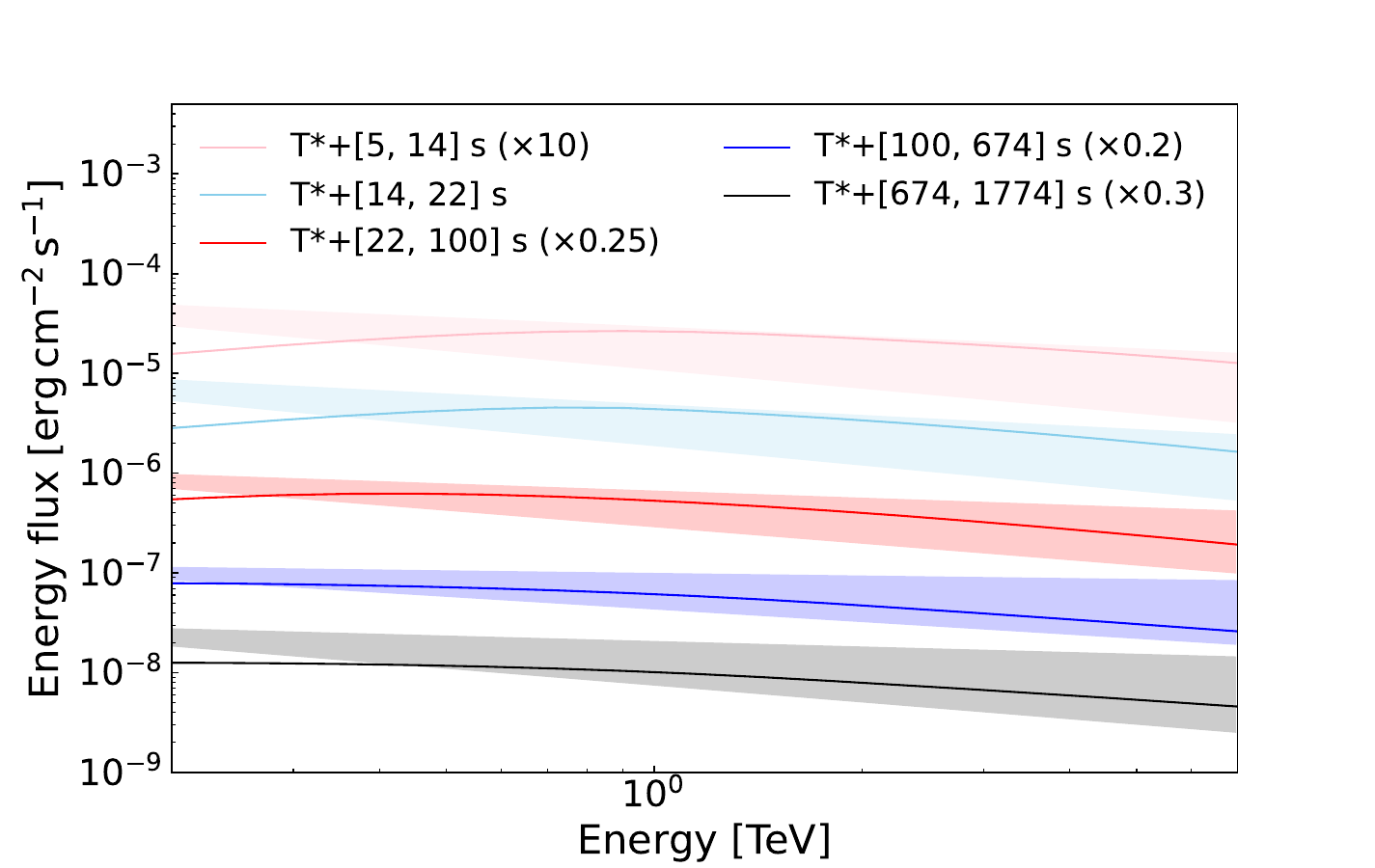}
    \caption{The figure shows synthetic light curves and spectra from the isotropic EIC model. Parameters: $E_{k0, \rm iso} = 5 \times 10^{54} \, \rm erg$, $\Gamma_{\rm ext, 0} = 540$, $\theta_{\rm j} = 0.4^{\circ}$, $n = 0.28 \, \rm cm^{-3}$, $\epsilon_e = 3 \times 10^{-2}$, $\epsilon_B = 10^{-3}$, and $p = 2.2$. We set $h\nu_{\rm min} = 20 \, \rm keV$ (see footnote 1). A test with $h\nu_{\rm min} = 2 \, \rm eV$ shows minimal change. The top panel shows TeV afterglow from EIC and SSC, with blue/red vertical dashed lines for $t = t_{\rm d}$ and $t = t_{\rm SSC}$. The center panel presents multi-band light curves (SYN+SSC+EIC), with the black dotted line showing the early afterglow (see text). The bottom panel shows TeV spectra from EIC+SSC at different times. GeV/TeV data from \cite{LHAASO2023}, X-ray from \cite{An2023}.}
    \label{fig_lc_iso}
\end{figure}

\begin{figure} [h]
    %\centering
    \includegraphics[width = 1\linewidth]{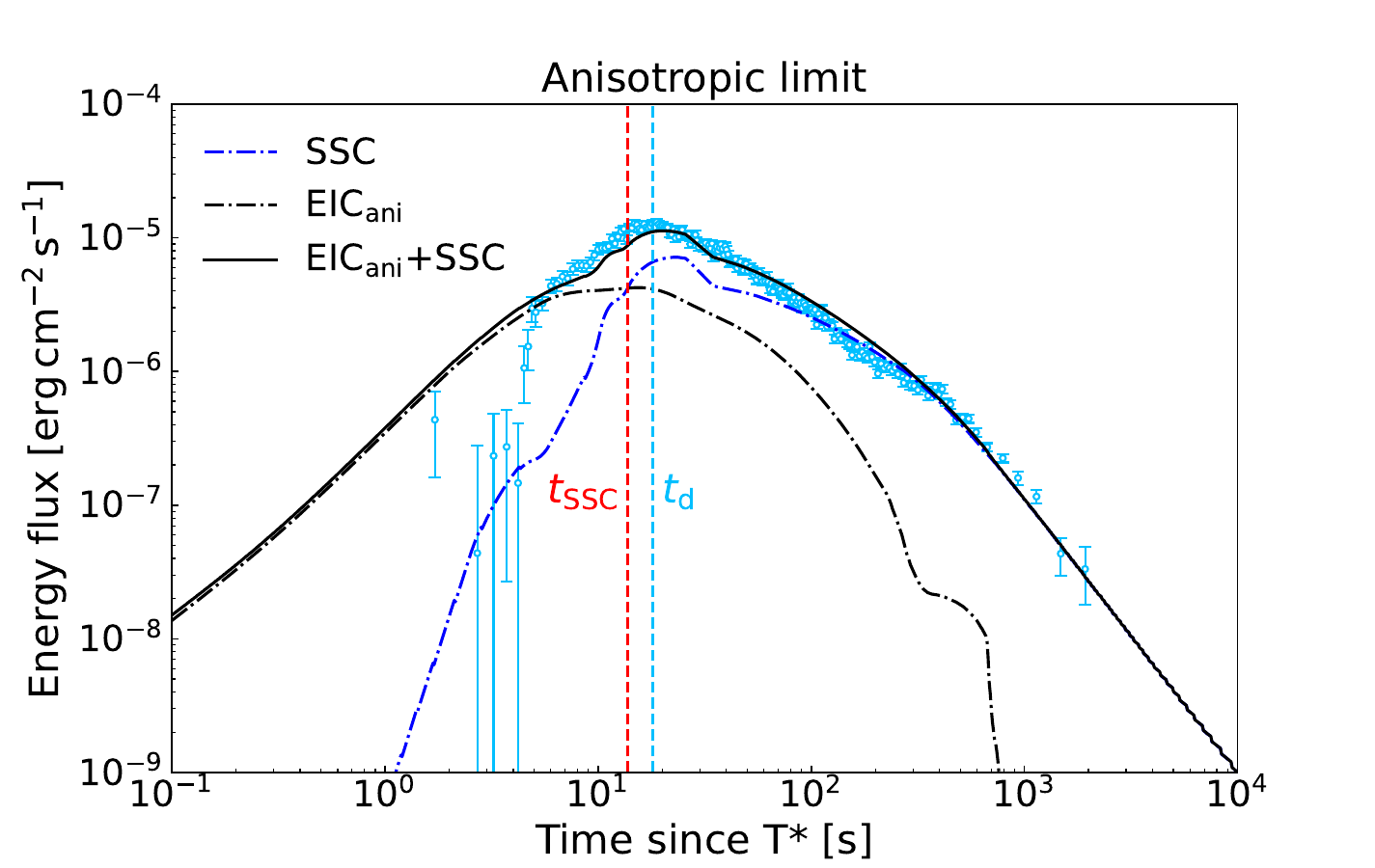} \\
    \includegraphics[width = 1\linewidth]{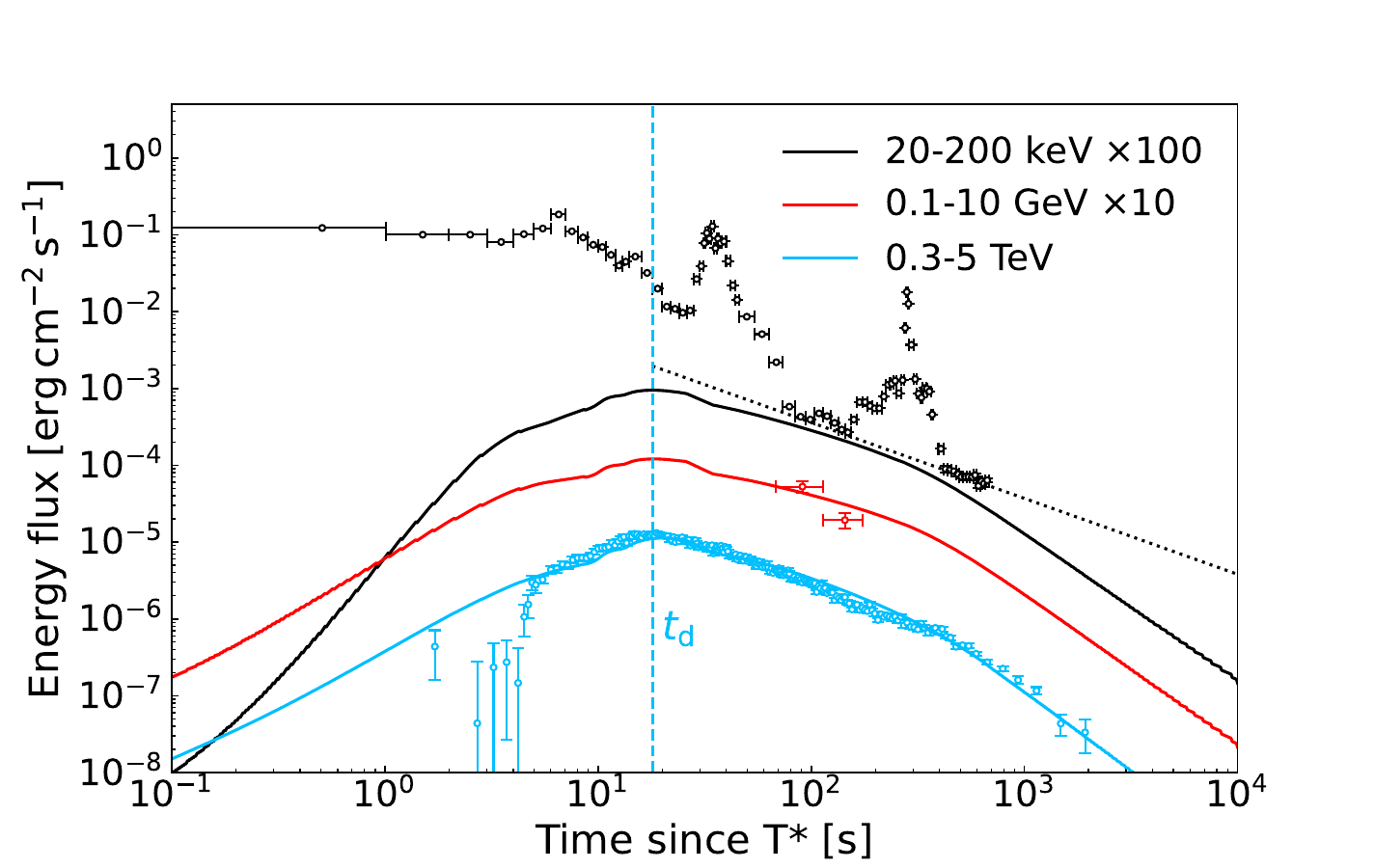} \\
    \includegraphics[width = 1\linewidth]{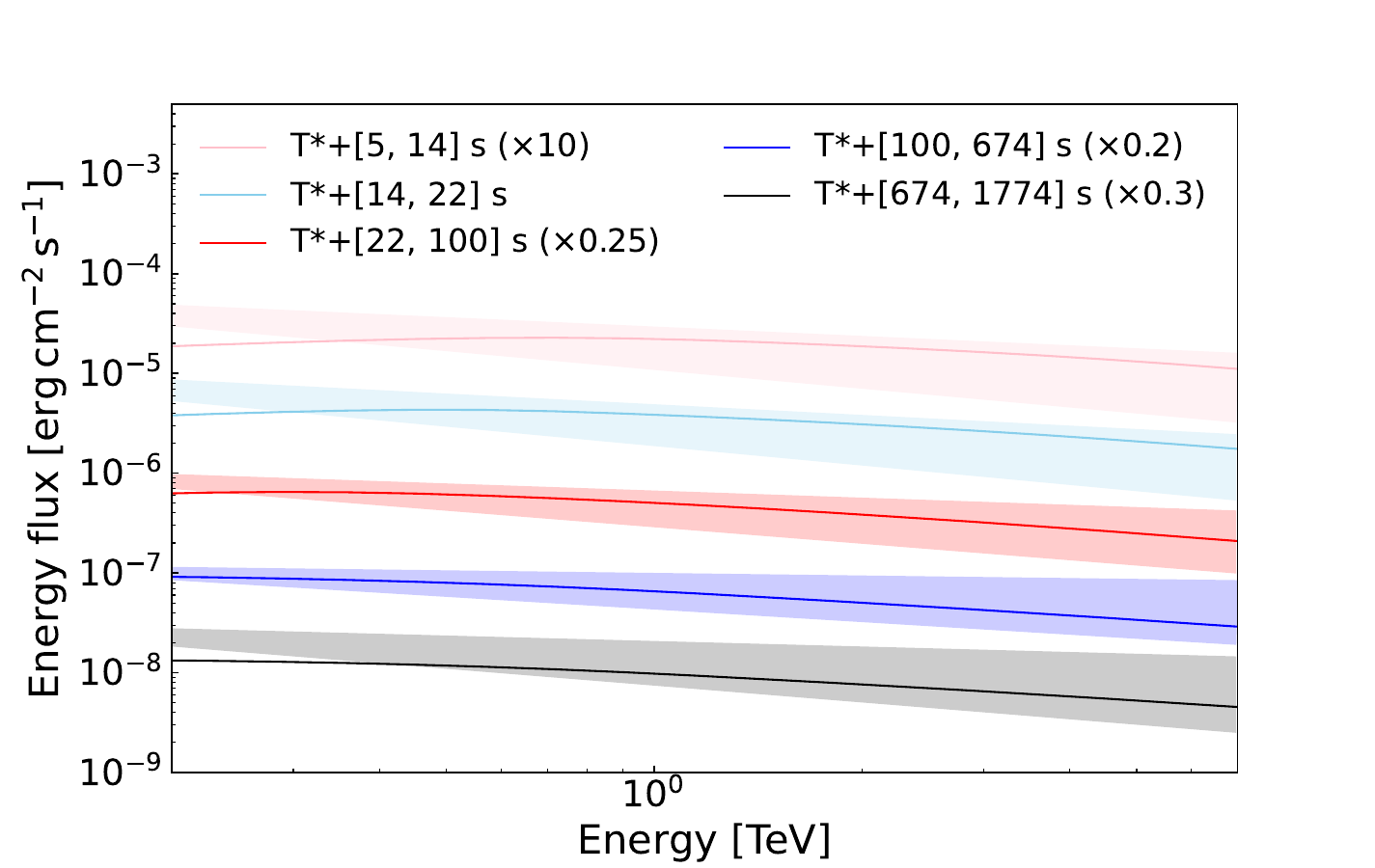} 
    \caption{Similar to Fig.\ref{fig_lc_iso}, but assuming the anisotropic limit for EIC. The modeling parameters are: $E_{k0, \rm iso} = 5 \times 10^{54} \, \rm erg$, $\Gamma_{\rm ext, 0} = 630$, $\theta_{\rm j} = 0.4^{\circ}$, $n = 0.4 \, \rm cm^{-3}$, $\epsilon_e = 3 \times 10^{-2}$, $\epsilon_B = 10^{-3}$, and $p = 2.2$.}
    \label{fig_lc_ani}
\end{figure}

The modeling results are presented in Fig.\ref{fig_lc_iso} and Fig.\ref{fig_lc_ani}, with approximations in the isotropic and anisotropic EIC limits, respectively. 
%In this section, we have not yet considered the $\gamma\gamma$ absorption from prompt MeV photons, which will be addressed in Sect.\ref{sec_gg_abs}.
The upper panels of Fig.\ref{fig_lc_iso} and Fig.\ref{fig_lc_ani} show the TeV flux contributions from the EIC and SSC processes, with both light curves peaking around the deceleration time $t_{\rm d}\sim$\(T^* + 18 \, \rm s\). This peak is attributed to the deceleration of the external shock: before the shock's deceleration, the EIC flux rises with \(t^2\), assuming a uniform circumburst density (i.e., \(k=0\), with \(n \propto R_{\rm ext}^{-k}\)). After the shock's deceleration, the EIC flux declines over time, as discussed in Sect.\ref{sec_discussion} and Table.\ref{table_Feic}. Additionally, the SSC flux also peaks at the deceleration time, as noted in \cite{LHAASO2023}. Thus, regardless of whether the observed TeV flux is dominated by EIC or SSC, the peak of the TeV light curve always marks the onset of external shock deceleration for GRB 221009A. In our numerical results, the SSC peak occurs slightly later than the EIC peak. This is due to the rapid decrease in the luminosity of the prompt MeV emission near $t_{\rm d}$, which reduces electron cooling from EIC and enhances the synchrotron and SSC emission.

The upper panels of Fig.\ref{fig_lc_iso} and Fig.\ref{fig_lc_ani} also illustrate the transition time \(t_{\rm SSC}\) for the TeV flux, marking the transition from EIC dominance to SSC dominance. In the isotropic EIC scenario, the TeV flux from SSC surpasses that from EIC at \(t_{\rm SSC} \sim T^*+ 60 \, \rm s\). $t_{\rm SSC}$ can be roughly inferred from MeV observations.  Before shock deceleration, the energy density of the MeV afterglow, estimated from the black dotted line, is at least an order of magnitude lower than that of the prompt emission even considering the moderate suppression factor due to the KN effect (as discussed in Sect.\ref{sec_ana_shock_EIC_cooling}). Therefore, the EIC process is expected to be dominated in the TeV afterglow before deceleration, regardless of the exact external shock parameters. After deceleration, the luminosity of prompt emission in the MeV band decreases rapidly and is overtaken by the afterglow at \(T^* + 74 \, \rm s\) \citep{zhang2023}{\footnote { \cite{zhang2023} noted that the MeV light curve during \(T^*+[74, 150] \, \rm s\) and \(T^*+[400, 1000] \, \rm s\) follows a decaying power law (indicated by the black dotted line in the middle panels of Fig.\ref{fig_lc_iso} and Fig.\ref{fig_lc_ani}), likely representing the underlying afterglow emission rather than prompt emission.  Consequently, EIC is surpassed by SSC around this time.}}. In our numerical results (see the upper panel of Fig.\ref{fig_lc_iso}), the transition time \( t_{\rm SSC} \sim T^* + 60 \, \rm s \), which is not strictly equal to \( T^* + 74 \, \rm s \) for two reasons: first, the KN effect in the EIC process is more significant than in the SSC process during the deceleration phase, as the energy of incident photons in the comoving frame of the external shock increases as \(\Gamma_{\rm ext}\) decrease. However, this effect is small since the transition time is close to the deceleration time, and \(\Gamma_{\rm ext}\) decreases slowly with time (\(\Gamma_{\rm ext} \propto t^{-\frac{3}{8}}\)). The second reason is that the afterglow emission extends to photons below \(20 \, \rm keV\), so the bolometric afterglow luminosity is higher. These two factors lead to a transition time at \(t_{\rm SSC} \sim T^* + 60 \, \rm s\), which is slightly earlier than \(T^* + 74 \, \rm s\). In the anisotropic EIC scenario, the TeV flux is dominated by SSC at an earlier time due to the anisotropic scattering effect, which leads to a suppression factor of about 0.3.
%which results in about \(60\%\) of the EIC flux being scattered out of LOS compared to isotropic scattering, as discussed in the previous section.

The middle panel illustrates the multi-band modeling of the afterglow light curves. As shown in the figure, our model explains the multi-band afterglows, with MeV and GeV emissions primarily attributed to the synchrotron processes, while the TeV afterglow results from contributions by EIC and SSC.

The lower panel displays the modeling of the TeV spectra at different observation times. Overall, our model successfully explains the TeV spectra. In the early time intervals (\(T^*+[5,14] \, \rm s\) and \(T^*+[14,22] \, \rm s\), corresponding to the pink and sky blue lines), the spectrum is dominated by EIC and in the late time intervals, the SSC emission is dominated. Since the EIC peak is higher than \(\sim 0.2 \, \rm TeV\), the EIC flux falls slightly below the observed values by a factor of \(\sim 0.5\) at \(\sim 0.2 \, \rm TeV\). This may be contributed to by additional components, such as EIC from thermal electrons.

Finally, we would like to point out that the modeling parameters for the multi-band data of GRB 221009A are not significantly affected by the inclusion of the EIC process. To justify this, we compare the TeV flux in the two scenarios: the first one includes EIC, where the flux \(F^{\rm EIC}_{\nu_{\rm out}}\) dominates the TeV flux at \(t_{\rm d}\); the second one ignores EIC (i.e, the TeV flux is only contributed by the SSC process) represented by \(F^{\rm SSC}_{\nu_{\rm out}}\). The ratio between \(F^{\rm EIC}_{\nu_{\rm out}}\) and \(F^{\rm SSC}_{\nu_{\rm out}}\) at \(t_{\rm d}\) is given by (assuming \(k = 0\) and \(p = 2.2\), see Sect.\ref{sec_EIC_effect}):
\begin{align}
    \frac{F^{\rm EIC}_{\nu_{\rm out}}}{F^{\rm SSC}_{\nu_{\rm out}}} &= 0.9 \left(\frac{n}{0.4 \, \rm cm^{-3}}\right)^{-0.05} \left( \frac{\Gamma_{\rm ext, 0}}{630}\right)^{-0.4} \notag \\
    &\times \left( \frac{\epsilon_e}{3\times 10^{-2}}  \right)^{-0.2} \left( \frac{\epsilon_B}{10^{-3}} \right)^{-0.05}.
\end{align}
Thus,  we can see that the TeV flux at \(t_{\rm d}\) are similar in the two scenarios and remains insensitive to shock parameters, provided that \(1 \,\rm TeV > \max(\nu^{\rm EIC}_{\rm m}, \nu^{\rm EIC}_{\rm c})\) in the EIC case and \(1 \,\rm TeV > \max(\nu^{\rm SSC}_{\rm m}, \nu^{\rm SSC}_{\rm c})\) in the case without EIC at \(t_{\rm d}\). Furthermore, if EIC is included, its flux dominates the TeV light curve before the shock deceleration, rising as \(t^2\), which is consistent with the behavior predicted in the SSC scenario  \citep{LHAASO2023}. Therefore, EIC does not significantly alter the expected TeV light curve before \(t_{\rm d}\). The X-ray and GeV afterglows are observed at relatively late times when EIC has been surpassed by SSC. Consequently, the model parameters are primarily constrained by multi-band fitting during the SSC-dominant phase,  regardless of whether the EIC is included or not.

\section{The $\gamma\gamma$ annihilation between the prompt emission and the afterglow}

\label{sec_gg_abs}

Since the external shock is immersed by prompt MeV photons during the early stages, the absorption of the TeV afterglow by these photons should be considered. In Sect.\ref{Sec_abs_twozone} and Sect.\ref{Sec_abs_twozone_R}, we study the $\gamma\gamma$ absorption within the two-zone model and explore how the optical depth $\tau_{\gamma\gamma}$ for photons from different latitudes $\theta_{\rm L}$ is influenced by the ratio $R_{\rm dis}/R_{\rm ext}$, which determines the angular distribution of the incident photon field. We then explore the $\gamma \gamma$ optical depth from the prompt emission in both the internal shock scenario and the magnetic field dissipation/turbulence scenario to assess its impact on the early TeV afterglow in Sect.\ref{sec_abs_multi_zone}.

\subsection{$\gamma \gamma$ absorption in the two-zone model}
\label{Sec_abs_twozone}
To estimate the \(\gamma \gamma\) absorption during the early afterglow, we first consider a simplified scenario: the prompt MeV photon field, generated as a "flash" at \(R_{\rm dis}\), interacts with afterglow photons, also emitted as a "flash" at \(R_{\rm ext}\) (see Fig.\ref{fig_gmgm_abs}). This model is consistent with the two-zone model used in the EIC calculation, with the primary difference being that the EIC process occurs only at \(R_{\rm ext}\) (see Fig.\ref{fig_jet}), while the afterglow photons injected at \(R_{\rm ext}\) are continuously absorbed by prompt emission photons as they propagate outward. In this scenario, the optical depth for \(\gamma \gamma\) absorption is given by
\begin{align}
\label{Eq_tau_gm_gm}
    &\tau_{\gamma\gamma}(\nu_{\rm ext}) = \int \, dl \int \, d\Omega_{\rm in} \\
    &\times \int_{\nu_{\rm c}}^\infty d\nu_{\rm dis} \, n_{\rm ph}(\nu_{\rm dis}, \Omega_{\rm in}, l) \sigma_{\gamma\gamma}(\nu_{\rm ext}, \nu_{\rm dis}, \psi)(1-\cos\psi), \notag
\end{align}
where \(n_{\rm ph}(\nu_{\rm dis}, \Omega_{\rm in}, l) \, (\rm photon \, Hz^{-1} \, sr^{-1} \, cm^{-3})\) is the photon density at interaction location, i.e., point $M$ in Fig.\ref{fig_gmgm_abs}. \(\nu_{\rm ext}\) and \(\nu_{\rm dis}\) represent the frequencies of the afterglow and prompt emission photons, respectively. The energy threshold is given by \(h\nu_{\rm c} = \frac{2(m_e c^2)^2}{h\nu_{\rm ext}(1-\cos \psi)}\), where \(\psi\) is the interaction angle between the prompt emission and afterglow photons. \(l\) is the path length of the afterglow photons, and \(\sigma_{\gamma \gamma}\) is the \(\gamma\gamma\) cross-section. The differential solid angle is expressed as \(d\Omega_{\rm in} = \sin \theta_{\rm in} d\theta_{\rm in} d\phi_{\rm in}\), where \(\theta_{\rm in}\) and \(\phi_{\rm in}\) denote the polar and azimuthal angles of the incident photon direction (see Fig.\ref{fig_gmgm_abs}). Given the photon number density \(n_{\rm ph}(\nu_{\rm dis}, \Omega_{\rm in}, l) \, (\rm photon \, Hz^{-1} \, sr^{-1} \, cm^{-3})\) and the interaction angle \(\psi\) from Eq.(\ref{A_Eq_n1_gg}) and Eq.(\ref{A_Eq_cospsi}), the optical depth \(\tau_{\gamma\gamma}(\nu_{\rm ext})\) can be obtained using Eq.(\ref{Eq_tau_gm_gm}).

\subsection{The impact of  $R_{\rm dis}/R_{\rm ext}$  on $\tau_{\gamma \gamma}$}
\label{Sec_abs_twozone_R}

As discussed in Sect.\ref{sec_EIC_in_jet}, the scattering rate of EIC is influenced by the angular distribution of the photon field. Similar to the EIC process, \(\gamma \gamma\) absorption is also affected by the degree of anisotropy in the prompt emission photons, which depends on \(R_{\rm dis}/R_{\rm ext}\).

\begin{figure} [h]
    \centering
    \includegraphics[width = 1\linewidth]{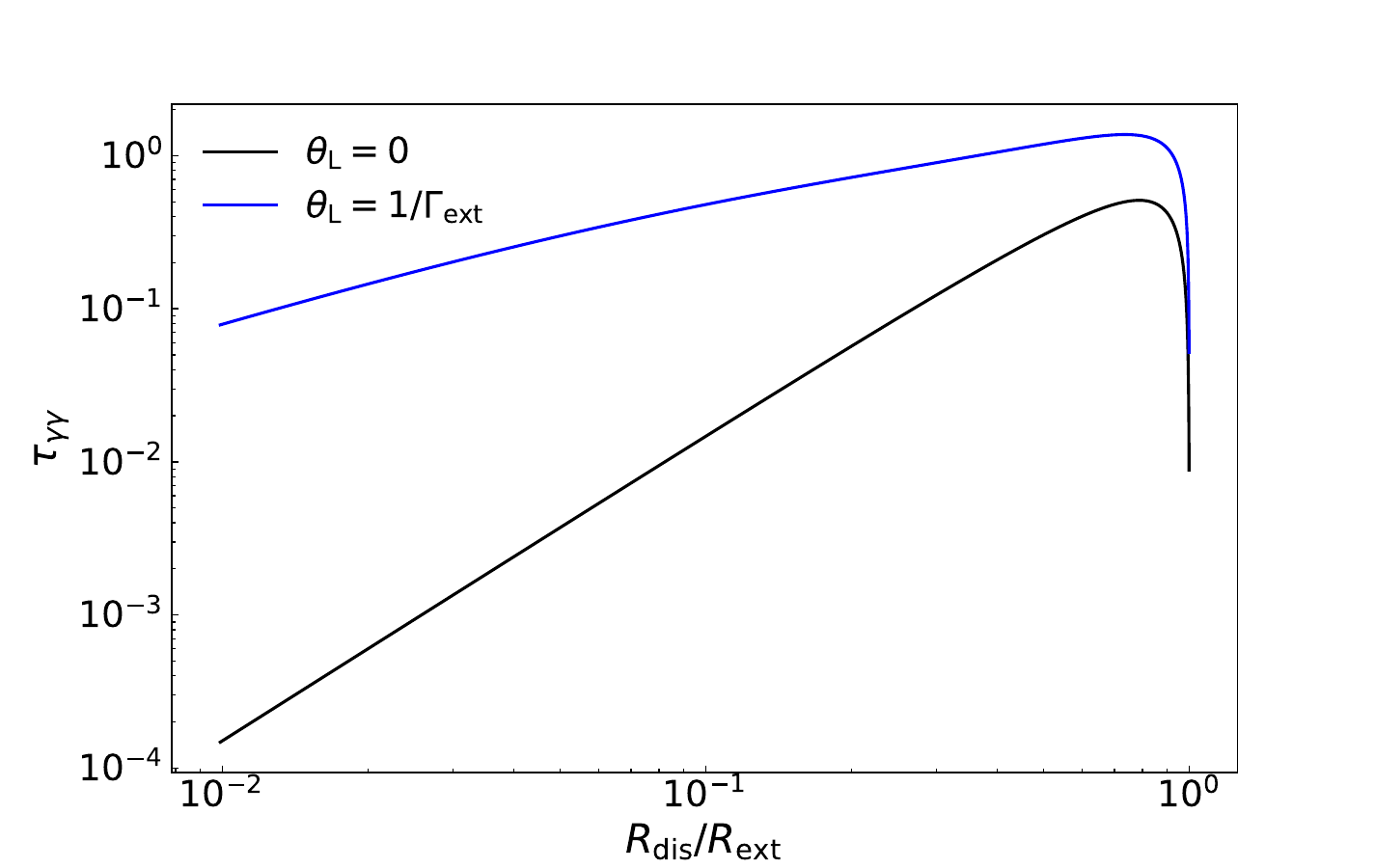}
    \caption{The $\gamma \gamma$ absorption at $1 \, \rm TeV$ in the two-zone model. We set $\alpha_{\rm l} = 1$, $\alpha_{\rm h} = 2.2$, and $h\nu_{\rm p} = 1 \, \rm MeV$ for the prompt emission spectrum. The Lorentz factors are set as $\Gamma_{\rm dis} = \Gamma_{\rm ext} = 500$, with $R_{\rm ext}$ fixed at $10^{17} \, \rm cm$ and $L_{\gamma, 0} = 10^{53} \, \rm erg \, s^{-1}$.}
    \label{fig_tau_flash}
\end{figure}

As a case study, we examine a specific scenario within the two-zone model (see Fig.\ref{fig_gmgm_abs}): the afterglow photons are injected at point $O$, while the prompt emission photons, originating from $R_{\rm dis}$ and propagating from $\zeta_1 = 1/\Gamma_{\rm dis}$, reach point $O$ at the same time as the afterglow photons are injected. Subsequently, the afterglow photons propagate outward, continuously interacting with the prompt emission photons arriving from different latitudes $\zeta_2$. As discussed in Sect.\ref{sec_exa_EIC}, the geometry of the interaction depends on the path length \(l\) of the afterglow photons (see Fig.\ref{fig_gmgm_abs}). The prompt emission photons always incident at a fixed angle at the interaction location, which implies \(n_{\rm ph} \propto \delta(\cos\theta_{\rm in} - \cos\zeta_3)\). We maintain the photon number density $n_{\rm ph}(\nu_{\rm dis}, \Omega_{\rm in}, l)$ (see Eq.(\ref{A_Eq_n1_gg})) of the prompt emission at point $O$ (i.e., $l = 0$) and only vary $R_{\rm dis}$ to observe how the anisotropy of the prompt emission photon field affects $\tau_{\gamma \gamma}$, as shown in Fig.(\ref{fig_tau_flash}): (1) For \(\theta_{\rm L} = 1/\Gamma_{\rm ext}\), \(R_{\rm dis}/R_{\rm ext}\) primarily influences the photon number density at the interaction location as afterglow photons propagate outward, while the interaction angle \(\psi\) remains insensitive to \(R_{\rm dis}/R_{\rm ext}\). When \(R_{\rm dis} \ll R_{\rm ext}\), the Doppler factor \(D_{\rm dis} = \frac{1}{\Gamma_{\rm dis}(1 - \beta_{\rm dis}\cos \zeta_1)}\) decreases more rapidly as afterglow photons move outward, suppressing the photon number density \(n_{\rm ph}(\nu_{\rm dis}, \Omega_{\rm in}, l)\) and resulting in weaker \(\gamma\gamma\) absorption. Conversely, if \(R_{\rm dis}/R_{\rm ext}\) approaches 1, the incident photon density at the interaction location will also decrease rapidly as afterglow photons propagate outward. This is because the incident photon density is normalized at \(l = 0\), so for \(l > 0\), the photon density \(n_{\rm ph}(\nu_{\rm dis}, \Omega_{\rm in}, l) \propto s_0/s\) (see Eq.(\ref{A_Eq_n2_gg})), where \(s_0\) and \(s\) represent the path lengths of prompt emission photons that interact with afterglow photons at points \(O\) and \(M\), respectively (see Fig.\ref{fig_gmgm_abs}). Hence, \(n_{\rm ph}(\nu_{\rm dis}, \Omega_{\rm in}, l)\) is concentrated around locations where \(s < s_0\). As \(R_{\rm dis}/R_{\rm ext}\) approaches 1, this results in a small \(s_0\), causing the photon density of prompt emission to be more concentrated around point \(O\) and decreasing rapidly as afterglow photons propagate outward, weakening \(\gamma\gamma\) absorption. (2) For \(\theta_{\rm L} = 0\), the interaction angle \(\psi\) is sensitive to \(R_{\rm dis}/R_{\rm ext}\), similar to the EIC scenario. When \(R_{\rm dis} \ll R_{\rm ext}\), the interaction angle is significantly suppressed. As a result, \(\tau_{\gamma\gamma}\) decreases more rapidly with \(R_{\rm dis}\) compared to the case where \(\theta_{\rm L} = 1/\Gamma_{\rm ext}\).

Hence, $\tau_{\gamma \gamma}$ reaches its maximum when $R_{\rm dis} \sim R_{\rm ext}$. As $R_{\rm dis}$ decreases, the incident photon field becomes more anisotropic, resulting in a reduction of $\tau_{\gamma \gamma}$, especially for the case with angles $\theta'_{\rm L} \sim 0$.

\subsection{$\gamma \gamma$ absorption in the internal shock and magnetic field dispassion/turbulence scenarios}
\label{sec_abs_multi_zone}

As previously discussed, the radius of internal dissipation, where the prompt emission is generated, is highly uncertain and can affect the optical depth of \(\gamma \gamma\) absorption. In the internal shock scenario, the typical dissipation radius \(R_{\rm dis} \ll R_{\rm ext}\), whereas in magnetic field dissipation/turbulence scenarios, \(R_{\rm dis}\) is typically larger and may even be comparable to the deceleration radius. Here, we calculate the \(\gamma\gamma\) absorption in both scenarios:

(1) Internal shock scenario: \cite{Dai2023} calculates the $\gamma \gamma$ optical depth from the prompt TeV photons by modeling the internal shock process using the method outlined in \cite{Kobayashi1997}. Here, we calculate the optical depth of TeV afterglow photons assuming a simplified scenario: the prompt emission is generated by the wind collision at \(R_{\rm dis} = 2\Gamma_{\rm dis}^2 c \Delta t_{\rm v}\) and at time \(t = \Delta t_{\rm v}\), with the photon field having a width of approximately $c\Delta t_{\gamma}$. This assumption is based on the fact that, in the internal shock model, the Lorentz factors of the relativistic winds are highly variable. The collision of these winds is concentrated around \(R_{\rm dis} = 2\Gamma_{\rm dis}^2 c \Delta t_{\rm v}\) at time \(t \sim \Delta t_{\rm v}\). We set the parameters according to observations: \cite{Lesage2023} shows that the minimum variability time is about $0.2 \, \rm s$ just before and after the main burst. Additionally, the Lorentz factor of $\sim 500$ at the deceleration time of the external shock \citep{LHAASO2023} reflects the Lorentz factor of the winds that have merged as a result of the collision. Thus, we set $R_{\rm dis} \sim 2\Gamma_{\rm dis}^2 c \Delta t_{\rm v} \sim 2\times 10^{15} \, \rm cm$. Additionally, we assume a prompt emission luminosity of $L_{\gamma} = 10^{54} \, \rm erg \, s^{-1}$ over a duration of $\Delta t_{\gamma} = 10 \, \rm s$, adopting a Band function with $h\nu_{\rm p} = 1 \, \rm MeV$, $\alpha_{\rm l} = 1.2$, and $\alpha_{\rm h} = 2.2$. This assumption is based on prompt MeV observations showing that the MeV flux remains nearly constant during the first pulse, which lasts about \(10 \, \rm s\) with a luminosity of approximately \(10^{54} \, \rm erg \, s^{-1}\) \citep{An2023, Lesage2023}.

\begin{figure} [h]
    \centering
    \includegraphics[width = 1\linewidth]{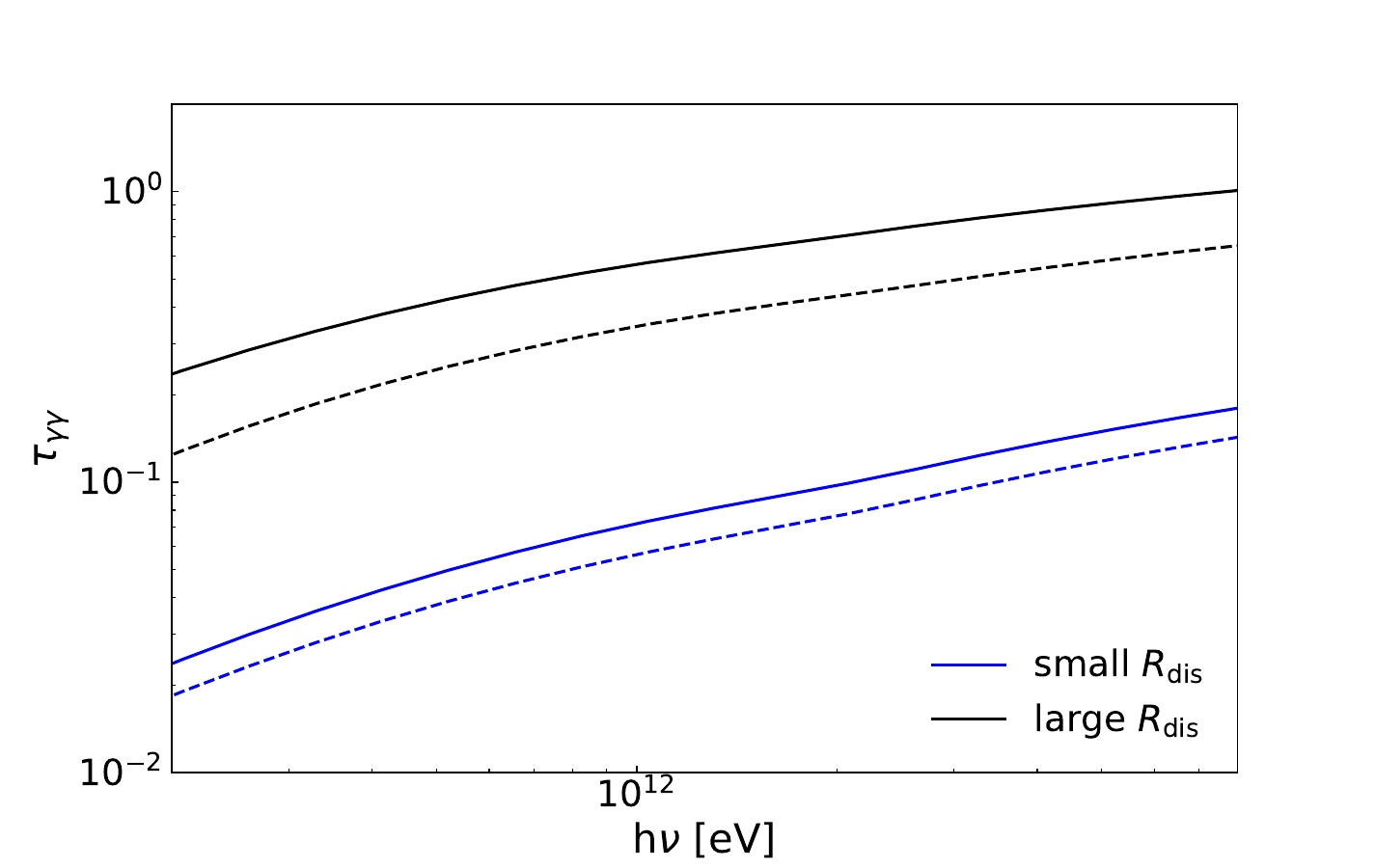}\\
    \includegraphics[width = 1\linewidth]{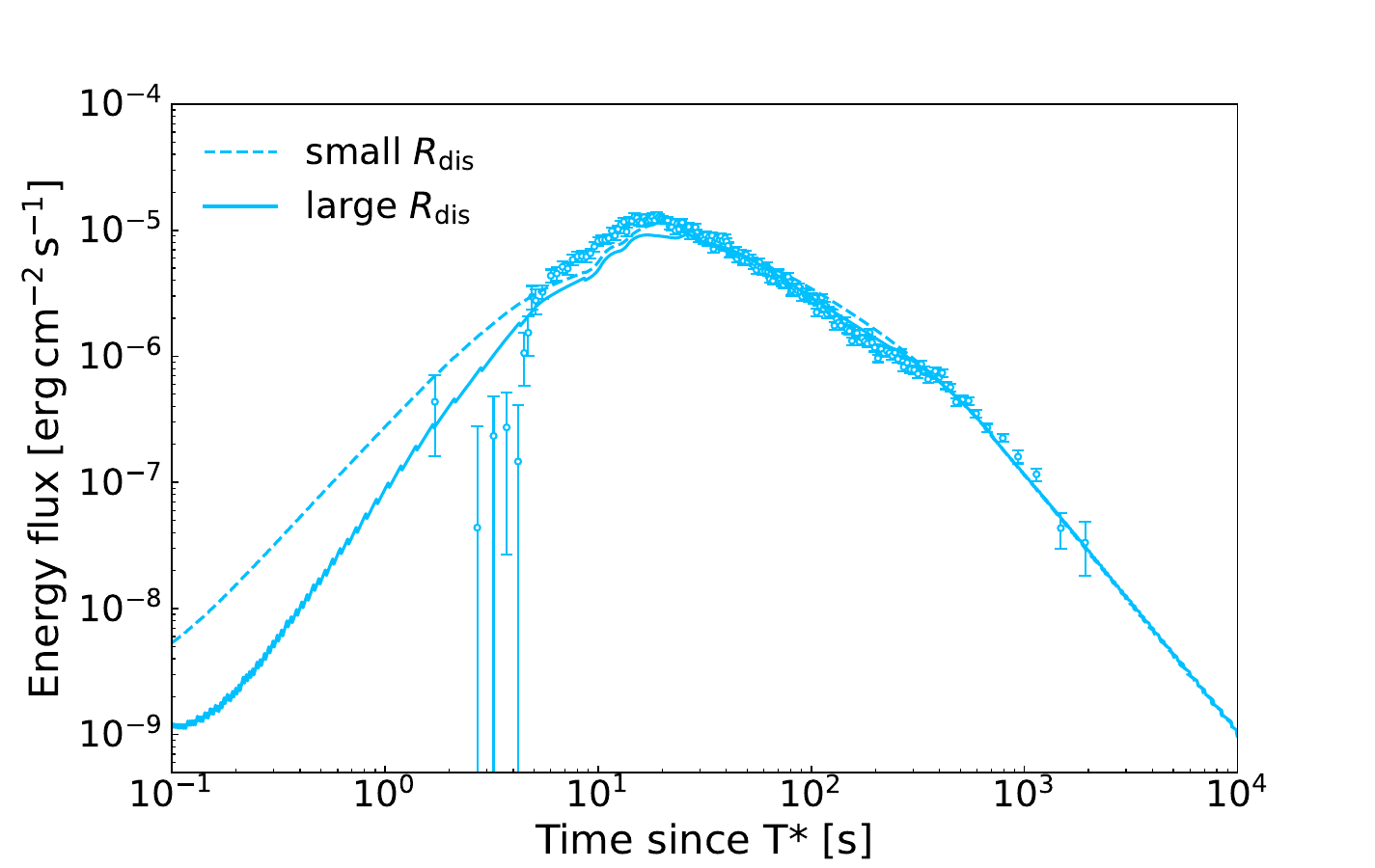}
    \caption{The upper panel shows the optical depth from the prompt emission at \(t = 3 \, \text{s}\). We use the internal shock and magnetic field dissipation/turbulence scenarios from Sect.\ref{sec_abs_multi_zone} to represent small and large \(R_{\rm dis}\), respectively. The dashed and solid lines are derived from external shock parameters based on multi-band afterglow modeling under anisotropic and isotropic EIC limits, respectively (see captions of Fig.\ref{fig_lc_ani} and Fig.\ref{fig_lc_iso}). The lower panel presents the TeV afterglow corrected for \(\gamma\gamma\) absorption. The dashed line represents the TeV afterglow for small \(R_{\rm dis}\), using parameters from multi-band afterglow modeling in the anisotropic EIC limit, with \(\tau_{\gamma\gamma}\) calculated in the internal shock scenario. The solid line corresponds to large \(R_{\rm dis}\), using parameters from the multi-band modeling in the isotropic EIC limit, where \(\tau_{\gamma\gamma}\) is determined in the magnetic field dissipation/turbulence scenario.}
    \label{fig_lc_abs}
\end{figure}

After determining the emission time and radius of the photon field, we numerically calculate the $\gamma\gamma$ opacity by extending the two-zone model described earlier to a multi-zone scenario \footnote{We decompose the photon field, uniformly distributed from \(R_{\rm dis} = 2\Gamma_{\rm dis}^2 c \Delta t_{\rm v} - c\Delta t_{\gamma}\) to \(R_{\rm dis} = 2\Gamma_{\rm dis}^2 c \Delta t_{\rm v}\), into a series of "flashes". These flashes are injected simultaneously at \(t = \Delta t_{\rm v}\). The \(\gamma\gamma\) absorption from each "flash" is then calculated using a two-zone model.}. The absorption derived from this scenario is shown in Fig.\ref{fig_lc_abs}. Our results indicate that $\gamma\gamma$ absorption is quite weak when the internal shock occurs at a radius significantly smaller than the external shock. This finding aligns with the earlier discussion and the results shown in Fig.\ref{fig_tau_flash}, since if the emission radius of the prompt MeV photons is much smaller than that of the TeV photons from the external shock, the MeV photon field becomes significantly anisotropic. This strong anisotropy reduces the interaction angle, leading to weak $\gamma\gamma$ absorption, particularly at lower latitude $\theta_{\rm L}$. 

(2) Magnetic field dispassion/turbulence scenario: simulating the light curve in a magnetic field dispassion/turbulence scenario is beyond the scope of this paper. However, we can approximate these models by assuming a single thin shell that continues to radiate while expanding outward with a constant Lorentz factor. We turn on the prompt emission at \(R_{\rm dis} = R_{\rm ext} = 2 \times 10^{15} \, \rm cm\) with a luminosity of \(L_{\gamma} = 10^{54} \, \rm erg \, s^{-1}\), sustaining it for a duration of $2\Gamma^2_{\rm ext} \Delta t_{\gamma}$, where $\Delta t_{\gamma} = 10 \, \rm s$. In this scenario, $\gamma \gamma$ absorption is greater than in the internal shock scenario, where \(R_{\rm dis} \ll R_{\rm ext}\). However, it remains insufficient to account for the rapid rise during \(T^*+[0, 5] \, \text{s}\), as illustrated in Fig.\ref{fig_lc_abs}.

\section{Discussion}

\label{sec_discussion}

\subsection{Analytical estimate of the EIC flux}
\label{sec_ana_eic_flux}

In this section, we attempt to give an analytical description of the TeV light curve from the EIC process, which may help to understand its properties and contribution to the TeV afterglow better. As noted previously, in the precise EIC calculation, the scattering rate differs by less than a factor of two across different $R_{\rm dis}/R_{\rm ext}$ values. Therefore, we adopt the isotropic limit for EIC in the following analytical discussion. As discussed in Sect.\ref{sec_ana_shock_EIC_cooling}, the KN effect is not significant for EIC before the deceleration phase of GRB 221009A. Although the KN effect becomes more significant after external shock deceleration, with $\nu'_{\rm p} \propto \Gamma_{\rm ext}$ and $\nu'_{\rm out} \propto \Gamma_{\rm ext}$, this increase occurs slowly over time, as $\Gamma_{\rm ext} \propto t^{-\frac{3}{8}}$. For most GRBs, where the typical $\nu'_{\rm p} < 1 \, \rm MeV$ and the observational energy band is below the TeV range, the KN effect becomes even less significant. Therefore, we neglect the KN effect and assume that the energy density of the prompt emission $U'_{\gamma} \gg \max(U'_B, U'_{\rm SYN})$, meaning $1 + Y' \simeq U'_{\gamma}/U'_B$, which applies to most GRBs.

We approximate the spectrum of the prompt emission as mono-energetic photons with energy \(h\nu_{\rm p}\), so the frequency of the up-scattered photons is $\nu_{\rm out} = 2{\gamma'_e}^2 \nu_{\rm p}$. The electron spectrum $N_e(\gamma'_e)$ follows a broken power-law with a break at $\max(\gamma'_{\rm m}, \gamma'_{\rm c})$ and a cutoff at $\min(\gamma'_{\rm m}, \gamma'_{\rm c})$. Consequently, the EIC spectrum follows a broken power-law, with breaks at $\nu^{\rm EIC}_{\rm m} = 2{\gamma'_{\rm m}}^2 \nu_{\rm p}$ and $\nu^{\rm EIC}_{\rm c} = 2{\gamma'_{\rm c}}^2 \nu_{\rm p}$. The EIC flux $F^{\rm EIC}_{\nu'_{\rm out}}$ $(\rm erg \, s^{-1} \, Hz^{-1} \, cm^{-2})$ can be derived from energy conservation: $F^{\rm EIC}_{\nu'_{\rm out}} d\nu'_{\rm out} = N_e[\gamma'_e(\nu'_{\rm out})] P_e^{\rm EIC}[\gamma'_e(\nu'_{\rm out})] d\gamma'_e$, where $P_e^{\rm EIC}(\gamma'_e)$ is the the EIC power, which scales as $P_e^{\rm EIC}(\gamma'_e) \propto {\gamma'_e}^2$. Substituting the electron spectrum from Sect.\ref{sec_ana_shock_EIC_cooling} into the energy conservation equation, we obtain:
\begin{equation}
\label{Eq_Feic_ana}
\frac{F^{\rm EIC}_{\nu_{\rm out}}}{F^{\rm EIC}_{\rm m}} = \begin{cases} 
       \left(\frac{\nu_{\rm out}}{\nu_{\rm m}^{\rm EIC}}\right)^{-\frac{p-1}{2}}, & \nu_{\rm m}^{\rm EIC} < \nu_{\rm out} < \nu_{\rm c}^{\rm EIC} \\
       \left(\frac{\nu_{\rm out}}{\nu_{\rm c}^{\rm EIC}}\right)^{-\frac{1}{2}}, & \nu_{\rm c}^{\rm EIC} < \nu_{\rm out} < \nu_{\rm m}^{\rm EIC} \\
       \left(\nu_{\rm m}^{\rm EIC}\right)^{\frac{p-1}{2}} \left(\nu_{\rm c}^{\rm EIC}\right)^{\frac{1}{2}} \nu_{\rm out}^{-\frac{p}{2}}. & \nu_{\rm out} > \max(\nu_{\rm m}^{\rm EIC}, \nu_{\rm c}^{\rm EIC})
   \end{cases}
\end{equation}

The evolution of \(F^{\rm EIC}_{\rm m}\), \(\nu^{\rm EIC}_{\rm m}\), and \(\nu^{\rm EIC}_{\rm c}\) can be expressed as follows:

(1) Before deceleration: in this phase, the Lorentz factor of the external shock, \(\Gamma_{\rm ext}\), remains constant, so \(R_{\rm ext} \sim 2\Gamma_{\rm ext}^2 c t \propto t\). The optical depth for EIC scattering is \(\tau_{\rm EIC} \propto N_e/R^2_{\rm ext} \propto t^{1-k}\), where \(N_e \propto R_{\rm ext}^{3-k}\) is the total number of swept-up electrons, and \(k\) is the index of the circumburst density, \(n \propto R_{\rm ext}^{-k}\). Assuming that the prompt emission flux evolves as \(F_{\gamma} \propto L_{\gamma} \propto t^w\), the peak EIC flux becomes \(F^{\rm EIC}_{\rm m} = \tau_{\rm EIC} F_{\gamma} \propto t^{1-k+w}\). The minimum Lorentz factor, \(\gamma'_{\rm m}\), remains constant as \(\gamma'_{\rm m} = \epsilon_e \frac{m_p}{m_e} \frac{p-2}{p-1} \Gamma_{\rm ext} \propto t^0\). The critical Lorentz factor is \(\gamma'_{\rm c} = \frac{6\pi m_e c}{(1+Y')\sigma_{\rm T}\Gamma_{\rm ext} {B'}^2 t}\), where \(B' = \sqrt{32 \pi m_p \epsilon_B n} \Gamma_{\rm ext} c\), \(1+Y' \simeq Y' \simeq U'_{\gamma}/U'_B\), and the energy density of the prompt emission is \(U'_{\gamma} = \frac{L_{\gamma}}{16\pi \Gamma^2_{\rm ext} R^2_{\rm ext} c}\). From this, we find \(\nu^{\rm EIC}_{\rm m} \propto t^0\) and \(\nu^{\rm EIC}_{\rm c} \propto t^{2-2w}\).

(2) After deceleration: in this phase, $\Gamma_{\rm ext} \propto t^{\frac{-3+k}{8-2k}}$. From this, we obtain $F^{\rm EIC}_{\rm m} \propto t^{\frac{1-k}{4-k} + w}$, $\nu^{\rm EIC}_{\rm m} \propto t^{\frac{k-3}{4-k}}$, and $\nu^{\rm EIC}_{\rm c} \propto t^{\frac{3k-7}{4-k}-2w}$. Substituting these into Eq.(\ref{Eq_Feic_ana}), we derive $F^{\rm EIC}_{\nu_{\rm out}}$, as summarized in Table.\ref{table_Feic}.

Next, we discuss the EIC flux of GRB 221009A using the analytical expressions derived above: for GRB 221009A, the electrons are in the fast cooling regime, causing the TeV light curve to rise as \(t^{2-k}\), consistent with the observed \(t^{1.82^{+0.21}_{-0.18}}\) \citep{LHAASO2023} for the uniform density case \((k=0)\). The correction factor for the TeV afterglow due to the KN effect is on the order of a few before shock deceleration, as discussed in Sect.\ref{sec_ana_shock_EIC_cooling}. Notably, the KN correction remains constant over time, since the prompt emission spectrum does not evolve in the comoving frame of the external shock before deceleration. After deceleration, the SSC surpasses EIC in contributing to the TeV afterglow as discussed in Sect.\ref{sec_modeling_result}. We do not consider the transition phase where EIC and SSC contributions are comparable. Once SSC dominates the TeV afterglow, the flux can be described as in \citep{LHAASO2023}. Additionally, as the shock decelerates, $h\nu'_{\rm out} = 1 \, \rm TeV / (2\Gamma_{\rm ext})$ increases over time, making the KN effect more significant. However, this increase is slowly, as $\Gamma_{\rm ext} \propto t^{-\frac{3}{8}}$. At later times, the KN effect should be incorporated into the analysis \citep{Nakar2009}.

\begin{table*}
\centering
\caption{Analytical expression for the EIC flux $F^{\rm EIC}_{\nu_{\rm out}}$}
\label{best_fit_parameter}
\begin{tabular}{l@{\hspace{5cm}}l} % 这里的@{\hspace{1cm}} 增加了1cm的间距
\hline\hline
Frequency Range & $F^{\rm EIC}_{\nu_{\rm out}}$
\\
\hline
$\boldsymbol{t \leqslant t_{\rm d}}:$
\\
$\nu^{\rm EIC}_{\rm m} < \nu_{\rm out} < \nu^{\rm EIC}_{\rm c}$ & $F^{\rm EIC}_{\rm m} \left( \frac{\nu_{\rm out}}{\nu^{\rm EIC}_{\rm m}} \right)^{\frac{1-p}{2}} \propto \nu_{\rm out}^{\frac{1-p}{2}}  t^{1-k+w}$
\\
$\nu^{\rm EIC}_{\rm c} < \nu_{\rm out} < \nu^{\rm EIC}_{\rm m}$ & $F^{\rm EIC}_{\rm m} \left( \frac{\nu_{\rm out}}{\nu^{\rm EIC}_{\rm c}} \right)^{-\frac{1}{2}} \propto \nu_{\rm out}^{-\frac{1}{2}} t^{2-k} $
\\
$\max(\nu^{\rm EIC}_{\rm m}, \nu^{\rm EIC}_{\rm c}) < \nu_{\rm out}$ & $F^{\rm EIC}_{\rm m} \left( \nu^{\rm EIC}_{\rm m} \right)^{\frac{p-1}{2}} \left( \nu^{\rm EIC}_{\rm c} \right)^{\frac{1}{2}} \nu_{\rm out}^{-\frac{p}{2}} \propto \nu_{\rm out}^{-\frac{p}{2}} t^{2-k} $
\\
\hline
$\boldsymbol{t > t_{\rm d}}:$
\\
$\nu^{\rm EIC}_{\rm m} < \nu_{\rm out} < \nu^{\rm EIC}_{\rm c}$ & $F^{\rm EIC}_{\rm m} \left( \frac{\nu_{\rm out}}{\nu^{\rm EIC}_{\rm m}} \right)^{\frac{1-p}{2}} \propto \nu_{\rm out}^{\frac{1-p}{2}} t^{\frac{5-3k+kp-3p}{8-2k}+w}$
\\
$\nu^{\rm EIC}_{\rm c} < \nu_{\rm out} < \nu^{\rm EIC}_{\rm m}$ & $F^{\rm EIC}_{\rm m} \left( \frac{\nu_{\rm out}}{\nu^{\rm EIC}_{\rm c}} \right)^{-\frac{1}{2}} \propto \nu_{\rm out}^{-\frac{1}{2}} t^{\frac{k-5}{8-2k}}$
\\
$\max(\nu^{\rm EIC}_{\rm m}, \nu^{\rm EIC}_{\rm c}) < \nu_{\rm out}$ & $F^{\rm EIC}_{\rm m} \left( \nu^{\rm EIC}_{\rm m} \right)^{\frac{p-1}{2}} \left( \nu^{\rm EIC}_{\rm c} \right)^{\frac{1}{2}} \nu_{\rm out}^{-\frac{p}{2}} \propto \nu_{\rm out}^{-\frac{p}{2}} t^{\frac{-2+kp-3p}{8-2k}}$
\\
\hline\hline
\end{tabular}
\label{table_Feic}
\end{table*}

\subsection{the synchrotron flux in EIC dominance phase}

As discussed in Sect.\ref{sec_ana_shock_EIC_cooling}, the EIC process plays a significant role in electron cooling and is unaffected by the angular distribution of prompt emission. EIC cooling reduces $\gamma'_{\rm c}$, significantly suppressing the synchrotron flux. Predicting the synchrotron flux during the EIC-dominant phase allows us to test the EIC model through early afterglow observations, especially in the optical or GeV bands, where the prompt emission may have cut off.

Following the method in \cite{Sari98} and the previous discussion, the synchrotron flux can be expressed as
\begin{equation}
\label{Eq_Fsyn_ana}
\frac{F^{\rm SYN}_{\nu_{\rm out}}}{F^{\rm SYN}_{\rm m}} = \begin{cases} 
        \left(\frac{\nu_{\rm out}}{\nu_{\rm m}^{\rm SYN}}\right)^{\frac{1}{3}}, & \nu_{\rm out} < \nu_{\rm m}^{\rm SYN}  < \nu_{\rm c}^{\rm SYN} \\
        \left(\frac{\nu_{\rm out}}{\nu_{\rm c}^{\rm SYN}}\right)^{\frac{1}{3}}, & \nu_{\rm out} < \nu_{\rm c}^{\rm SYN}  < \nu_{\rm m}^{\rm SYN} \\
       \left(\frac{\nu_{\rm out}}{\nu_{\rm m}^{\rm SYN}}\right)^{-\frac{p-1}{2}}, & \nu_{\rm m}^{\rm SYN} < \nu_{\rm out} < \nu_{\rm c}^{\rm SYN} \\
       \left(\frac{\nu_{\rm out}}{\nu_{\rm c}^{\rm SYN}}\right)^{-\frac{1}{2}}, & \nu_{\rm c}^{\rm SYN} < \nu_{\rm out} < \nu_{\rm m}^{\rm SYN} \\
       \left(\nu_{\rm m}^{\rm SYN}\right)^{\frac{p-1}{2}} \left(\nu_{\rm c}^{\rm SYN}\right)^{\frac{1}{2}} \nu_{\rm out}^{-\frac{p}{2}}. & \nu_{\rm out} > \max(\nu_{\rm m}^{\rm SYN}, \nu_{\rm c}^{\rm SYN})
   \end{cases}
\end{equation}

The peak synchrotron flux is given by $F^{\rm SYN}_{\rm m} = \frac{N_e P^{\rm SYN}_{\rm m}}{4\pi D^2_{\rm L}}$, where $P^{\rm SYN}_{\rm m} = \frac{m_e c^2 \sigma_{\rm T}}{3e}\Gamma_{\rm ext}B'$. The synchrotron frequency for electrons with Lorentz factor $\gamma'_e$ in the observer frame is $\nu(\gamma'_e) = \Gamma_{\rm ext} {\gamma'_e}^2 \frac{eB}{2 \pi m_e c}$. The peak and cooling frequencies are $\nu^{\rm SYN}_{\rm m} = \nu(\gamma'_{\rm m})$ and $\nu^{\rm SYN}_{\rm c} = \nu(\gamma'_{\rm c})$, respectively.

The only difference between the cases with and without EIC cooling lies in $\gamma'_{\rm c}$, which is given by $\gamma'_{\rm c} = \frac{6\pi m_e c}{(1+Y') \sigma_{\rm T} \Gamma_{\rm ext} {B'}^2 t}$. Without EIC, $1+Y' \simeq Y' \simeq \sqrt{\frac{\epsilon_e}{\epsilon_B}}$ for $\epsilon_e \gg \epsilon_B$ \citep{Sari2001}. Including EIC, $1+Y' \simeq U'_{\gamma}/U'_B$, as discussed in Sect.\ref{sec_ana_eic_flux}. Using these two forms of $\gamma'_{\rm c}$, we can derive the synchrotron flux for both cases, as shown in the Table.\ref{table_syn_ana}.

\begin{table*}
\centering
\caption{Analytical expression for the synchrotron flux $F^{\rm SYN}_{\nu_{\rm out}}$}
\label{best_fit_parameter}
\begin{tabular}{l@{\hspace{3cm}}l@{\hspace{3cm}}l}
\hline\hline
Frequency Range & $F^{\rm SYN}_{\nu_{\rm out}}$ (including EIC) & $F^{\rm SYN}_{\nu_{\rm out}}$ (excluding EIC)
\\
\hline
$\boldsymbol{t \leqslant t_{\rm d}}:$
\\
$\nu_{\rm out} < \nu^{\rm SYN}_{\rm m} < \nu^{\rm SYN}_{\rm c}$ & $ \propto \nu_{\rm out}^{\frac{1}{3}} t^{3-\frac{4k}{3}}$ & $ \propto \nu_{\rm out}^{\frac{1}{3}} t^{3-\frac{4k}{3}}$
\\
$\nu_{\rm out} < \nu^{\rm SYN}_{\rm c} < \nu^{\rm SYN}_{\rm m}$ & $ \propto \nu_{\rm out}^{\frac{1}{3}} t^{\frac{7-4k+2w}{3}}$ & $ \propto \nu_{\rm out}^{\frac{1}{3}} t^{\frac{11}{3} - 2k}$
\\
$\nu^{\rm SYN}_{\rm m} < \nu_{\rm out}  < \nu^{\rm SYN}_{\rm c}$ & $ \propto \nu_{\rm out}^{\frac{1-p}{2}} t^{3-\frac{k(5+p)}{4}}$ & $ \propto \nu_{\rm out}^{\frac{1-p}{2}} t^{3-\frac{k(5+p)}{4}}$
\\
$\nu^{\rm SYN}_{\rm c} < \nu_{\rm out}  < \nu^{\rm SYN}_{\rm m}$ & $ \propto \nu_{\rm out}^{-\frac{1}{2}} t^{4-\frac{7k}{4}-w}$ & $ \propto \nu_{\rm out}^{-\frac{1}{2}} t^{2-\frac{3k}{4}}$
\\
$\max(\nu^{\rm SYN}_{\rm m}, \nu^{\rm SYN}_{\rm c}) < \nu_{\rm out}$ & $ \propto \nu_{\rm out}^{-\frac{p}{2}} t^{4-\frac{k(6+p)}{4}-w}$ & $ \propto \nu_{\rm out}^{-\frac{p}{2}} t^{2-\frac{k(2+p)}{4}}$
\\
\hline
$\boldsymbol{t > t_{\rm d}}:$
\\
$\nu_{\rm out} < \nu^{\rm SYN}_{\rm m} < \nu^{\rm SYN}_{\rm c}$ & $ \propto \nu_{\rm out}^{\frac{1}{3}} t^{1+\frac{2}{-4+k}}$ & $ \propto \nu_{\rm out}^{\frac{1}{3}} t^{1+\frac{2}{-4+k}}$
\\
$\nu_{\rm out} < \nu^{\rm SYN}_{\rm c} < \nu^{\rm SYN}_{\rm m}$ & $ \propto \nu_{\rm out}^{\frac{1}{3}} t^{\frac{5(-2+k)}{3(-4+k)} + \frac{2w}{3}}$ & $ \propto \nu_{\rm out}^{\frac{1}{3}} t^{1+\frac{10}{3(-4+k)}}$
\\
$\nu^{\rm SYN}_{\rm m} < \nu_{\rm out}  < \nu^{\rm SYN}_{\rm c}$ & $ \propto \nu_{\rm out}^{\frac{1-p}{2}} t^{\frac{5}{4} + \frac{2}{-4+k} - \frac{3p}{4}}$ & $ \propto \nu_{\rm out}^{\frac{1-p}{2}} t^{\frac{5}{4}+\frac{2}{-4+k}-\frac{3p}{4}}$
\\
$\nu^{\rm SYN}_{\rm c} < \nu_{\rm out}  < \nu^{\rm SYN}_{\rm m}$ & $ \propto \nu_{\rm out}^{-\frac{1}{2}} t^{-\frac{5}{4}-w}$ & $ \propto \nu_{\rm out}^{-\frac{1}{2}} t^{-\frac{1}{4}}$
\\
$\max(\nu^{\rm SYN}_{\rm m}, \nu^{\rm SYN}_{\rm c}) < \nu_{\rm out}$ & $ \propto \nu_{\rm out}^{-\frac{p}{2}} t^{\frac{-2-3p-4w}{4}}$ & $ \propto \nu_{\rm out}^{-\frac{p}{2}} t^{\frac{1}{2} - \frac{3p}{4}}$
\\
\hline\hline
\end{tabular}
\label{table_syn_ana}
\end{table*}

\subsection{The origin of the rapid rising of the observed TeV afterglow during $T^*+[0, 5] \, \rm s$}

As discussed in Sect.\ref{sec_gg_abs}, $\gamma \gamma$ absorption from prompt MeV photons is insufficient to account for the rapid rise in the TeV afterglow. We do not consider absorption from the prompt emission before \(T^*\). During \(T^*+[-6, 0] \, \text{s}\), the prompt emission's luminosity increases, but the energy is significantly lower (see Fig.3 in \cite{An2023}) compared to \(T^*+[0, 10] \, \text{s}\). Therefore, the absorption during \(T^*+[-6, 0] \, \text{s}\) is not significant. 
%We also test with the larger luminosity \(L_{\gamma} = 2\times 10^{54} \, \rm erg \, s^{-1}\) in the absorption model described in Sect.\ref{sec_abs_multi_zone}, which is about twice the observed value; however, \(\gamma \gamma\) absorption still cannot explain the rapid rise of the TeV afterglow at \(T^*+[0, 5] \, \text{s}\). 
\cite{Gao2024} used a two-zone model to calculate $\tau_{\gamma \gamma}$ from the prompt emission. They found that $\gamma \gamma$ absorption may cause a rapid rise in the early TeV afterglow  assuming $\Gamma_{\rm dis} = 100$ and $\Gamma_{\rm ext} \simeq 400$. The small value of $\Gamma_{\rm dis}$ employed in their calculation is different from our choice of $\Gamma_{\rm dis}\approx  \Gamma_{\rm ext}$, leading to a less anisotropic prompt emission and consequently a stronger $\gamma \gamma$ absorption. While our model may also achieve a stronger $\gamma\gamma$ absorption at the early time by assuming a smaller $\Gamma_{\rm dis}$, we'd like to point out other possibilities that may cause the rapid rise of the TeV afterglow:

(1) density jump in the circumburst medium: GRB 221009A exhibits a weak precursor that starts at the trigger time and lasts for approximately $10 \, \rm s$, with a luminosity four orders of magnitude lower than the prompt emission. The external shock associated with this precursor sweeps through the surrounding medium, potentially creating a low-density bubble.  When the external shock catches up with the edge of the bubble, it encounters a density jump, leading to a rapid rise in the afterglow \citep{Dai2002, Dai2003}.

(2) continuous energy injection: as described in \cite{Derishev2024}, the initial phase of the afterglow is produced by a blast wave with intermittent energy supply, where the central engine continuously provides energy to the inner part (shocked ejecta material) via the reverse shock.

\subsection{the origin of the TeV excess observed between $T^* + [250, 270] \, \rm s$}

According to \cite{An2023}, the MeV light curve shows a flare at $t_{\rm flare} \sim T^* + 250 \, \rm s$, lasts for $\Delta t_{\rm flare} \sim 20 \, \rm s$ (see the black data in the middle panel of Fig.\ref{fig_lc_ani}). Simultaneously, the TeV light curve also shows an excess \citep{LHAASO2023}. This TeV excess cannot be easily explained by the EIC of the MeV flare, as EIC leads to lengthened emission under the anisotropic EIC limit, as discussed in Sect.\ref{sec_EIC_properties} (during \(T^* + [250, 270] \, \text{s}\), \(R_{\rm dis} \ll R_{\rm ext}\), making the anisotropic EIC limit reasonable). The duration of the EIC radiation for a single "flash" injected at a radius $R_{\rm flare} \simeq 5 \times 10^{17} \left(\frac{E_{\rm k0, iso}}{5 \times 10^{54} \, \rm erg} \right)^{\frac{1}{4}} \left(\frac{n}{0.4 \, \rm cm^{-3}} \right)^{-\frac{1}{4}} \left(\frac{t_{\rm flare}}{250 \, \rm s} \right)^{\frac{1}{4}} \, \rm cm$ with $\Gamma_{\rm flare} \simeq 135 \left(\frac{E_{\rm k0, iso}}{5 \times 10^{54} \, \rm erg} \right)^{\frac{1}{8}} \left(\frac{n}{0.4 \, \rm cm^{-3}} \right)^{-\frac{1}{8}} \left(\frac{t_{\rm flare}}{250 \, \rm s} \right)^{-\frac{3}{8}}$ would last for $\sim \min[R_{\rm flare}/(\Gamma_{\rm flare}^2c), R_{\rm flare}\theta_{\rm j}^2/c] = R_{\rm flare}\theta_{\rm j}^2/c = 875 \left(\frac{E_{\rm k0, iso}}{5 \times 10^{54} \, \rm erg} \right)^{\frac{1}{4}} \left(\frac{n}{0.4 \, \rm cm^{-3}} \right)^{-\frac{1}{4}} \left(\frac{t_{\rm flare}}{250 \, \rm s} \right)^{\frac{1}{4}} \left( \frac{\theta_{\rm j}}{0.4 \, \rm deg} \right)^2 \, \rm s$, which is much longer than  $\Delta t_{\rm flare} \sim 20 \, \rm s$.

This is clearly shown in the upper panel of Fig.\ref{fig_lc_ani}, where the EIC light curve exhibits a plateau as the prompt MeV emission rises, starting at approximately \(T^* + 250 \, \rm s\) and lasting for a few hundred seconds before cutting off as the emission latitude reaches \(\theta_{\rm j}\).

%{\bf One possible explanation for the short duration of the TeV flare is that the EIC delay is shortened by a small $\theta_{\rm j}$, where the delay of EIC from high latitudes only lasts for $R_{\rm flare}\theta_{\rm j}^2/c$. In this case, $\theta_{\rm j} \sim \sqrt{c\Delta t_{\rm flare}/R_{\rm flare}} \sim 0.06 \left(\frac{E_{\rm k0, iso}}{5 \times 10^{54} \, \rm erg}\right)^{-\frac{1}{8}} \left(\frac{n}{0.4 \, \rm cm^{-3}} \right)^{\frac{1}{8}} \left(\frac{\Delta t_{\rm flare}}{20 \, \rm s} \right)^{\frac{1}{2}} \left(\frac{t_{\rm flare}}{250 \, \rm s} \right)^{-\frac{1}{8}} \, \rm deg$, which is unreasonable and contradicts the $\theta_{\rm j}$ derived from the jet break \citep{LHAASO2023}.  Remove this??}

Instead, we argue that the TeV flare could originate from later internal dissipation. It has been argued that the absence of TeV emission during $\sim T^* + [220, 230] \, \rm s$ suggests that the jet is magnetized, and SSC in the internal dissipation phase has been suppressed \citep{Dai2023}. Nevertheless, at later times, a significant amount of magnetic field energy may have been dissipated through magnetic reconnection induced by internal collisions and turbulence, causing the magnetization parameter, \(\sigma\), to decrease significantly \citep{zhang2011}. In this case, the SSC emission from the internal dissipation is no longer suppressed, allowing TeV emission to be observed.

\section{Conclusion}

\label{sec_conclusion}

The BOAT GRB 221009A, which exhibits a distinct TeV afterglow during its prompt emission phase, presents a unique opportunity to investigate how the prompt emission affects the early afterglow through processes such as EIC and $\gamma \gamma$ absorption.

Due to the uncertainty in the internal dissipation radius, $R_{\rm dis}$, which impacts the degree of anisotropy in the prompt emission photon field, we begin by analyzing the dependence of the EIC scattering rate on the ratio $R_{\rm dis}/R_{\rm ext}$. Our results show that when $R_{\rm dis}/R_{\rm ext} < 0.3$, the anisotropic limit produces EIC fluxes that differ from the precise calculation by less than 10\%. Similarly, the isotropic limit maintains comparable accuracy when $R_{\rm dis}/R_{\rm ext} > 0.85$. Note that for both limits, the largest deviation from the accurate calculation remains within a factor of two for any ratio of $R_{\rm dis}/R_{\rm ext}$.

For GRB 221009A, the energy density of the prompt emission significantly exceeds that of the magnetic field and the SSC radiation at an early time. The EIC cooling leads to a fast-cooling electron spectrum throughout nearly the entire rising phase of the TeV afterglow. We model the light curve and spectrum of the early afterglow of GRB 221009A with anisotropic and isotropic EIC limits separately, finding that the model parameters are almost the same in two cases. Our modeling indicates that the EIC process dominates over the SSC emission during the rising phase of the TeV afterglow, while the SSC process gradually becomes dominated after the deceleration of the external shock. The transition from EIC to SSC dominance in the TeV afterglow can also be roughly inferred from the behavior of the early MeV emission, where a hidden afterglow component is found during the prompt emission phase \citep{zhang2023}. This transition occurs around the time when the MeV afterglow luminosity surpasses that of the prompt emission, at approximately \(T^* + 74 \, \rm s\) in the isotropic EIC limit, which applies to \(R_{\rm dis} \sim R_{\rm ext}\). In the anisotropic EIC limit (\(R_{\rm dis} \ll R_{\rm ext}\)), the EIC flux is suppressed by a factor of \(\sim 0.3\), and the transition occurs earlier. Moreover, we find that the peak of the TeV light curve for GRB 221009A still marks the onset of deceleration of the external shock.

Additionally, we find that $\gamma \gamma$ absorption becomes significant only when the radii of internal dissipation and the external shock are comparable. Otherwise, the strong anisotropy of the prompt MeV photon field suppresses the $\gamma \gamma$ absorption. We calculate the optical depth $\tau_{\gamma \gamma}$ for both relatively small internal dissipation radii (typically predicted by the internal shock model scenario) and large radii (typically predicted by the magnetic field dissipation/turbulence scenarios). We find that even in the latter case, where \(R_{\rm dis} \sim R_{\rm ext}\), \(\gamma \gamma\) absorption still cannot account for the observed rapid increase in the early TeV afterglow (during $T^*+[0, 5] \, \rm s$). The rapid rise may be attributed to other effects, such as additional energy injection or a sudden density jump.

\section*{Acknowledgements}
This study is supported by National Natural Science Foundation of China under grants No.~12393852, 12333006, 12121003, 12393811.

\clearpage

\appendix
\renewcommand*{\thefigure}{A\arabic{figure}}
\setcounter{figure}{0}  
\section{precise calculation of EIC}
\label{sec_exa_EIC}

As any continuous emission can be regarded as consistent of a series of "flash" in numerical calculation, we focus on the EIC process in two-zone scenario: a single "flash" photon field, injected at $R_{\rm dis}$ within an infinitesimally short time, is scattered by electrons accelerated by the external shock at point $O$ at $R_{\rm ext}$ (see Fig.\ref{fig_jet}). We set the photons emitted from point $Q$ to reach point $O$ at $t_{\rm s0} = 0$, and the photons emitted from point $P$ will then reach point $O$ at $t_{\rm s} = \int \frac{d|PO|}{c} = \int_{\cos \theta_2}^{1} \frac{R_{\rm dis} R_{\rm ext}}{c\sqrt{R_{\rm dis}^2+R_{\rm ext}^2 - 2R_{\rm dis}R_{\rm ext}\cos\theta}} \, d\cos\theta$. Due to the relativistic beaming, we set $\theta_{1, \rm max} = 1/\Gamma_{\rm dis}$. The duration of the EIC emission from the flash is then given by $\Delta t_{\rm s} = \frac{|PO|-|QO|}{c} < \frac{|PQ|}{c} \simeq \frac{R_{\rm dis} \sin \theta_2}{c} < \frac{R_{\rm dis}}{c\Gamma_{\rm dis}} \ll \frac{R_{\rm ext}}{c}$. Hence, the external shock can be considered quasi-steady, and $R_{\rm ext}$ can be assumed constant during the EIC process.

The geometry of the scattering process depends on the arrival time $t_{\rm s}$ of the incident photons at point $O$, which in turn determines two key quantities of the EIC process: the scattering angle $\theta'_{\rm s}(t'_{\rm s})$ and the angular distribution of the incident photon number density $n'_{\rm ph}(\nu_{\rm in}', \Omega_{\rm in}', t'_{\rm s}) \, ({\rm photon \, Hz^{-1} \, sr^{-1} \, cm^{-3}} )$. In the following calculation, the quantities with single prime are in the comoving frame of external shock at $R_{\rm ext}$, with double prime, is in the comoving frame of the internal dispassion region at $R_{\rm dis}$, otherwise is in the source frame. The internal dispassion in the comoving frame can be regarded as isotropic. Therefore, the photon flux $f''_{\rm dis}(\nu_{\rm in}'', \Omega_{\rm dis}'') \,\rm (photon \, Hz^{-1} \, sr^{-1} \, cm^{-2})$ emitted from the spherical surface at $R_{\rm dis}$ as a "flash" is given by:
\begin{equation}
\label{AEq_f1}
    f''_{\rm dis}(\nu_{\rm in}'', \Omega_{\rm dis}'') = \frac{N_{\gamma, \rm dis}''{\mathcal{B}}''_{\rm dis}\left(\frac{\nu_{\rm in}''}{\nu''_{\rm p}}\right)}{16\pi^2R_{\rm dis}^2 \nu''_{\rm p}},
\end{equation}
where $N_{\gamma, \rm dis}''{\mathcal{B}}''_{\rm dis}\left(\frac{\nu_{\rm in}''}{\nu''_{\rm p}}\right) /\nu''_{\rm p}$ represents the photon spectrum $\rm (ph \, Hz^{-1})$ emitted from the spherical surface, $N''_{\gamma, \rm dis}$ is the total number of photons emitted from the spherical surface, and $\int \mathcal{B}''_{\rm dis}(x) \, dx = 1$. We denote the photon flux reaching point $O$ at time $t_{\rm s}$ as $F_{\rm in}(\nu_{\rm in}, t_{\rm s}) \, \rm (photon \, Hz^{-1} \, s^{-1} \, cm^{-2})$. According to the conservation of photon number, we have

\begin{equation}
\label{AEq_f2}
    F_{\rm in}(\nu_{\rm in}, t_{\rm s})dA_2dt_{\rm s}d\nu_{\rm in} = f_{\rm dis}(\nu_{\rm in},\Omega_{\rm dis})dA_1d\nu_{\rm in} d\Omega_{\rm dis},
\end{equation}
where $dt_{\rm s} = \frac{d|PO|}{c}$, $d\Omega_{\rm dis} = \frac{dA_2}{|PO|^2}$, $dA_1 = -2\pi R_{\rm dis}^2 d\cos\theta_2$, and $dA_2$ is the receiving area. At any $t_{\rm s}$, the incident photons penetrate through point $O$  with a fixed incident angle $\theta_{\rm in} = \theta_3$, hence the photon number density $n_{\rm ph}(\nu_{\rm in}, \Omega_{\rm in}, t_{\rm s})  \propto \delta(\cos\theta_{\rm in} - \cos\theta_3)$. Using $\int n_{\rm ph}(\nu_{\rm in}, \Omega_{\rm in}, t_{\rm s}) \, d\Omega_{\rm in} = F_{\rm in}(\nu_{\rm in}, t_{\rm s})/c$ and combining Eq.(\ref{AEq_f1}) and Eq.(\ref{AEq_f2}), we have

\begin{equation}
    \label{A_Eq_Bv}
    n'_{\rm ph}(\nu'_{\rm in}, \Omega'_{\rm in}, t'_{\rm s}) = -\frac{D_{\rm dis} N''_{\gamma, \rm dis}\mathcal{B}''_{\rm dis}\left( \frac{\nu'_{\rm in}}{\nu'_{\rm p}} \right)\delta(\cos\theta'_{\rm in} - \cos\theta'_3)}{16\pi^2 R_{\rm dis}R_{\rm ext}|PO| \nu''_{\rm p}},
\end{equation}
where we have used the Doppler transformation $f_{\rm dis}(\nu_{\rm in}, \Omega_{\rm in}, t_{\rm s}) = D_{\rm dis} f''_{\rm dis}(\nu''_{\rm in}, \Omega''_{\rm in}, t_{\rm s})$ and $n_{\rm ph}(\nu_{\rm in}, \Omega_{\rm in}, t_{\rm s}) = D_{\rm ext, in}^2 n'_{\rm ph}(\nu'_{\rm in}, \Omega'_{\rm in}, t'_{\rm s})$. Here, $D_{\rm dis} = \frac{1}{\Gamma_{\rm dis}(1-\beta_{\rm dis} \cos\theta_1)}$ and $D_{\rm ext, in} = \frac{1}{\Gamma_{\rm ext}(1-\beta_{\rm ext}\cos\theta_{\rm in})}$. Furthermore, the scattering angle, another crucial factor in determining EIC, can be expressed as
\begin{equation}
\label{A_Eq_costhe_s}
    \cos\theta'_{\rm s} = \sin\theta'_{\rm L}\sin\theta'_{\rm in} \cos\phi'_{\rm in} + \cos\theta'_{\rm L} \cos\theta'_{\rm in},
\end{equation}
where $\theta'_{\rm in} = \theta'_3$ and $\phi'_{\rm in}$ represent the polar and azimuthal angles of the incident photon direction. By substituting the photon number density from Eq.(\ref{A_Eq_Bv}) and the scattering angle from Eq.(\ref{A_Eq_costhe_s}) into Eq.(\ref{Eq_eic_jet_total}), we can obtain the scattering rate per solid angle at time $t_{\rm s}$ in the source frame:

\begin{equation}
    \mathcal{S}(\nu_{\rm out}, \Omega_{\rm L}, t_{\rm s}) = \frac{D_{\rm ext, out}^2}{\Omega_{\rm j}} \int d\gamma_e' \, \int d\nu'_{\rm in} \, \frac{3\sigma_{\rm T} c}{16\pi{\gamma'_e}^{2}\nu'_{\rm in}}f_{\rm ani}(\nu'_{\rm in}, \nu'_{\rm out}, \gamma'_e, \theta'_{\rm s})  \frac{D_{\rm dis} N''_{\gamma, \rm dis}\mathcal{B}''_{\rm dis}\left( \frac{\nu'_{\rm in}}{\nu'_{\rm p}} \right)}{8\pi R_{\rm dis}R_{\rm ext}|PO| \nu''_{\rm p}} N'_{e, \rm j}(\gamma_e'),
\end{equation}
where $D_{\rm ext, out} = \frac{1}{\Gamma_{\rm ext}(1-\beta_{\rm ext}\cos\theta_{\rm L})}$. All geometric quantities are determined by $t_{\rm s}$: $|PO|$ can be written as
\begin{equation}
    |PO|^2 = R_{\rm dis}^2+R_{\rm ext}^2 - 2R_{\rm dis}R_{\rm ext}\cos\theta_2,
\end{equation}
where the angle $\theta_2$ can be determined from the definition of $t_{\rm s}$:
\begin{equation}
    t_{\rm s} = \int \frac{d|PO|}{c} = \int_{\cos \theta_2}^{1} \frac{R_{\rm dis} R_{\rm ext}}{c\sqrt{R_{\rm dis}^2+R_{\rm ext}^2 - 2R_{\rm dis}R_{\rm ext}\cos\theta}} \, d\cos\theta
\end{equation}

In addition, 
\begin{equation}
    \cos\theta_1 = \frac{R_{\rm ext}^2-R_{\rm dis}^2-|PO|^2}{2R_{\rm dis}|PO|},
\end{equation}

and

\begin{equation}
\label{A_Eq_costhe_3}
    \cos\theta_3 = \frac{|PO|^2+R_{\rm ext}^2-R_{\rm dis}^2}{2R_{\rm ext} |PO|}.
\end{equation}

\section{The approximation of IC scattering spectrum.}
\label{sec_fani_approximation}

The scattering spectrum $f_{\rm ani}$ can be written as

\begin{equation}
    f_{\rm ani} = 1 + \frac{{\xi'}^2}{2(1-\xi')} - \frac{2\xi'}{b'_\theta(1-\xi')} + \frac{2{\xi'}^2}{{b'_\theta}^2(1-\xi')^2},
\end{equation}
where $\xi' = h\nu'_{\rm out}/(\gamma'_e m_e c^2)$ and $b'_{\theta} = 2(1-\cos\theta'_{\rm s})\gamma'_e h\nu'_{\rm in}/(m_e c^2)$. Given $h\nu'_{\rm in} \ll h\nu'_{\rm out} \leqslant \gamma'_e m_e c^2 b'_{\theta}/(1+b'_{\theta})$, we can derive $0 < \frac{\xi'}{b'_{\theta}(1-\xi')} \leqslant 1$. Thus, we have $f_{\rm ani} = \frac{{\xi'}^2}{2(1-\xi')} + h(\theta'_{\rm s})$, where $h(\theta'_{\rm s}) \in [0.5, 1]$. Therefore, $f_{\rm ani}$ is insensitive to $\theta'_{\rm s}$ and can be approximated as $f_{\rm ani} \simeq \frac{{\xi'}^2}{2(1-\xi')} + 1$.

%\section{the scattering rate of EIC from different viewing angles $ \theta_{\rm L} $}
\section{the dependence of the scattering rate on $\theta_{\rm L}$ in the anisotropic and isotropic EIC limits}
\label{sec_EIC_diff_angle_derivation}

%Before proceeding further, let's first discuss the EIC scattering rate from different viewing angles $\theta_{\rm L}$ to help us understand the properties of EIC.

\subsection{anisotropic EIC limit}
\label{sec_th_anisotropic}
In the anisotropic EIC limit, we have $\theta_{\rm s} = \theta_{\rm L}$, and the spectrum of incident photons is given by $n'_{\rm ph}(\nu'_{\rm in}, \Omega'_{\rm in}) = -n'_{\rm ph0}\mathcal{B}'\left(\frac{\nu'_{\rm in}}{\nu'_{\rm p}}\right)\delta(\cos\theta'_{\rm in} - 1)/(2\pi \nu'_{\rm p})$. Substituting this spectrum into Eq.(\ref{Eq_eic_jet_total}), we can derive the EIC scattering rate in the anisotropic EIC limit at $\theta_{\rm L}$:
\begin{equation}
    \label{Eq_eic_ani}
    \mathcal{S}_{\rm ani}(\nu_{\rm out}, \Omega_{\rm L}) = \frac{D_{\rm ext, out}^2}{\Omega_{\rm j}}\int_{\gamma'_{e, \rm min}} \, d\gamma'_e \int_{\nu'_{\rm in, min}(\theta'_{\rm L})} \, d\nu'_{\rm in}  \frac{3\sigma_{\rm T} c B'\left( \frac{\nu'_{\rm in}}{\nu'_{\rm p}} \right) N'_{e, \rm j}(\gamma'_e) n'_{\rm ph0} }{16\pi {\gamma'_e}^2 \nu'_{\rm in} \nu'_{\rm p}}  f_{\rm ani}(\nu'_{\rm in}, \nu'_{\rm out}, \gamma'_e, \theta'_{\rm L}),
\end{equation}
where $\gamma'_{e, \rm min}m_e c^2 = h\nu'_{\rm out}$ and $h\nu'_{\rm in, min}(\theta'_{\rm L}) = \frac{m_e c^2}{2\gamma'_e(1-\cos\theta'_{\rm L})}\frac{\xi'}{1-\xi'}$ (derived from the conservation of energy and momentum conditions of Eq.(\ref{Eq_f_ani_spec})). From Eq.(\ref{Eq_eic_ani}), we can see that $\mathcal{S}_{\rm ani}(\nu_{\rm out}, \Omega_{\rm L})$ will change with $\theta_{\rm L}$ in three ways: the minimum scattering angle $\nu'_{\rm min}$, the spectrum of scattering $f_{\rm ani}(\nu'_{\rm in}, \nu'_{\rm out}, \gamma'_e, \theta'_{\rm L})$, and the Doppler factor $D_{\rm ext, out}$. Since $f_{\rm ani}(\nu'_{\rm in}, \nu'_{\rm out}, \gamma'_e, \theta'_{\rm s})$ is not sensitive to $\theta'_{\rm s}$ (see Sect.\ref{sec_fani_approximation}), we will focus on how $\mathcal{S}_{\rm ani}(\nu_{\rm out}, \Omega_{\rm L})$ changes with $\theta_{\rm L}$ through $\nu'_{\rm min}$ and $D_{\rm ext, out}$: 

(1) $0 \leqslant \theta_{\rm L} \leqslant 1/\Gamma_{\rm ext}$: in this case, $D_{\rm ext, out} \sim \Gamma_{\rm ext}$ can be treated as a constant. Consequently, $\mathcal{S}_{\rm ani}(\nu_{\rm out}, \Omega_{\rm L})$ primarily depends on $\nu'_{\rm in, min}$. A larger $\theta_{\rm L}$ (in the anisotropic limit, $\theta_{\rm s} = \theta_{\rm L}$) leads to a lower $\nu'_{\rm min}$ ($h\nu'_{\rm in, min} = \frac{m_e c^2}{2\gamma_e(1-\cos\theta'_{\rm s})} \frac{\xi'}{1-\xi'}$), allowing more incident photons to participate in the scattering, thereby increasing $\mathcal{S}_{\rm ani}(\nu_{\rm out}, \Omega_{\rm L})$.

(2) $\theta_{\rm L} > 1/\Gamma_{\rm ext}$: in this scenario, $\nu'_{\rm in}$ becomes less sensitive to $\theta_{\rm L}$, as $\theta_{\rm L} = 1/\Gamma_{\rm ext}$ corresponds to $\theta'_{\rm L} = \pi/2$ and $\nu'_{\rm in, min}(\pi) = 0.5 \nu'_{\rm in, min}\left( \frac{\pi}{2} \right)$. Therefore, $\mathcal{S}_{\rm ani}(\nu_{\rm out}, \Omega_{\rm L})$ primarily varies with $D_{\rm ext, out}$. A larger $\theta_{\rm L}$ leads to a smaller $D_{\rm ext, out}$, which increases $\nu'_{\rm out}$  ($\nu'_{\rm out} = \nu_{\rm out}/D_{\rm ext, out}$). This requires more energetic electrons or incident photons to participate in the scattering, resulting in a decrease in $\mathcal{S}_{\rm ani}(\nu_{\rm out}, \Omega_{\rm L})$. Additionally, the Doppler boosting of the scattering rate decreases as $\theta_{\rm L}$ increases ($\mathcal{S}_{\rm ani}(\nu_{\rm out}, \Omega_{\rm L}) = D_{\rm ext, out}^2 \mathcal{S'}_{\rm ani}(\nu'_{\rm out}, \Omega'_{\rm L})$), further reducing $\mathcal{S}_{\rm ani}(\nu_{\rm out}, \Omega_{\rm L})$.

If both the electrons and incident photons follow a power-law spectrum, a semi-analytical solution to Eq.(\ref{Eq_eic_ani}) can be obtained:
\begin{align}
    \label{Eq_eic_ani_ana}
    \mathcal{S}_{\rm ani}(\nu_{\rm out}, \Omega_{\rm L}) &\propto D^2_{\rm ext, out}\int_{\gamma'_{e, \rm min}} d\gamma'_e \, \int_{\nu'_{\rm in, min}} d\nu'_{\rm in} \, {\nu'_{\rm in}}^{-\alpha -1} {\gamma'_e}^{-2-p} f_{\rm ani}(\nu'_{\rm in}, \nu'_{\rm out}, \gamma'_e, \theta'_{\rm L}) \\ \notag
    &\propto D^2_{\rm ext, out}\int_{\gamma'_{e, \rm min}} d\gamma'_e \, \int_{\nu'_{\rm in, min}} d\nu'_{\rm in} \, {\nu'_{\rm in}}^{-\alpha -1} {\gamma'_e}^{-2-p} \left[ \frac{{\xi'}^2}{2(1-\xi')} + 1  \right] \\ \notag
    &\propto D^2_{\rm ext, out}\int_{\gamma'_{e, \rm min}} d\gamma'_e \,  \nu_{\rm in, min}^{'-\alpha} {\gamma'_e}^{-2-p} \left[ \frac{{\xi'}^2}{2(1-\xi')} + 1  \right] \\ \notag
    &\propto (1-\cos \theta'_{\rm L})^{\alpha} D_{\rm ext, out}^{3+p-\alpha} \int^0_1 \xi'^{p-\alpha} \left( \frac{\xi'}{1-\xi'} \right)^{-\alpha} \left[ \frac{{\xi'}^2}{2(1-\xi')} +1 \right] \, d\xi',
\end{align}
where $\alpha$ and $p$ represent the spectral indices of the incident photons and electrons, with $h\nu'_{\rm in, min} = \frac{\xi'}{1-\xi'}\frac{m_e c^2}{2\gamma'_e(1-\cos \theta'_{\rm L})}$ and $\gamma'_{e, \rm min} m_e c^2 = h\nu'_{\rm out}$. The approximation for $f_{\rm ani}$ is applied in Eq.(\ref{Eq_eic_ani_ana}), as described in Sect.\ref{sec_fani_approximation}.

In summary, as $\theta_{\rm L}$ increases, the scattering rate $\mathcal{S}_{\rm ani}(\nu_{\rm out}, \Omega_{\rm L})$ initially increases until $\theta_{\rm L} \sim 1/\Gamma_{\rm ext}$ due to the decrease in $\nu'_{\rm in, min}$, and then decreases with further increases in $\theta_{\rm L}$ as a result of the reduction in $D_{\rm ext, out}$.

\subsection{isotropic EIC limit}
\label{sec_th_isotropic}

In the isotropic EIC limit, the spectrum of incident photons is given by $n'_{\rm ph}(\nu'_{\rm in}, \Omega'_{\rm in}) = n'_{\rm ph0}\mathcal{B}'\left(\frac{\nu'_{\rm in}}{\nu'_{\rm p}}\right)/(4\pi \nu'_{\rm p})$. The EIC scattering rate can then be written as:
\begin{equation}
    \label{Eq_eic_iso}
    \mathcal{S}_{\rm iso}(\nu_{\rm out}, \Omega_{\rm L}) = \frac{D_{\rm ext, out}^2}{\Omega_{\rm j}}\int_{\gamma'_{e, \rm min}} \, d\gamma'_e \int_{\nu'_{\rm in, min}(\pi)} \, d\nu'_{\rm in}  \frac{3\sigma_{\rm T} c B'\left( \frac{\nu'_{\rm in}}{\nu'_{\rm p}} \right) N'_{e, \rm j}(\gamma'_e) n'_{\rm ph0} }{16\pi {\gamma'_e}^2 \nu'_{\rm in} \nu'_{\rm p}}  f_{\rm iso}(\nu'_{\rm in}, \nu'_{\rm out}, \gamma'_e),
\end{equation}
where $f_{\rm iso}(\nu'_{\rm in}, \nu'_{\rm out}, \gamma'_e) = \int^{\cos\theta'_{\rm s, min}}_{-1} f_{\rm ani}(\nu'_{\rm in}, \nu'_{\rm out}, \gamma'_e, \theta'_{\rm s}) \, d\cos\theta'_{\rm s}/2$ and $\cos\theta'_{\rm s, min} = 1 - \frac{m_e c^2}{2\gamma'_{e} h\nu'_{\rm in}} \frac{\xi'}{1-\xi'}$. From Eq.(\ref{Eq_eic_iso}), we observe that the scattering rate of isotropic EIC limit varies only through the Doppler factor $ D_{\rm ext, out} $, which affects both $\nu'_{\rm out}$ and the boosting of the scattering rate, similar to the discussion in anisotropic EIC limit scenario. The difference is that $\theta'_{\rm s}$ is no longer dependent on $\theta'_{\rm L}$ in the isotropic EIC limit, but instead ranges from $\theta'_{\rm s, min}$ to $\pi$. Consequently, the isotropic scattering rate exhibits a slow decrease for $ \theta_{\rm L} \leqslant 1/\Gamma_{\rm ext} $ and a rapid decrease for $ \theta_{\rm L} > 1/\Gamma_{\rm ext} $.

\section{The ratio $\mathcal{S}_{\rm ani}(\nu_{\rm out})/\mathcal{S}_{\rm iso}(\nu_{\rm out})$}

\label{sec_compare_ani_iso}

In this section, we compare the total scattering rate $\mathcal{S}(\nu_{\rm out})$ integrated over $\Omega_{\rm L}$ in both the anisotropic and isotropic EIC limits. As discussed in Sect.\ref{sec_EIC_diff_angle_derivation}, the majority of the scattering rate is contributed by electrons from $\theta_{\rm L} \leqslant 1/\Gamma_{\rm ext}$. Therefore, for simplicity, we focus on $\mathcal{S}_{\rm ani}(\nu_{\rm out})$ and $\mathcal{S}_{\rm iso}(\nu_{\rm out})$ integrated over $0 \leqslant \theta_{\rm L} \leqslant 1/\Gamma_{\rm ext}$ in this section.

We begin by considering the simplest case, where both the electrons and incident photons are monoenergetic. In this scenario, the incident photon spectrum is given by $n'_{\rm ph}(\nu'_{\rm in}, \Omega'_{\rm in}) = -n'_{\rm ph0} \delta(\nu'_{\rm in} - \nu'_{\rm in0}) \delta(\cos\theta'_{\rm in} - 1)/(2\pi)$ for the anisotropic case, and by $n'_{\rm ph}(\nu'_{\rm in}, \Omega'_{\rm in}) = n'_{\rm ph0} \delta(\nu'_{\rm in} - \nu'_{\rm in0})/(4\pi)$ for the isotropic case. The electron spectrum is $N'_{e, \rm j}(\gamma'_e) = N'_{e, \rm j0} \delta(\gamma'_e - \gamma'_{e0})$. Considering that $\theta_{\rm s} = \theta_{\rm L}$ in the anisotropic EIC limit and using Eq.(\ref{Eq_eic_jet_total}), we obtain
\begin{align}
    \label{eq_eic_me_mph_ani}
    \mathcal{S}_{\rm ani}(\nu_{\rm out}) &= \int \mathcal{S}_{\rm ani}(\nu_{\rm out}, \Omega_{\rm L}) \, d\Omega_{\rm L} \\ \notag
    &= \int  \frac{D^2_{\rm ext}}{\Omega_{\rm j}} \frac{3\sigma_{\rm T} c N'_{e, \rm j0} n'_{\rm ph0}}{16\pi {\gamma'_{e0}}^2 \nu'_{\rm in0}} f_{\rm ani}(\nu'_{\rm in0}, \nu'_{\rm out}, \gamma'_{e0}, \theta'_{\rm L}) \, d\Omega_{\rm L} \\ \notag
    &\sim -\frac{\Gamma^2_{\rm ext}}{\Omega_{\rm j}} \frac{3\sigma_{\rm T} c N'_{e, \rm j0} n'_{\rm ph0}}{8{\gamma'_{e0}}^2 \nu'_{\rm in0}} \int^{\cos\frac{1}{\Gamma_{\rm ext}}}_{\cos\theta_{\rm min}}  f_{\rm ani}(\nu'_{\rm in0}, \nu'_{\rm out}, \gamma'_{e0}, \theta'_{\rm L}) \, d\cos\theta_{\rm L}, \,\,\,\,\,\,\,\, \nu'_{\rm in0} \geqslant \nu'_{\rm in, min}\left( \frac{\pi}{2} \right).
\end{align}

\begin{figure*} [h]
    %\centering
    \includegraphics[width = 0.5\linewidth]{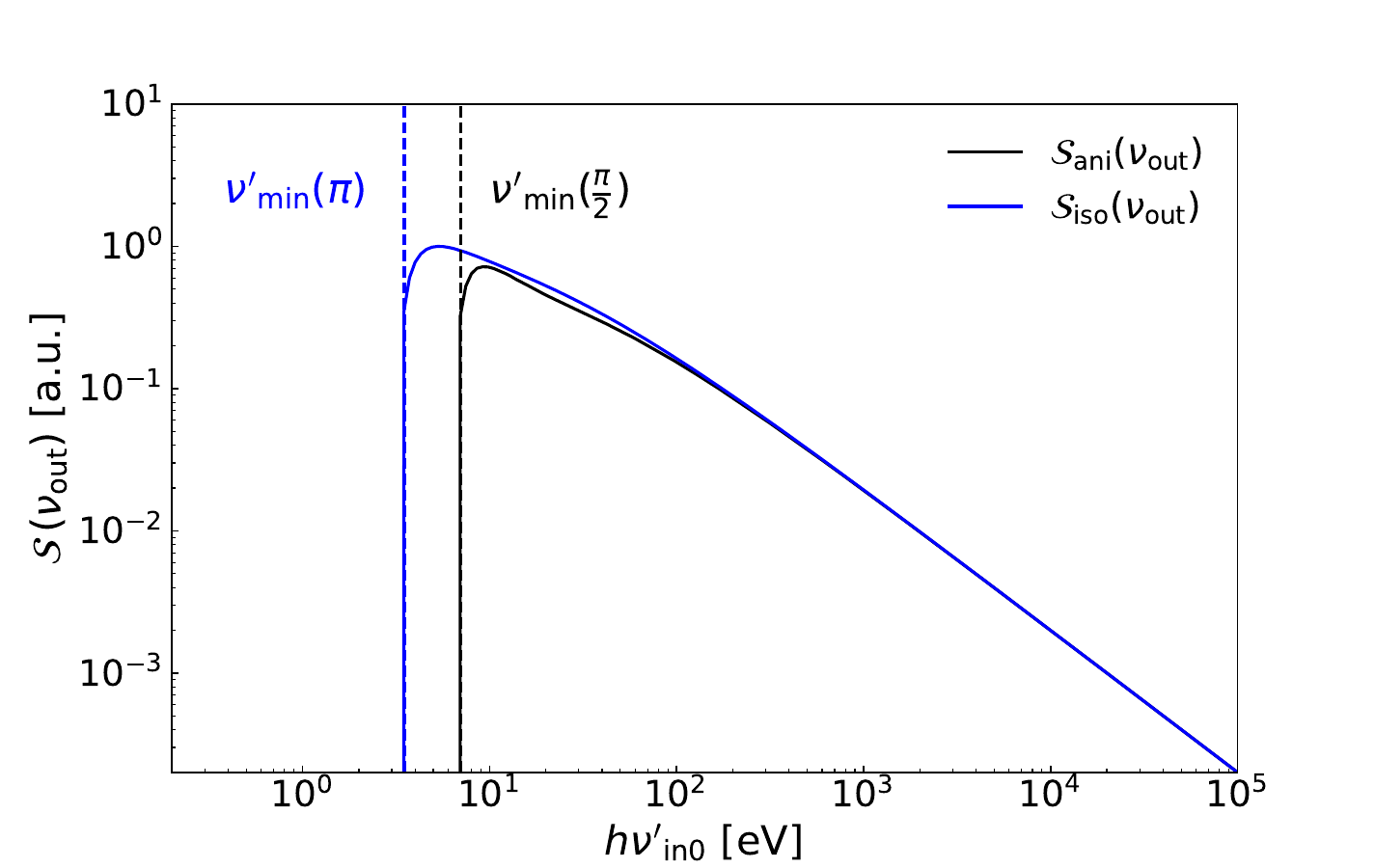}
    \includegraphics[width = 0.5\linewidth]{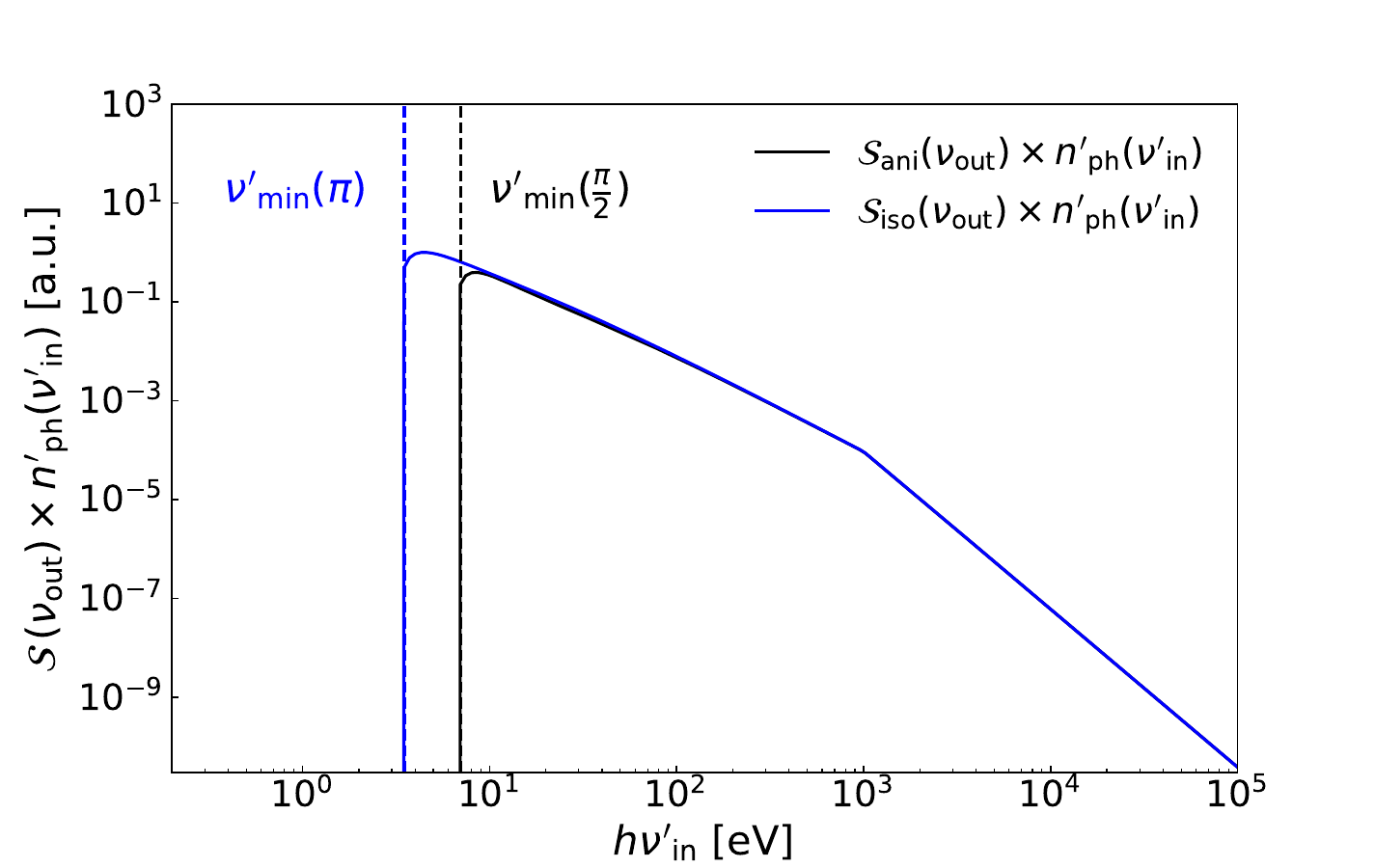}
    \caption{The left panel shows the scattering rate $\mathcal{S}(\nu_{\rm out})$ calculated using Eq.(\ref{eq_eic_me_mph_ani}) and Eq.(\ref{eq_eic_me_mph_iso}), assuming $h\nu'_{\rm out} = 1\, \rm TeV/(2\Gamma_{\rm ext})$, $\gamma'_{e0} m_e c^2 = 5\, \rm TeV/(2\Gamma_{\rm ext})$, and $\Gamma_{\rm ext} = 500$. The right panel shows $\mathcal{S}(\nu_{\rm out})$ (Eq.(\ref{eq_eic_me_mph_ani}) and Eq.(\ref{eq_eic_me_mph_iso})) multiplied by the Band function, assuming $\alpha_{\rm l} = 1$, $\alpha_{\rm h} = 2$, and $h\nu_{\rm p} = 1 \, \rm MeV$. Note that $\mathcal{S}(\nu_{\rm out})$ is calculated under the assumption that both electrons and photons are mono-energetic.}
    \label{fig_12}
\end{figure*}

We note that the condition $\nu'_{\rm in0} \geqslant\nu'_{\rm in, min}\left( \frac{\pi}{2} \right)$ must be ensured, as we are only considering contributions from electrons within $\theta_{\rm L} \leqslant 1/\Gamma_{\rm ext}$, which implies $\theta'_{\rm s} = \theta'_{\rm L} \leqslant \frac{\pi}{2}$. In addition, in the isotropic EIC limit, the spectrum of incident photons is described by $n'_{\rm ph}(\nu'_{\rm in}, \Omega'_{\rm in}) = n'_{\rm ph0} \delta(\nu'_{\rm in} - \nu'_{\rm in0})/(4\pi)$, leading to

\begin{equation}
    \label{eq_eic_me_mph_iso}
    \mathcal{S}_{\rm iso}(\nu_{\rm out}) \sim -\frac{\Gamma^2_{\rm ext}}{\Omega_{\rm j}} \frac{3\sigma_{\rm T} c N'_{e, \rm j0} n'_{\rm ph0} }{8{\gamma'_{e0}}^2 \nu'_{\rm in0}} \int^{\cos\frac{1}{\Gamma_{\rm ext}}}_1  f_{\rm iso}(\nu'_{\rm in0}, \nu'_{\rm out}, \gamma'_{e0}) \, d\cos\theta_{\rm L}, \,\,\,\,\,\,\,\, \nu'_{\rm in0} \geqslant \nu'_{\rm in, min}(\pi).
\end{equation}
Eq.(\ref{eq_eic_me_mph_ani}) and Eq.(\ref{eq_eic_me_mph_iso}) are shown in the left panel of Fig.\ref{fig_12}. Then, for $\nu'_{\rm in0} \geqslant \nu'_{\rm in, min}\left( \frac{\pi}{2} \right)$, we can derive the ratio of the scattering rates between anisotropic and isotropic scattering as follows:
\begin{equation}
\label{Aeq_chi}
    \chi \equiv \frac{\mathcal{S}_{\rm ani}(\nu_{\rm out})}{\mathcal{S}_{\rm iso}(\nu_{\rm out})} \sim \frac{\int_{\cos\theta_{\rm s, min}}^{\cos\frac{1}{\Gamma_{\rm ext}}} f_{\rm ani}(\nu'_{\rm in0}, \nu'_{\rm out}, \gamma'_{e0}, \theta'_{\rm L}) \, d\cos\theta_{\rm L}}{\int_1^{\cos\frac{1}{\Gamma_{\rm ext}}} f_{\rm iso}(\nu'_{\rm in0}, \nu'_{\rm out}, \gamma'_{e0}) \, d\cos\theta_{\rm L}}.\\
\end{equation}

% The suppression factor $\chi$ is displayed in the left panel of Fig.\ref{fig_12}. The right panel shows the ratio of Eq.(\ref{eq_eic_me_mph_ani}) and Eq.(\ref{eq_eic_me_mph_iso}), both multiplied by the Band function, which is used to calculate the suppression factor $\chi$ for mono-energetic electrons scattering with photons that follow the Band function, as explained in Sect.\ref{sec_ratio_ani_iso_EIC}.

For a more realistic estimate, we assume the incident photons follow the Band function while still considering mono-energetic electrons. The scattering rates are then proportional to Eq.(\ref{eq_eic_me_mph_ani}) and Eq.(\ref{eq_eic_me_mph_iso}), both multiplied by the Band function (as shown in the right panel of Fig.\ref{fig_12}) and integrated over $\nu'_{\rm in}$ (see Sect.\ref{sec_ratio_ani_iso_EIC}).

\clearpage

\section{The ratio $F^{\rm EIC}_{\nu_{\rm out}}/F^{\rm SSC}_{\nu_{\rm out}}$ at $t_{\rm d}$}

\label{sec_EIC_effect}

In this section, we assume a uniform circumburst environment to compare the TeV flux at deceleration time between two scenarios: one excludes EIC, with TeV flux produced only through the SSC process (\(F^{\rm SSC}_{\nu_{\rm out}}\)); the other includes EIC, where \(F^{\rm EIC}_{\nu_{\rm out}}\) dominates the TeV emission at \(t_{\rm d}\).

(1) $F^{\rm SSC}_{\nu_{\rm out}}$ (excluding EIC): before shock deceleration, the magnetic field is given by $B' = \sqrt{32\pi m_p\epsilon_B n}\Gamma_{\rm ext, 0}c$, and the external radius by $R_{\rm ext} = 2\Gamma_{\rm ext, 0}^2 ct$. Using the description in \cite{Sari2001}, we derive the SSC spectrum, which has two breaks at $\nu^{\rm SSC}_{\rm m}$ and $\nu^{\rm SSC}_{\rm c}$. The break frequency $\nu^{\rm SSC}_{\rm m} = 2{\gamma'_{\rm m}}^2 \nu_{\rm m}$, where $\gamma'_{\rm m} = \epsilon_e \left( \frac{p-2}{p-1} \right) \frac{m_p}{m_e} \Gamma_{\rm ext, 0}$ and $\nu_{\rm m} = \frac{eB'}{2\pi m_e c} \Gamma_{\rm ext, 0}{\gamma'_{\rm m}}^2$. Similarly, $\nu^{\rm SSC}_{\rm c} = 2{\gamma'_{\rm c}}^2 \nu_{\rm c}$, where $\gamma'_{\rm c} = \frac{6\pi m_e c}{(1+Y') \sigma_{\rm T} \Gamma_{\rm ext, 0} {B'}^2 t}$, and $\nu_{\rm c} = \frac{eB'}{2\pi m_e c} \Gamma_{\rm ext, 0}{\gamma'_{\rm c}}^2$. 

    In the absence of EIC, the SSC process is marginally in the fast cooling regime at \(t_{\rm d}\), and for \(\epsilon_e \gg \epsilon_B\) (see Sect.\ref{sec_ana_shock_EIC_cooling}), we have \(1+Y' \simeq Y' \simeq \sqrt{\frac{\epsilon_e}{\epsilon_B}}\) \citep{Sari2001}. The peak flux of SSC is $F^{\rm SSC}_{\rm m} = \tau_{\rm IC} F^{\rm SYN}_{\rm m}$, where the optical depth for IC scattering is $\tau_{\rm IC} = \frac{\sigma_{\rm T} N_e}{4\pi R^2_{\rm ext}} = \frac{\sigma_{\rm T} n R_{\rm ext}}{3}$, and the peak flux of synchrotron is $F^{\rm SYN}_{\rm m} = \frac{N_e P^{\rm SYN}_{\rm m}}{4 \pi D_{\rm L}^2}$. The peak synchrotron spectral power is $P^{\rm SYN}_{\rm m} = \frac{m_e c^2 \sigma_{\rm T}}{3e}\Gamma_{\rm ext, 0} B'$ \citep{Sari98}. Thus, we derive:

    \begin{align}
        \nu^{\rm SSC}_{\rm m} &\propto n^{\frac{1}{2}} \Gamma_{\rm ext, 0}^6 \epsilon_e^4 \epsilon_B^{\frac{1}{2}} \left( \frac{p-2}{p-1} \right)^4, \\ \notag
        \nu^{\rm SSC}_{\rm c} &\propto n^{-\frac{7}{2}} \Gamma_{\rm ext, 0}^{-10} \epsilon_e^{-2} \epsilon_B^{-\frac{3}{2}} t^{-4}, \\ \notag
        F^{\rm SSC}_{\rm m} &\propto n^{\frac{5}{2}} \Gamma_{\rm ext, 0}^{10} \epsilon_B^{\frac{1}{2}} t^4.
    \end{align}

    According to \cite{LHAASO2023}, the TeV light curve lies in the regime where $\nu_{\rm out} > \max(\nu^{\rm SSC}_{\rm m}, \nu^{\rm SSC}_{\rm c})$. Therefore,

    \begin{align}
        F^{\rm SSC}_{\nu_{\rm out}} &= F^{\rm SSC}_{\rm m} \left( \nu^{\rm SSC}_{\rm m} \right)^{\frac{p-1}{2}} \left( \nu^{\rm SSC}_{\rm c} \right)^{\frac{1}{2}} \nu_{\rm out}^{-\frac{p}{2}} \\ \notag 
        &\propto E_{\rm k0, iso}^{\frac{2}{3}} n^{\frac{-2+3p}{12}} \Gamma_{\rm ext, 0}^{\frac{-10+9p}{3}} \epsilon_e^{-3+2p} \epsilon_B^{\frac{-2+p}{4}} \left( \frac{p-2}{p-1} \right)^{-2+2p} \nu_{\rm out}^{-\frac{p}{2}}
    \end{align}
    at $t = t_{\rm d} = \left( \frac{3E_{\rm k0, iso}}{32\pi n m_p c^5 \Gamma_{\rm ext, 0}^8} \right)^{1/3}$.

(2) $F^{\rm EIC}_{\nu_{\rm out}}$ (including EIC): in this scenario, we derive $F^{\rm EIC}_{\nu_{\rm out}}$ at $t_{\rm d}$ using similar steps as above. Since the electrons are in the fast cooling regime at $t_{\rm d}$, as discussed in Sect.\ref{sec_ana_shock_EIC_cooling}, and using the results from Table.\ref{table_Feic}, we obtain

    \begin{equation}
        F^{\rm EIC}_{\nu_{\rm out}} \propto E_{\rm k0, iso}^{\frac{2}{3}} n^{\frac{1}{3}} \Gamma_{\rm ext, 0}^{\frac{2+3p}{3}} \epsilon_e^{-1+p} \left( \frac{p-2}{p-1} \right)^{-1+p} \nu_{\rm out}^{-\frac{p}{2}}
    \end{equation}
    at $t_{\rm d}$. Therefore, we have
    \begin{equation}
        \frac{F^{\rm EIC}_{\nu_{\rm out}}}{F^{\rm SSC}_{\nu_{\rm out}}} \propto n^{\frac{2-p}{4}} \Gamma_{\rm ext, 0}^{4-2p} \epsilon_e^{2-p} \epsilon_B^{\frac{2-p}{4}} \left( \frac{p-2}{p-1} \right)^{1-p}
    \end{equation}
    at $t_{\rm d}$ for $h\nu_{\rm out} = 1 \, \rm TeV$.

\section{Calculation of optical depth in the two-zone model}
\label{sec_tau_gg_2zone}

\begin{figure} [h]
\centering
\includegraphics[width = 1\linewidth]{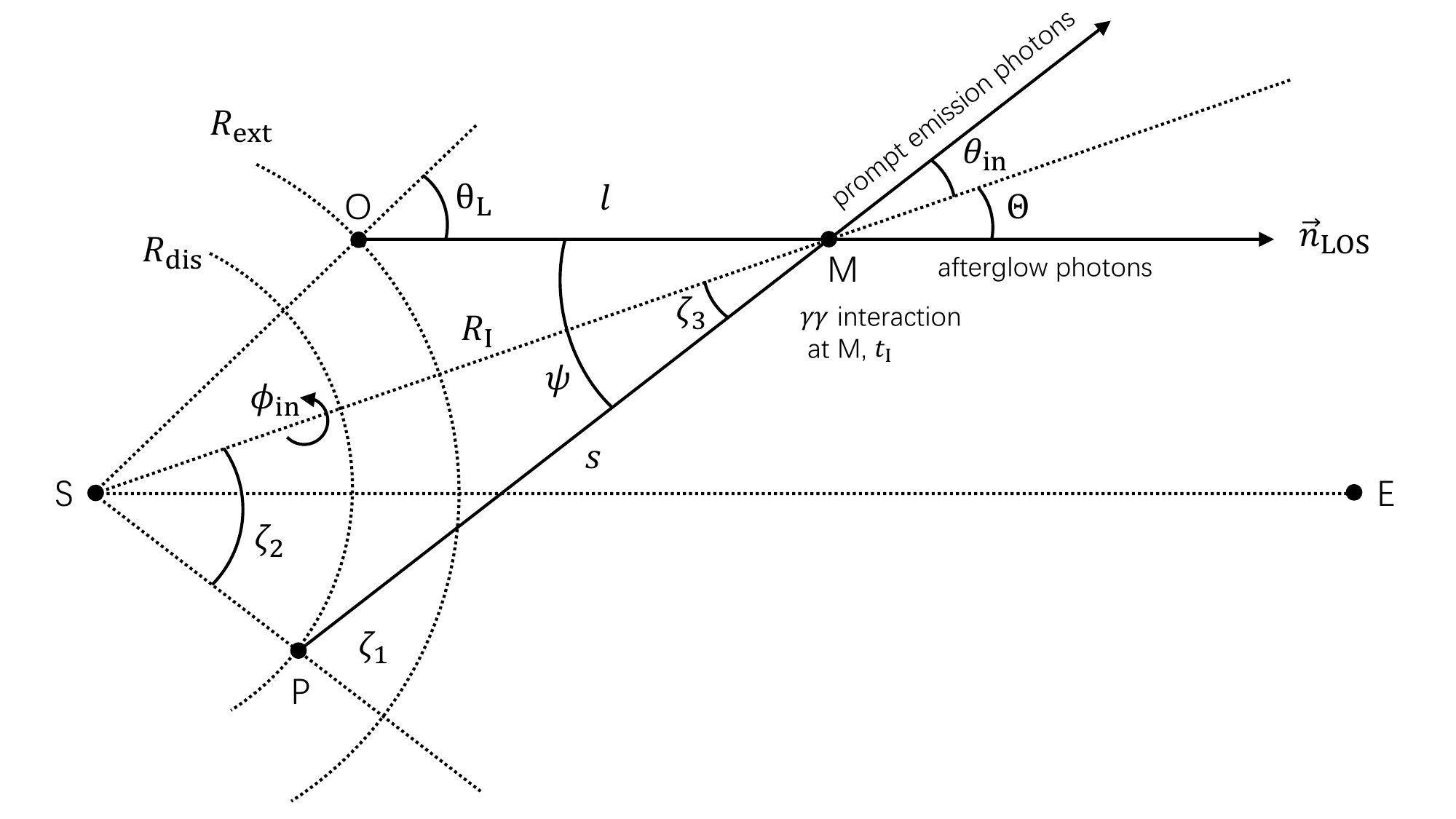}
\caption{The geometry of $\gamma\gamma$ absorption in the two-zone model at a given time $t_{\rm I}$, illustrates the interaction between afterglow and prompt emission photons (see also \cite{Hascoet2012}). Afterglow photons, emitted from $R_{\rm ext}$ at time $t_{\rm e}$ with latitude $\theta_{\rm L}$, travel along LOS and interact with prompt emission photons at point $M$, which are injected from $R_{\rm dis}$ at time $t_0$. Denoting $s$ and $l$ as the path lengths from the injection point to the interaction point for the prompt emission photons and afterglow photons, respectively, we derive the relations $s = c(t_{\rm e} - t_0) + l$ and $t_{\rm I} = t_{\rm e} + l/c = t_0 + s/c$. Note that points $S$, $O$, $M$, and $P$ do not necessarily lie in the same plane.}
\label{fig_gmgm_abs}
\end{figure}

In this section, we calculate the $\gamma \gamma$ absorption in the two-zone model. In this scenario, both prompt and afterglow photons are emitted as "flashes" at $R_{\rm dis}$ and $R_{\rm ext}$, respectively. The optical depth $\tau_{\gamma\gamma}$ of afterglow photons (see Eq.(\ref{Eq_tau_gm_gm})) depends on two key factors: the photon density of the prompt emission $n_{\rm ph}$ and the interaction angle $\cos\psi$ over time (or $l$). The geometry of $\gamma \gamma$ absorption can be referred to Fig.\ref{fig_gmgm_abs}. As afterglow photons move outward, $n_{\rm ph}$ and $\cos\psi$ at the interaction point $M$ change. At any time (or $l$), the photon number density of prompt emission at point $M$ can be given by \citep{Hascoet2012}
\begin{equation}
\label{A_Eq_n1_gg}
    n_{\rm ph}(\nu_{\rm dis}, \Omega_{\rm in}, l) = -\frac{D_{\rm dis}N''_{\gamma, \rm dis} \mathcal{B}''_{\rm dis}\left( \frac{\nu_{\rm dis}}{\nu_{\rm p}} \right)}{16 \pi^2 \nu''_{\rm p}} \frac{1}{sR_{\rm I} R_{\rm dis}} \delta(\cos\theta_{\rm in} - \cos\zeta_3),
\end{equation}
where the quantities $N''_{\gamma, \rm dis}$ and $\mathcal{B}''_{\rm dis} \left( \frac{\nu_{\rm dis}}{\nu_{\rm p}} \right)$ are defined as in Sect.\ref{sec_exa_EIC}. In Eq.(\ref{A_Eq_n1_gg}), $D_{\rm dis} = \frac{1}{\Gamma_{\rm dis}(1-\beta_{\rm dis} \cos\zeta_1)}$, and $\theta_{\rm in}$ refers to the angle between the direction of the prompt emission photons and the vector $\overrightarrow{SM}$.

% where $D_{\rm dis}$, $s$, $R_{\rm I}$, and $\cos\theta_{\rm in}$ depend on $t_{\rm I} = t_{\rm e}+l/c$, see \cite{Hascoet2012}.

To examine how the optical depth is influenced by the anisotropy of the prompt emission photon field, we normalize the photon number density at $l = 0$ (see Sect.\ref{Sec_abs_twozone_R}), i.e., $\int d\nu_{\rm dis}  \int d\Omega_{\rm in} \, h\nu_{\rm dis} n_{\rm ph}(\nu_{\rm dis}, \Omega_{\rm in}, 0) = \frac{L_{\gamma, 0}}{4\pi R_{\rm ext}^2c}$, where $L_{\gamma, 0} \,(\rm erg \, s^{-1})$ represents the luminosity of the incident photons at $l = 0$. Then we can get $N''_{\gamma, \rm dis}$ from the normalization at $l = 0$, and bring it into Eq.(\ref{A_Eq_n1_gg}), we can derive 

% If we observe that the prompt emission overlap with TeV afterglow during early time, then $\theta_1$ should within $1/\Gamma_{\rm dis}$. Hence we simply set $D_{\rm dis} = 1/\Gamma_{\rm dis}$ at $l = 0$.

% \begin{equation}
% \label{A_Eq_n2_gg}
%     n_{\rm ph}(\nu_{\rm dis}, \Omega_{\rm in}, l) = -\frac{ D_{\rm dis}^2 L_{\gamma} B''\left( \frac{\nu_{\rm dis}}{D_{\rm dis}\nu_{\rm p}''} \right)  }{ 8\pi^2c D_{\rm dis, 0}^3 h\nu_{\rm dis} \nu''_{\rm p}  } \frac{s_0}{sR_{\rm I}R_{\rm ext}} \delta(\cos\theta_{\rm in} - \cos\theta_3).
% \end{equation}

\begin{equation}
\label{A_Eq_n2_gg}
    n_{\rm ph}(\nu_{\rm dis}, \Omega_{\rm in}, l) = -\frac{D_{\rm dis} L_{\gamma, 0} \mathcal{B}''_{\rm dis}\left( \frac{\nu_{\rm dis}}{\nu_{\rm p}} \right) }{ 8 \pi^2 c D_{\rm dis, 0} \int h\nu_{\rm dis} B''_{\rm dis} \left( \frac{\nu_{\rm dis}}{\nu_{\rm p}} \right) \, d\nu_{\rm dis}} \frac{s_0}{s R_{\rm I} R_{\rm ext}} \delta(\cos \theta_{\rm in} - \cos \zeta_3),
\end{equation}
where $D_{\rm dis, 0} = D_{\rm dis}(l = 0)$ and $s_0 = s(l = 0)$.

Another key factor affect $\tau_{\gamma \gamma}$ is the scattering angle $\psi$, which can be expressded as

\begin{equation}
\label{A_Eq_cospsi}
    \cos\psi =  \sin\Theta \sin\theta_{\rm in} \cos\phi_{\rm in} + \cos\Theta \cos\theta_{\rm in},
\end{equation}

% where $\Theta$, $\theta_{\rm in}$ depend on $t_{\rm I} = t_{\rm e}+l/c$ (see \cite{Hascoet2012}).
All the geometric quantity need in Eq.(\ref{A_Eq_n2_gg}) and Eq.(\ref{A_Eq_cospsi}) can be determined by $R_{\rm dis}$, $R_{\rm ext}$, $\theta_{\rm L}$ and $l$:

\begin{align}
\label{A_Eq_geo_gg}
    &R_{\rm I} = \sqrt{R_{\rm ext}^2+l^2+2R_{\rm ext}l\cos\theta_{\rm L}},\\ \notag
    &\sin\Theta = \frac{R_{\rm ext}}{R_{\rm I}} \sin\theta_{\rm L},\\ \notag
    &s = c(t_{\rm e}-t_0 +l/c),\\ \notag
    &\cos\theta_{\rm in} = \frac{R_{\rm I}^2+s^2-R_{\rm dis}^2}{2R_{\rm I}s},\\ \notag
    &\cos\zeta_1 = \frac{R_{\rm I}^2 - R_{\rm dis}^2 - s^2}{2R_{\rm dis} s}.\\ \notag
\end{align}

% With Eq.(\ref{A_Eq_n2_gg}), Eq.(\ref{A_Eq_cospsi}) and Eq.(\ref{A_Eq_geo_gg}), we can calculate how $R_{\rm dis}/R_{\rm ext}$ and $\Gamma_{\rm dis}$ affect $\tau_{\gamma\gamma}$ by changing the anisotropic degree of prompt emission photons field.

\clearpage

% \clearpage
% \appendix 
% %\appendices \label{Asec_appendix}
% %\section*{APPENDIX A} \label{Asec_appendix}
% \setcounter{equation}{0}
% \renewcommand\theequation{A\arabic{equation}}
% %\section{method} \label{Asec_appendix}

% \section{}

\clearpage

\bibliography{sample631}{}
\bibliographystyle{aasjournal}

\end{document}